\documentclass[journal]{new-aiaa}

\usepackage{graphicx}
\usepackage{color}
\usepackage{amsmath}
\usepackage[tight,footnotesize]{subfigure}
\usepackage{float}
\usepackage{multirow}
\usepackage{booktabs}
\usepackage{subeqnarray}
\usepackage{epstopdf}
\usepackage{wrapfig}
\usepackage{soul}
\usepackage{siunitx}
\usepackage{longtable,tabularx}
\usepackage{natbib}
\usepackage{algorithm}
\usepackage[noend]{algorithmic}

\usepackage{tikz}
\usetikzlibrary{shapes,arrows,positioning,calc}
\tikzset{
	block/.style = {draw, fill=white, rectangle, minimum height=3em, minimum width=4em},
	alertblock/.style = {draw, color = red, 
		fill=white, rectangle, minimum height=3em, minimum width=4em},
	tmp/.style  = {coordinate}, 
	sum/.style= {draw, fill=white, circle, node distance=1cm},
	input/.style = {coordinate},
	output/.style= {coordinate},
	pinstyle/.style = {pin edge={to-,thin,black}}
}

\usepackage{setspace}

\newtheorem{prob}{Problem}
\newcommand{\real}[1][]{\mathbb{R}^{#1}}                                
\newcommand{\nat}[1][]{\mathbb{N}^{#1}}                                 

\newcommand{\defeq}{:=}                                                 
\newcommand{\msub}[1]{_\mathrm{#1}}                                     

\newcommand{\myfracB}[2]{{#1}/{#2}}                                     

\newcommand{\df}{\mathrm{d}}                                            
\newcommand{\parder}[2]{\frac{\partial #1}{\partial #2}}                

\newcommand{\clint}[2]{\left[#1, #2\right]}                             

\renewcommand{\leq}{\leqslant}                                          

\def\engE{\textsc{~e}}

\newcommand{\E}[2]{\mathbb{E}_{#2}\left[ #1 \right]}

\newcommand{\mydef}[1]{{\textit{#1}}}

\newcommand{\revrmv}[1]{\textcolor{red}{\st{#1}}}
\newcommand{\revnew}[1]{\textcolor{red}{#1}}

\newcommand{\revrpl}[2]{\st{#1}\textcolor{red}{{#2}}}

\newcommand{\revhl}[1]{\hl{#1}}

\newcommand{\matlab}{MATLAB\textsuperscript{\textregistered}}

\newcommand{\figpath}{../Figures}

\newcommand{\eqnnt}[1]{\hyperref[#1]{(\ref*{#1})}}
\newcommand{\eqnsnt}[2]{\hyperref[#1]{(\ref*{#1})}
	and~\hyperref[#2]{(\ref*{#2})}}
\newcommand{\eqnsernt}[2]{\hyperref[#1]{(\ref*{#1})}--\hyperref[#2]{(\ref*{#2})}}

\newcommand{\eqn}[1]{\hyperref[#1]{Eqn.~(\ref*{#1})}}
\newcommand{\eqns}[2]{\hyperref[#1]{Eqns.~(\ref*{#1})} and~\hyperref[#2]{(\ref*{#2})}}
\newcommand{\eqnser}[2]{\hyperref[#1]{Eqns.~(\ref*{#1})}--\hyperref[#2]{(\ref*{#2})}}
\newcommand{\eqnf}[1]{\hyperref[#1]{Equation~(\ref*{#1})}}
\newcommand{\eqnfs}[2]{\hyperref[#1]{Equations~(\ref*{#1})} and~\hyperref[#2]{(\ref*{#2})}}

\newcommand{\scn}[1]{\hyperref[#1]{Sec.~\ref*{#1}}}
\newcommand{\scns}[2]{\hyperref[#1]{Secs.~\ref*{#1}} and~\hyperref[#2]{\ref*{#2}}}
\newcommand{\scnser}[2]{\hyperref[#1]{Secs~.\ref*{#1}}--\hyperref[#2]{\ref*{#2}}}

\newcommand{\fig}[1]{\hyperref[#1]{Fig.~\ref*{#1}}}
\newcommand{\figs}[2]{\hyperref[#1]{Figs.~\ref*{#1}} and~\hyperref[#2]{\ref*{#2}}}
\newcommand{\figser}[2]{\hyperref[#1]{Figs.~\ref*{#1}}--\hyperref[#2]{\ref*{#2}}}
\newcommand{\figf}[1]{\hyperref[#1]{Figure~\ref*{#1}}}
\newcommand{\figfs}[2]{\hyperref[#1]{Figures~\ref*{#1}} and~\hyperref[#2]{\ref*{#2}}}
\newcommand{\figfser}[2]{\hyperref[#1]{Figures~\ref*{#1}}--\hyperref[#2]{\ref*{#2}}}

\newcommand{\tbl}[1]{\hyperref[#1]{Table~\ref*{#1}}}
\newcommand{\tbls}[2]{\hyperref[#1]{Tables~\ref*{#1}} and~\hyperref[#2]{\ref*{#2}}}
\newcommand{\tblser}[2]{\hyperref[#1]{Tables~\ref*{#1}}--\hyperref[#2]{\ref*{#2}}}

\newcommand{\apx}[1]{\hyperref[#1]{Appendix~\ref*{#1}}}

\newcommand{\prb}[1]{\hyperref[#1]{Problem~\ref*{#1}}}
\newcommand{\prp}[1]{\hyperref[#1]{Prop.~\ref*{#1}}}
\newcommand{\prpf}[1]{\hyperref[#1]{Proposition~\ref*{#1}}}

\newcommand{\algoref}[1]{\hyperref[#1]{Algorithm~\ref*{#1}}}

\newcommand{\thmref}[1]{\hyperref[#1]{Theorem~\ref*{#1}}}
\newcommand{\thmsref}[2]{\hyperref[#1]{Theorems~\ref*{#1}} and~\hyperref[#2]{\ref*{#2}}}
\newcommand{\thmserref}[2]{\hyperref[#1]{Theorems~\ref*{#1}}--\hyperref[#2]{\ref*{#2}}}

\newcommand{\algline}[1]{\hyperref[#1]{Line~\ref*{#1}}}
\newcommand{\alglines}[2]{\hyperref[#1]{Lines~\ref*{#1}} and~\hyperref[#2]{\ref*{#2}}}
\newcommand{\alglineser}[2]{\hyperref[#1]{Lines~\ref*{#1}}--\hyperref[#2]{\ref*{#2}}}

\renewcommand{\vec}[1]{\boldsymbol{#1}}

\newcommand{\probab}[1]{\mathbb{P}(#1)}

\newcommand{\threat}{c}
\newcommand{\threatState}{\vec{\Theta}}

\newcommand{\gen}{G_{\nnParamGen}}
\newcommand{\dsc}{D_{\nnParamDsc}}

\newcommand{\enc}{E_{\nnParamEnc}}
\newcommand{\dec}{G_{\nnParamDec}}
\def\encoder{E}
\def\decoder{G}

\def\scaleConstant{\alpha}

\newcommand{\nnParam}{\theta}
\newcommand{\nnParamWt}[1][]{\nnParam_{\mathrm{w}#1}}
\newcommand{\nnParamBs}[1][]{\nnParam_{\mathrm{b}#1}}
\newcommand{\nnParamEnc}{\phi}
\newcommand{\nnParamDec}{\theta}
\newcommand{\nParam}{m}

\newcommand{\nnParamDsc}{\phi}
\newcommand{\nnParamGen}{\theta}

\newcommand{\xState}{\xi}

\newcommand{\wParamsVec}{\vec{\eta}}
\newcommand{\wParams}{\eta}
\newcommand{\unoise}{\omega}
\newcommand{\linstate}{\boldsymbol{q}}
\newcommand{\pdDec}{\mathbb{P}_\nnParamDec(\datum \mid \latentVector)}
\newcommand{\pdEnc}{\mathbb{P}_\nnParamDec(\latentVector \mid \datum) }

\newcommand{\f}{f}
\newcommand{\h}{h}

\newcommand{\tFinal}{T}

\newcommand{\nSample}{K}
\newcommand{\nData}{N\msub{D}}
\newcommand{\nGen}{N\msub{S}}

\newcommand{\datum}{x}

\newcommand{\dataset}{\mathcal{X}}
\newcommand{\datasetGen}{\tilde{\mathcal{X}}}
\newcommand{\datasetSup}{\mathcal{X}\msub{s}}

\newcommand{\yaw}{u}
\newcommand{\dyaw}{\dot{u}}

\newcommand{\pos}{\vec{r}}

\newcommand{\hamiltonian}{H}
\newcommand{\coState}{p}
\newcommand{\costateden}{\nu}

\newcommand{\yOutput}{y}

\newcommand{\latentSpace}{\mathcal{Z}}
\newcommand{\latentVector}{z}
\newcommand{\latentVectorA}{\zeta_1}
\newcommand{\latentVectorB}{\zeta_2}
\newcommand{\latentSpaceBatch}{\mathcal{B}_z}

\newcommand{\loss}{L}
\newcommand{\lossSim}{L\msub{sim}}

\newcommand{\klDiv}[1]{D\msub{KL}\left(#1\right)}

\def\indicator{\mathcal{I}}

\newcommand{\nEpoch}{M_e}
\newcommand{\batchSize}{M_b}
\newcommand{\batch}{\mathcal{B}_x}

\newcommand{\simMet}{\delta}

\newcommand{\nDimData}{N_x}
\newcommand{\posState}{r}

\title{Case Studies of Generative Machine Learning Models
	\\ for Dynamical Systems}

\author{Nachiket U. Bapat,\footnote{Graduate Research Assistant, Aerospace
		Engineering Department. Email: \texttt{nubapat@wpi.edu}} 
		Randy C. Paffenroth,\footnote{Associate Professor,
		Mathematical Sciences Department, Computer Science Department, and Data Science
		Program.} and Raghvendra V. Cowlagi\footnote{Associate Professor, Aerospace
		Engineering Department. AIAA Senior Member. Corresponding Author. 
		Email: \texttt{rvcowlagi@wpi.edu}} }
\affil{Worcester Polytechnic Institute, Worcester, MA, USA.}

\renewcommand{\revrpl}[2]{#2}
\renewcommand{\revnew}[1]{#1}
\renewcommand{\revhl}[1]{#1}
\renewcommand{\revrmv}[1]{}

\def\figpath{.}

\begin{document}

\maketitle

\begin{abstract}
	\vspace{-\baselineskip}
	Systems like aircraft and spacecraft are expensive to operate in the real
	world. The design, validation, and testing for such systems therefore relies on
	a combination of mathematical modeling, abundant numerical simulations, and a
	relatively small set of real-world experiments. Due to modeling errors,
	simplifications, and uncertainties, the data synthesized by simulation models
	often does not match data from the system's real-world operation. We consider
	the broad research question of whether this model mismatch can be significantly
	reduced by generative artificial intelligence models (GAIMs). Loosely speaking,
	a GAIM learns to transform a set of uniformly
	\revnew{or normally} distributed vectors to a
	set of outputs with a distribution similar  to that of a training dataset.
	Unlike text- or image-processing, where generative models have attained recent
	successes, GAIM development for aerospace engineering applications must not
	only train with scarce operational data, but their outputs must also
	\revrpl{conform to}{satisfy} governing equations based on natural laws, e.g.,
	conservation laws. With this motivation, we study GAIMs for dynamical systems.
	The scope of this paper \revrpl{is restricted to}{primarily focuses on} two case studies
	of optimally controlled systems that are commonly understood and employed in
	aircraft guidance, namely: minimum-time navigation in a wind field and
	minimum-exposure navigation in a threat field. For these case studies, we
	report GAIMs that are trained with a relatively small set, \revnew{of the order
	of a few hundred,} of examples and with underlying governing equations. By
	focusing on optimally controlled systems, we formulate \revrpl{governing
	equations}{training loss functions} based on \revrpl{properties}{invariance} of
	the Hamiltonian function along system trajectories. \revnew{As an additional 
		case study, we
	consider GAIMs for high-dimensional linear time-invariant (LTI) systems with
	process noise of unknown statistics. LTI dynamical systems are widely used for
	control design in aerospace engineering.} We investigate three GAIM
	architectures, namely: the generative adversarial network (GAN) and two
	variants of the variational autoencoder (VAE). We provide architectural details
	and thorough performance analyses of these models. The main finding is that our
	new models, especially the VAE-based models, are able to synthesize data that
	\revrpl{conform to}{satisfy} the governing equations and are statistically 
	similar to the training data 
	{despite small volumes of training data.}
\end{abstract}
 
\onehalfspacing
\section*{Nomenclature}

\begin{longtable}{p{0.1\columnwidth} p{0.4\columnwidth}  p{0.1\columnwidth} p{0.3\columnwidth}}
	\toprule
	\textbf{Symbol} & \textbf{Meaning} &
	\textbf{Symbol} & \textbf{Meaning} \\
	\midrule
	GAIM & Generative Artificial Intelligence Model & 
	$\nnParamGen,\nnParamDsc$ & ANN parameters \\
	
	ANN & Artificial Neural Networks & 
	$\nnParamWt, \nnParamBs$ & ANN weights and biases \\
	
	VAE & Variational autoencoder &  
	$\encoder_\nnParamEnc, \decoder_\nnParamDec $ & Encoder and Decoder  \\
	
	GAN & Generative Adversarial Network &  
	$\dsc, \gen$ & Discriminator and Generator \\
	
	$\dataset$ & Observed Training Dataset (OTD) & 	
	$\nData$ & Number of training data points \\
	
	$\datasetGen$ &	 Generated synthetic dataset & 
	$\nGen$ & Number of generated data points \\
	
	$\datum$ & Training datapoint &  $\latentVector$ & Latent vector \\
	
	$\simMet_1,\simMet_2,\simMet_3$ & Performance indices \\
	\bottomrule
\end{longtable}
\setcounter{table}{0}

\doublespacing
\section{Introduction}

Systems like aircraft and spacecraft are expensive to operate in the real world.
The design, validation, and testing of controllers for such systems therefore relies on a
combination of mathematical modeling, abundant numerical simulations, and a
relatively small set of real-world experiments. Simulations are developed by
executing mathematical models, e.g., solutions of state-space differential- or
difference equations of the system, to computationally synthesize data of the
system's operation. These synthetic data are essential due to the scarcity of
real-world operational data. In typical model-based control design methods,
synthetic data may be used for preliminary validation of the controller. 
\revrmv{More recent model-free} 
Reinforcement learning-based control methods need large volumes of synthetic
data during the training phase \cite{kiumarsi2018optimal,kuutti2021survey}.
Other machine learning (ML) methods, such as vision-based object detectors and
classifiers widely used in various aerospace guidance and control applications,
also need large volumes of training
data~\cite{Gupta2022,Sisson2023,Sprockhoff2024}.

The mathematical models used for simulations encode an understanding of the
system's behavior, e.g., geometric constraints and the laws of physics. Almost
without exception, these models involve some simplifications, approximations,
and epistemic uncertainties such as inexact knowledge of the system's
properties. Aleatoric uncertainties such as environmental disturbances may also
be present, and are sometimes approximated within the simulation model.
Nevertheless, due to all of these discrepancies, the data synthesized by
simulation models does not match data from the system's real-world operation,
which is the fundamental problem of \emph{model mismatch.}
\revnew{In closed-loop controlled systems, model mismatch may also arise
due to unknown or partially known objective functions of blackbox controllers.}

On the one hand, system identification (ID) methods alleviate this problem by
tuning various parameters in the simulation model using real-world
data~\cite[pp. 97 -- 155]{Jategaonkar2006}. On the other hand, controllers can
stabilize the system \emph{despite}  model mismatch by robustness and/or online
adaptation~\cite{IoannouSun2012,hovakimyan2010L1}. The caveats are that the
accuracy of system ID relies on real-world data, the scarcity of which is the
root problem, whereas robustness and adaptation invariably degrade performance.
RL-based controllers are known to suffer from real-world performance degradation
due to what is known as a ``reality gap''~\cite{francois2018introduction}, i.e.,
the aforesaid model mismatch. A reduction in the mismatch between synthetic data
and real-world operational data can potentially deliver improvements not only in
control design, but also in other areas such as performance- and  reliability
analyses and digital twin development.

Recent years have witnessed explosive advances in computational data synthesis
through \mydef{generative artificial intelligence models} (GAIMs). Loosely
speaking, a GAIM learns to transform a set of uniformly \revnew{or normally} 
distributed latent
vectors to a set of output vectors with a distribution similar -- for example,
with a small Kullback-Liebler (KL) divergence or Wasserstein distance -- to that
of a training dataset~\cite{nikolenko2021synthetic}. Well-known examples of
GAIMs include GPT-3 and GPT-4 (which underlies the ChatGPT chatbot application),
the image generator DALL-E
\cite{ramesh2022hierarchicaltextconditionalimagegeneration}, the software code
generator GitHub Copilot~\cite{nguyen2022empirical}, and the human face
generator StyleGAN~\cite{melnik2024face}.

\revnew{The application of GAIMs is noted in the area of
	robotics for motion planning, including diffusion models
	to synthesize realistic trajectory distributions. For example, an
	approach to learning visual-motor control policies by modeling action
	sequences using diffusion models is reported~\cite{chi2023diffusion}.
	Instead of directly predicting actions, the policy learns to
	iteratively refine noisy action samples toward expert-like behavior,
	enabling high-quality, multi-modal action generation from raw visual
	observations. Similarly, diffusion-based generative models have been
	explored in aerospace applications for trajectory generation under
	constraints~\cite{presser2024diffusion}.}

Considering the success \revnew{and promise} of GAIMs \revnew{not only in}
image- and natural language processing (NLP), \revnew{but also in robotics,} 
one may consider the broad research
question of whether GAIMs may be developed to reduce the mismatch between
synthetic and real-world data. To investigate this question further, there are
two main issues where GAIM development for aerospace engineering systems
contrasts GAIMs in the image processing and NLP domains: (1) as previously
mentioned, training data is scarce for aerospace systems of our interest, and
(2) these systems are governed by underlying physical and algorithmic
principles, namely, natural laws and control laws.

\paragraph{Contributions:}
With this motivation, in this paper we study the development of generative
artificial intelligence models for dynamical systems. \revnew{Specifically,}
\revnew{we focus on two case studies of optimally controlled systems} that are
commonly understood and employed in aircraft guidance, namely: minimum-time
navigation in a wind field and minimum-exposure navigation in a threat field.
\revnew{Furthermore, we study GAIMs for linear time-invariant systems with 
unknown process noise. LTI dynamical systems with noise are commonly used 
in the design of controllers and estimators for various aerospace engineering 
applications.}

For these case studies, we report GAIMs that are trained with a relatively small
set of examples, \revnew{of the order of a few hundred data points,}
and with underlying governing equations. By choosing to focus on
optimally controlled systems, we formulate governing equations based on
\revnew{invariance} properties of the Hamiltonian function along system 
trajectories.  
\revnew{For these two case studies, the 
first-order variational necessary conditions for optimality state 
that the Hamiltonian remains constant along optimal trajectories.
We study GAIMs for a case where the optimal control \emph{objective}
differs between the model and trajectory examples, i.e., the data do not
\emph{exactly} satisfy the governing equations due to a mismatch in the
underlying controller optimization objective.}

GAIMs are quite recent even within the mainstream domains of NLP and image
processing. Within GAIMs, the sub-topic of physics-informed generative models,
to which this work belongs, is even more recent and under-explored. To the best
of our knowledge, our approach of exploiting the governing optimality properties
of controlled systems in developing GAIMs is novel. \revrpl{We find that
training the GAIMs using these governing equations requires significantly less
data compared to standard methods, which makes these GAIMs suitable for
synthetic data generation using scarce real-world operational data for training.
}{In general, training GAIMs requires large volumes of data, e.g., at least
tens- to hundreds- of thousands of data points for image GAIMs. By contrast, our
approach of incorporating governing equations allows for training GAIMs with
merely hundreds, i.e., two- to three orders of magnitude less than typical, of
training data points.} \revrpl{Therefore,}{Due to their ability to learn from
data as well as governing equations,} the \revrpl{work}{GAIMs} reported in this
paper can \revrpl{lead to the development of}{be used in the future for}
computational data synthesis with lower model mismatch than the state of the art
in aerospace engineering applications.

We investigate three GAIM architectures, namely: the \mydef{generative
	adversarial network} (GAN) and two variants of the \mydef{variational
	autoencoder} (VAE). In what follows we will provide details of these
architectures. We find that the VAE  are better-suited for the desired GAIMs.
However, the GAN -- and its relations to the VAE -- is also worth studying for
other similar applications in future work. To train these GAIMs, we develop new
loss functions 
\revnew{as dicsussed in \scn{sec-gaims}}.

In~\cite{BapatPaffenrothCowlagi2024ACC}, we reported preliminary results about
GAN development for the minimum-time navigation problem. The VAE results
reported in this paper are new and previously unreported.

For the remainder of this paper, we assume that the reader is familiar with the
idea of developing artificial neural networks (ANN) as universal function
approximators~\cite{HORNIK1989359,LESHNO1993861}. Briefly, a single-layer ANN
may be considered a nonlinear function of the form $f(\datum; \nnParam) =
\sigma(\nnParamWt^\intercal \datum + \nnParamBs),$ where $\datum$ is the input,
$\nnParam = (\nnParamWt, \nnParamBs)$ are parameters consisting of weights
$\nnParamWt$ and biases $\nnParamBs,$ and $\sigma$ is a nonlinear activation
function such as the sigmoid function. By extension, a multi-layer or deep ANN
may be considered a sequential composition of nonlinear functions of the form
$f(\datum; \nnParam) = \sigma(\nnParamWt[d]^\intercal \latentVector_{d-1}+
\nnParamBs[d]),$ where $d \in \nat$ is the number of layers, $\latentVector_{1}
\defeq \sigma(\nnParamWt[1]^\intercal \datum + \nnParamBs[1]),$ and
$\latentVector_{k} \defeq \sigma(\nnParamWt[k]^\intercal \latentVector_{k -1} +
\nnParamBs[k])$ for $k = 2, \ldots, d-1.$ The neural network \mydef{learns} or
\mydef{is trained} over a dataset of input-output pairs $\{(\datum^i,
\yOutput^i)\}_{i=1}^{\nData}.$ Training is accomplished by finding parameters
$\nnParam^*$ that minimize a \mydef{loss function} $\loss$:
\begin{align*}
	\nnParam^* \defeq \arg \min_{\nnParam} \loss (\datum, \yOutput, \nnParam).
\end{align*}
The exact form of the loss function depends on the application. 
A common example is the mean square loss function $\ell (\datum, \yOutput, 
\nnParam) \defeq \frac{1}{\nData} \sum_{i=1}^{\nData} 
\| \yOutput^i - f(\datum^i; \nnParam) \|^2.$

\paragraph{\revnew{Background and} Related Work:}
Two widely used GAIM architectures are generative adversarial networks
(GANs)~\cite{goodfellow2014generative}  and variational autoencoders
(VAEs)~\cite{kingma2019introduction-vae} briefly described below.
\revnew{The reader altogether unfamiliar with generative models may refer,
for instance, to~\cite{lamb2021briefintroductiongenerativemodels}
for a tutorial introduction to the subject.}

The VAE is an artificial neural network (ANN)-based generative model that
consists of two multi-layer networks called the \mydef{encoder} and the
\mydef{decoder}, respectively. This architecture is similar to the
\mydef{autoencoder (AE)}, where the encoder $\encoder$ maps its input $\datum$
to a latent vector $\latentVector = \enc(\datum).$ The decoder $\dec$ maps its
input $\latentVector$ to a vector $\dec(\latentVector)$ in the same vector space
as the encoder's input $\datum,$ and the pair is trained to minimize the mean
squared error between $\datum$ and $\decoder(\encoder(\datum)).$ Informally, the
difference between an AE and a VAE is that the VAE encoder maps its input
$\datum$ to a probability distribution
$\probab{\latentVector\mid\datum}$~\cite{kingma2019introduction-vae}. The VAE is
trained to \mydef{regularize} this distribution by minimizing the KL divergence
from $\probab{\latentVector \mid \datum}$ to a multivariate Gaussian
distribution. The decoder is simultaneously trained to map the distribution
$\probab{\latentVector \mid \datum}$ to an output distribution matching that of
the training data.

The GAN consists of two ANNs called the generator $G$ and the discriminator $D$,
respectively. The generator learns to map a latent vector $\latentVector$
sampled from, say, a uniform distribution to an output $G(\latentVector)$ such
that the output distribution matches the training data distribution. The
discriminator is a classifier that maps its input $\datum = G(\latentVector)$ to
a binary output, say, $D(\datum) \in \{0,1\}$ depending on whether $\datum$
belongs to the training data distribution. \revhl{The two ANNs $G$ and $D$ are trained
simultaneously in a zero-sum
game}~\cite{goodfellow2014generative,creswell2018generative}.\revhl{An equilibrium of
this game is a discriminator $D$ that cannot distinguish whether its input is
generated by $G$ or sampled from the training data distribution.}

VAEs are commonly used for image generation \cite{vahdat2020nvae}, and are
recently reported for wheeled mobile robot trajectories~\cite{chen2021trajvae},
feature learning of supercritical airfoils \cite{LiZhangChen2022}, time series
anomaly detection \cite{lin2020anomaly}, and healthcare expert
systems~\cite{deng2019collaborative}. Similarly, GANs are reported for image
generation \cite{goodfellow2014generative} including human facial images and
video~\cite{melnik2024face,Yin2022}. Of direct relevance to this work are time
series generators such as the TimeGAN~\cite{yoon2019time-series-gan} and
TimeVAE~\cite{desai2021timevaevariationalautoencodermultivariate}, which
implement GAN and VAE architectures, respectively to synthesize data with
temporal patterns matching those of the training data. Synthetic data from such
GAIMs is reported in the training of \emph{other} machine learning methods,
e.g., a vision-based RL controller-\cite{rao2020rl-cyclegan}. A comparison
between the quality of synthetic data produced by a VAE and a GAN is reported
in~\cite{chen2021trajvae} for wheeled mobile robot trajectory data.
\revnew{Improvements in VAEs, in particular, are reported using amortized 
(learning-based) optimization techniques \cite{MAL-102,kim2018semi}
using iterative refinement to improve posterior approximation quality
\cite{marino2018iterative,donti2021dc3}. These works, however, focus
on image- and text data, and do not 
consider any underlying governing equations or knowledge of the physics
of the system being learned.
}

The aforesaid literature on VAEs and GANs reports training these GAIMs using
real-world data. Like many other ML models, GAIMs need large volumes of training
data. This is contrary to the situation of our interest, where real-world data
are scarce. A potential remedy to this problem is provided by recent research on
\mydef{physics-informed neural networks (PINNs)}~\cite{raissi2019physics}. PINNs
are ANNs trained to satisfy ordinary- or partial differential equations, which
enables the integration of physics-based equations with data. PINNs have shown
promise in diverse applications, including fluid dynamics, material science, and
biomedical engineering, offering a versatile tool for combining data-driven
insights with domain knowledge in scientific
computing~\cite{mahmoudabadbozchelou2022nn,bharadwaja2022physics,wong2023multiple}. 
More pertinent to this work, a PINN-based approach for vehicle longitudinal 
trajectory prediction is reported in~\cite{geng2023physics}.
\revnew{Additionally, improvement in generalization and physical
accuracy is reported using physics-informed learning.
For example, PINNs are reported for modeling and control of complex robotic systems
in combination with  model-based controllers~\cite{liu2024physics}.
}

\revnew{Physics-informed generative models are recently reported,
	primarily based on GAN architectures for flow-related applications
	e.g.,\cite{yang2020physics,yang2022generative,wu2022navier,WANG2025124812}.
	The key architectural detail in these works is the training of the 
	generator using physics-based loss functions.
	The proposed work falls under the category of physics-informed
	generative models, with the understanding that ``physics'' refers
	to any underlying algebraic or differential equations known to govern
	the system. In comparison to the existing literature, 
	we consider not only the GAN but also a new type of physics-informed
	VAE architecture. Our proposed split latent space architecture
	provides a new way of training GAIMs from data points that do not
	\emph{exactly} satisfy the governing equations.
}


{The rest of this paper is organized as follows. } In
\scn{sec-problem-formulation}, we introduce the problem formulation. In
\scn{sec-gaims}, we describe the proposed generative model architectures. In
\scn{sec-results}, we provide results and discussion on the proposed synthetic
data generating methods, and conclude the paper in \scn{sec-con}.

\section{Problem Formulation}
\label{sec-problem-formulation}

Consider a dynamical system modeled in the standard state space form
\begin{align}
	\dot{\xState}(t) = \f (\xState(t); \wParamsVec),
	\label{eq-state}
\end{align}
where $\xState(t) \in \real[n]$ is the state, $\wParamsVec \in \real[\nParam]$
is a parameter vector, and $\f:\real[n] \rightarrow \real[n]$ is at least
Lipschitz continuous to guarantee existence and uniqueness of solutions
to~\eqnnt{eq-state}. For a given value of $\wParamsVec$ and over a finite time
interval $\clint{0}{\tFinal},$ a \mydef{model trajectory} of this system is a
sufficiently smooth function $\xState: \clint{0}{\tFinal} \rightarrow \real[n]$
that satisfies~\eqnnt{eq-state}. Note that the \emph{system parameters}
$\wParamsVec$ (e.g., aircraft parameters such as mass and moment of inertia, or
environmental parameters such as wind speed) are distinct from the \emph{neural
	network parameters} $\nnParamWt,\nnParamBs$ previously introduced.

An \mydef{observed trajectory} of the system is an output signal $\yOutput(t) \in \real[\ell]$ 
measured during the real-world operation of the system. The distinction between the
\emph{model} trajectory and the \emph{observed} trajectory emphasizes that the real-world 
behavior of the system may differ from the model due to various reasons including 
unmodeled dynamics, unmodeled process noise, and measurement error. 
The output model is $\yOutput(t) = \h(\xState(t); \wParamsVec),$ where 
$\h: \real[n] \rightarrow \real[\ell].$

Consider a finite sequence of time samples $t_1 < t_2 < \ldots < t_{\nSample}$ 
within the interval 
$\clint{0}{\tFinal}.$ A \mydef{datum,} or \revhl{``data point,'' 
$\datum$ consists of the output values of an observed trajectory discretized 
at the aforesaid time samples
and appended with the parameter value~$\wParams$ at which the system is operated,
namely,} $\datum = 
\left(\yOutput(t_1), \yOutput(t_2), \ldots, \yOutput(t_{\nSample}), \wParamsVec \right) \in 
\real[\nDimData],$ where $\nDimData \defeq \ell \nSample + \nParam.$
An \mydef{observed training dataset (OTD)} -- informally, ``real-world'' data -- 
is a set $\dataset = \{\datum_i\}_{i=1}^{\nData},$ 
where $\nData$ is the number of data points in the dataset.
Practically speaking, $\dataset$ may be the outcome 
of experimental observations of the system's operation.

The problem of interest is then formulated as follows:
\begin{prob}
	Given a training dataset $\dataset$ containing $\nData$ data points, 
	synthesize a new dataset $\datasetGen = \{\datum^j\}_{j=1}^{\nGen}$ such that
	$\datasetGen$ is statistically similar to $\dataset.$
	\label{prob-general}
\end{prob}
Implicit in this problem statement is the desire that the synthesis of samples 
in $\datasetGen$ should be computationally efficient, so that $\nGen \gg \nData$
can be made as large as needed.

For purely data-driven GAIMs, statistical similarity between 
the ``real-world'' and generated datasets may be considered as a desired
measure, e.g., a low KL divergence from the distributions of $\datasetGen$ 
to $\dataset.$ 
Because we are interested also in an underlying governing equations,
namely \eqnnt{eq-state}, any similarity measure should also consider 
closeness of the generated dataset to the satisfaction of~\eqnnt{eq-state}.

We consider two specific instances of Problem~1, 
described in \scns{sec-min-time}{sec-min-threat}. \revnew{These problems are
selected due to their widespread applications in path-planning and guidance
for autonomous aircraft. The common salient feature of these problems is that
the optimal solution is exactly characterized by an invariant Hamiltonian.
The Zermelo minimum-time navigation problem (\scn{sec-min-time}) involves
a drift field, e.g., wind, that directly affects the vehicle's motion.
The minimum-exposure problem involves a \emph{threat} field, which affects
the vehicle's motion indirectly through the optimization objective.}

\revnew{In addition to these optimally controlled systems, we consider
LTI systems with unknown process noise. These systems do not have a Hamiltonian
or similar governing condition. Via these systems we demonstrate the broader
applicability and scalability of the proposed GAIMs to high-dimensional state spaces.}

%

\subsection{Zermelo Navigation Problem}
\label{sec-min-time}

Consider the minimum-time motion of a vehicle in a drift (wind) field. 
The vehicle's motion is modeled by simple planar particle kinematics:
\begin{align*}
	\dot{\posState}_1(t) &= V \cos\yaw(t) + w_1(\vec{\posState}), 
	& \dot{\posState}_2(t) &= V \sin\yaw(t) + w_2(\vec{\posState}),
\end{align*}
where we denote by $\vec{\posState} = (\posState_1, \posState_2)$ 
the position vector with coordinates in a prespecified 
inertial Cartesian coordinate axis system, by $\yaw$ the heading angle (direction of the velocity 
vector), by $V$ the constant speed of the vehicle, and by $\vec{w}(\vec{\posState}) = (w_1(\vec{\posState}), 
w_2(\vec{\posState}))$ the position-dependent wind velocity vector field.
For prespecified initial and 
final points $\vec{\posState}_0$ and $\vec{\posState}_1,$ 
we would like to find the time of travel $t^*$ and the 
desired heading profile $\yaw^*(t)$ over the entire interval 
$\clint{0}{t^*}$ such that $\vec{\posState}(0) = 
\vec{\posState}_0$ and $\vec{\posState}(t^*) = \vec{\posState}_1.$ 
The heading angle~$\yaw$ may be considered a control 
input in this model; a typical aircraft autopilot can track desired heading angles.

This problem is often called the \emph{Zermelo navigation problem.} Variational optimal control 
theory provides a semi-analytical solution to this problem, by analyzing 
first-order necessary conditions of optimality. We provide the key results here without
derivation; the reader unfamiliar with variational optimal control is referred to~\cite{Bryson1975}.
Per these conditions, the minimum-time trajectory 
and heading angle profile must satisfy the following differential equations:
\begin{align}
	\dot{\posState}_1^* &= V\cos\yaw^* + w_1(\vec{\posState}^*), \qquad
	\dot{\posState}_2^* = V\sin\yaw^* + w_2(\vec{\posState}^*), \\
	{\dyaw^*} &= \parder{w_2}{\posState_1}(\vec{\posState}^*)\sin^2\yaw^* 
	- \parder{w_1}{\posState_2}(\vec{\posState}^*)\cos^2\yaw^* 
	+ \left(  \parder{w_1}{\posState_1}(\vec{\posState}^*) -
	\parder{w_2}{\posState_2}(\vec{\posState}^*) \right)\sin\yaw^*\cos\yaw^*.
	\label{eq-zermelo-dynamics}
\end{align}

The boundary conditions of these differential equations are $\vec{\posState}(0) = \vec{\posState}_0,$ 
$\yaw(0) = \yaw_0,$ and $\vec{\posState}(t^*) = \vec{\posState}_1,$ where $\yaw_0$ and~$t^*$ 
are numerically determined. These 
conditions formulate a {two-point boundary value problem} (TPBVP), numerical 
solutions to which are well-known. The superscript $^*$ on any variable denotes optimal evolution 
of that variable.

Analysis of the variational necessary conditions and transversality conditions~\cite{Bryson1975}
in the Zermelo problem leads also to the important observation that the \mydef{Hamiltonian}
function $\hamiltonian$ remains zero along any optimal trajectory, namely,
\begin{align}
	\hamiltonian(\vec{\posState}^*, \yaw^*, \vec{\coState}^*) &\defeq 1 + 
	\coState_1^*(t) (V\cos\yaw^*(t) + w_1(\vec{\posState}^*(t))) 
	+ \coState_2^*(t) (V\sin\yaw^*(t) + w_2(\vec{\posState}^*(t))) = 0
	\label{eq-hamiltonian-mintime}
\end{align}
for all $t \in \clint{0}{t^*}.$ Here $\vec{\coState} = (\coState_1, \coState_2)$ are 
so-called \mydef{costate} variables that can be shown to satisfy~\cite{Bryson1975}
\begin{align}
	\tan \yaw^*(t) &= \myfracB{\coState_2^*(t)}{\coState_1^*(t)}, 
	\label{eq-costate-control}\\ 
	\coState_1^*(t) &= -\myfracB{\cos\yaw^*(t)}{\costateden}, &
	\coState_2^*(t) &= -\myfracB{\sin\yaw^*(t)}{\costateden},
	\label{eq-costate}
\end{align}
where $\costateden \defeq V + w_1(\vec{\posState}^*(t))\cos\yaw^*(t) + 
w_2(\vec{\posState}^*(t))\sin\yaw^*(t).$

To instantiate Problem~1, we consider optimal trajectories of the Zermelo
navigation problem described by the state $\xState = (\vec{\posState},\yaw)$ and the 
dynamics~\eqnnt{eq-zermelo-dynamics}. The wind velocity field $\vec{w}$ is a parameter, 
but we need a finite-dimensional representation of $\vec{w}.$ 
To this end, let 
$\{\vec{\posState}^1, \vec{\posState}^2, \ldots, \vec{\posState}^N \}$ be a set of 
locations; we then identify $\wParamsVec \defeq (w_1(\vec{\posState}^1), 
w_1(\vec{\posState}^2), \ldots, 
w_2(\vec{\posState}^1), \ldots, w_2(\vec{\posState}^N) ).$
Finally, the {output} is $\yOutput = (\xState, \vec{\coState}).$

This instantiation of the general
problem has a simplifying benefit that the optimal trajectories
are characterized not only by the differential equation~\eqnnt{eq-zermelo-dynamics},
but also by the \emph{algebraic} equation~\eqnnt{eq-hamiltonian-mintime}. Each datum 
$\datum$ in the OTD consists of output values along 
an optimal trajectory appended with the wind velocity parameter $\wParamsVec.$

We make another simplifying assumption: 
\emph{the model trajectories are identical to the observed trajectories,}
i.e., the model is perfect. At first glance, this assumption seems
to contradict the primary motivation of this study
(scarce experimental data and imperfect models), but we will demonstrate the proposed 
generative models are agnostic to the source of $\dataset.$ That is, whenever experimental
data do become available, we can simply replace our synthetic $\dataset$ with the
experimental data without changing the GAIMs. 
For now, this assumption 
allows for the development of a solution to \prb{prob-general}
without collecting experimental data. Collecting experimental flight test data in windy
conditions would be time-consuming and expensive, but if the model is assumed perfect,
we can synthesize data for the dataset $\dataset$ by numerically 
solving~\eqnnt{eq-zermelo-dynamics}.

Furthermore, this assumption allows us to define an
easier similarity measure than comparing the distributions of $\dataset$ and~$\datasetGen.$
Specifically, we define similarity based on the errors in
satisfying~\eqnser{eq-zermelo-dynamics}{eq-costate} by data points in $\datasetGen.$
For any $\datum \in \dataset,$ the costates, heading angle, and Hamiltonian
at the sample points are denoted $\coState[\datum],$  $\yaw[\datum],$ and
$\hamiltonian[\datum].$

\def\thisfigwidth{0.35\columnwidth}
\def\thisfigspace{0.05\columnwidth}
\begin{figure}
	\centering
	\subfigure{\includegraphics[width=\thisfigwidth]{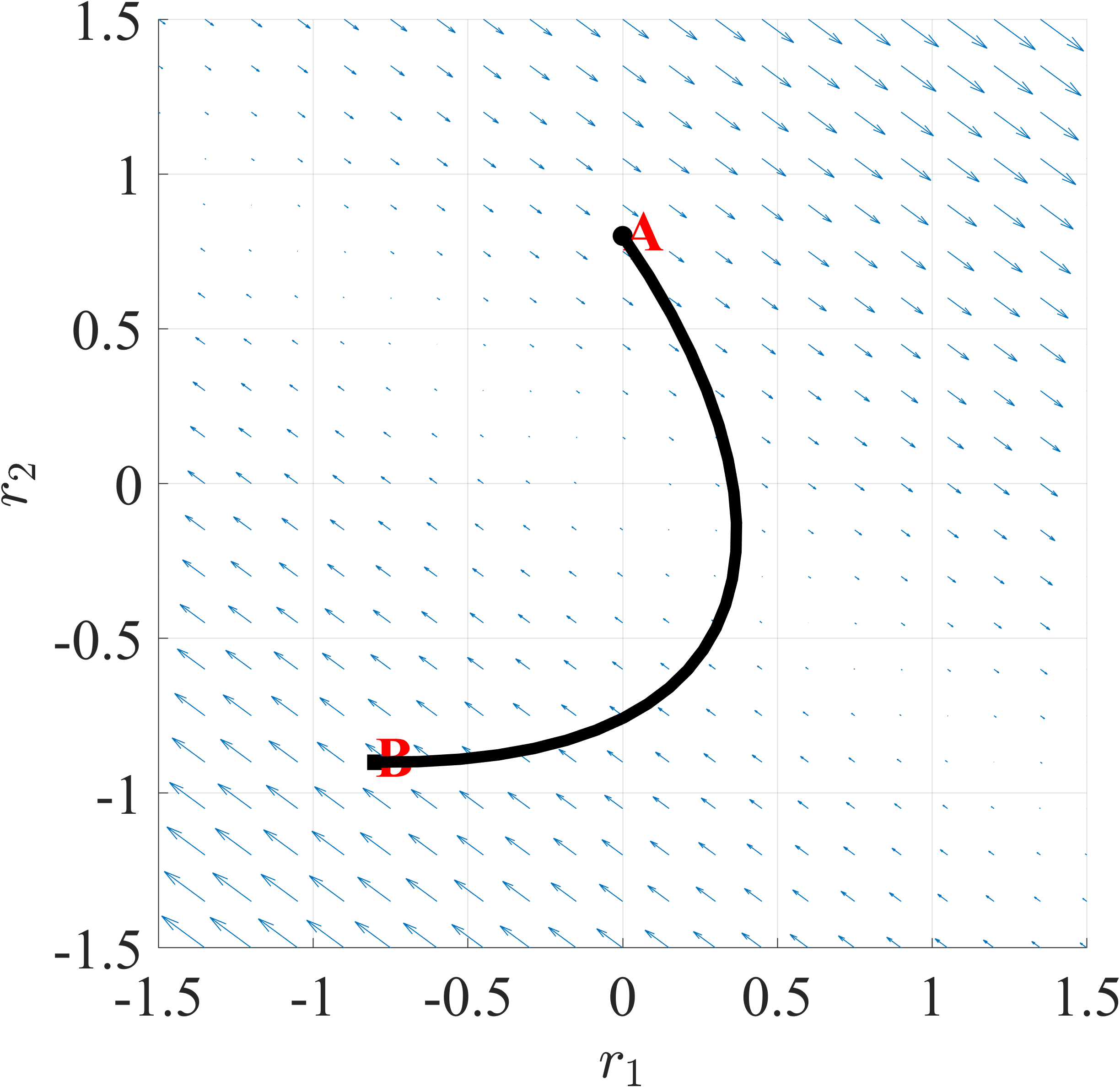}}
	\hspace{\thisfigspace}
	\subfigure{\includegraphics[width=\thisfigwidth]{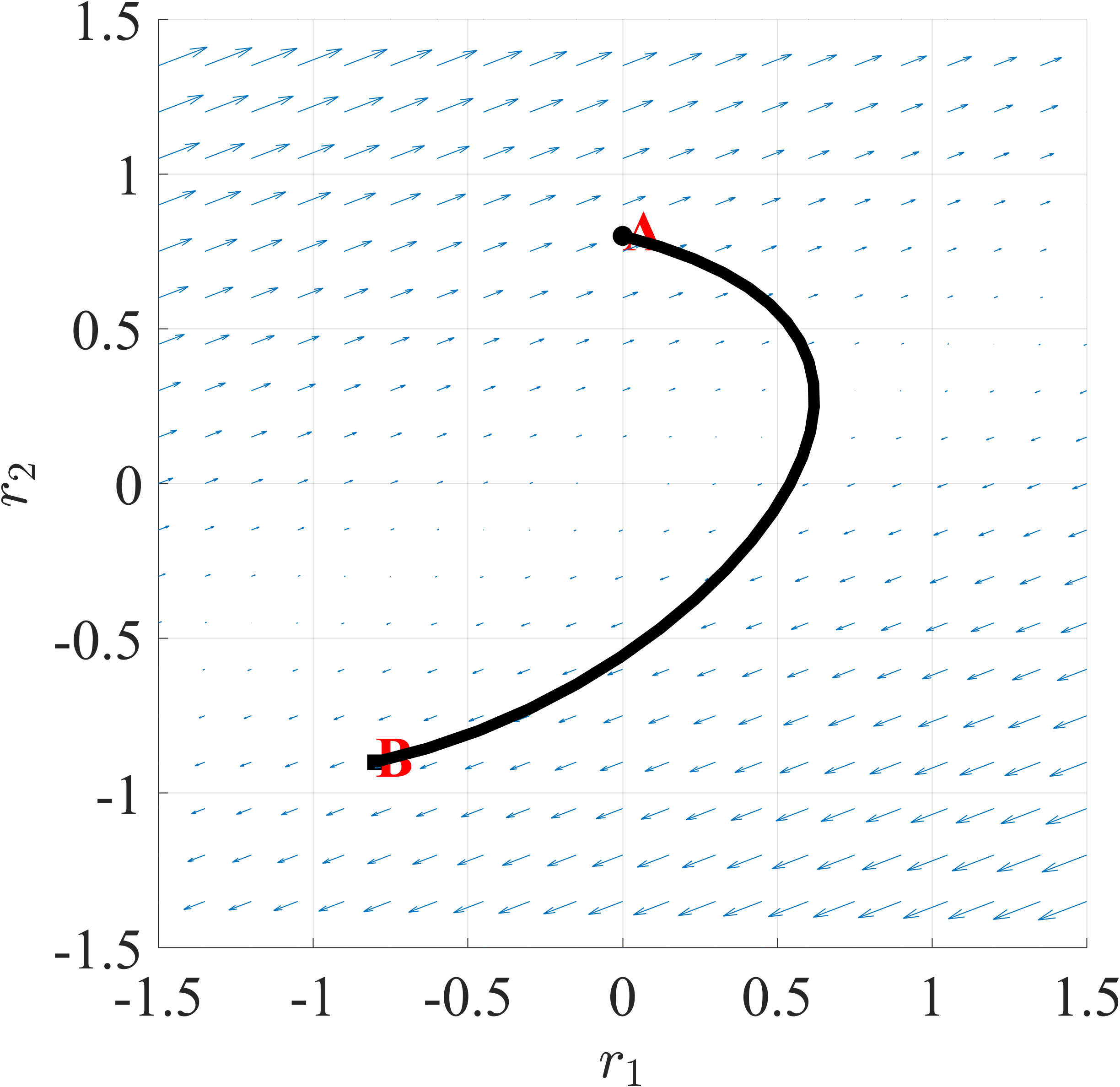}}
	\hspace{\thisfigspace}
	\subfigure{\includegraphics[width=\thisfigwidth]{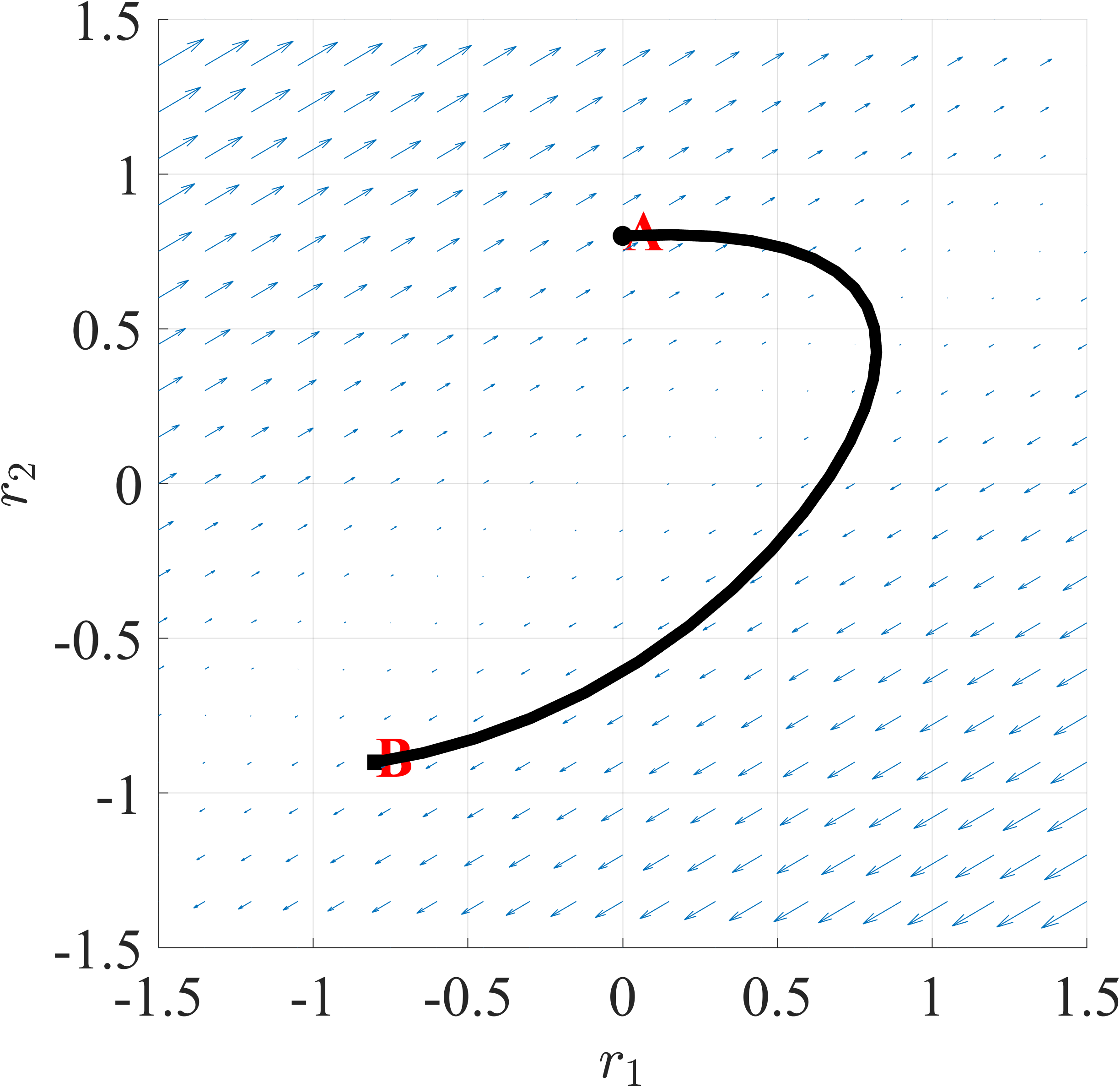}}
	\hspace{\thisfigspace}
	\subfigure{\includegraphics[width=\thisfigwidth]{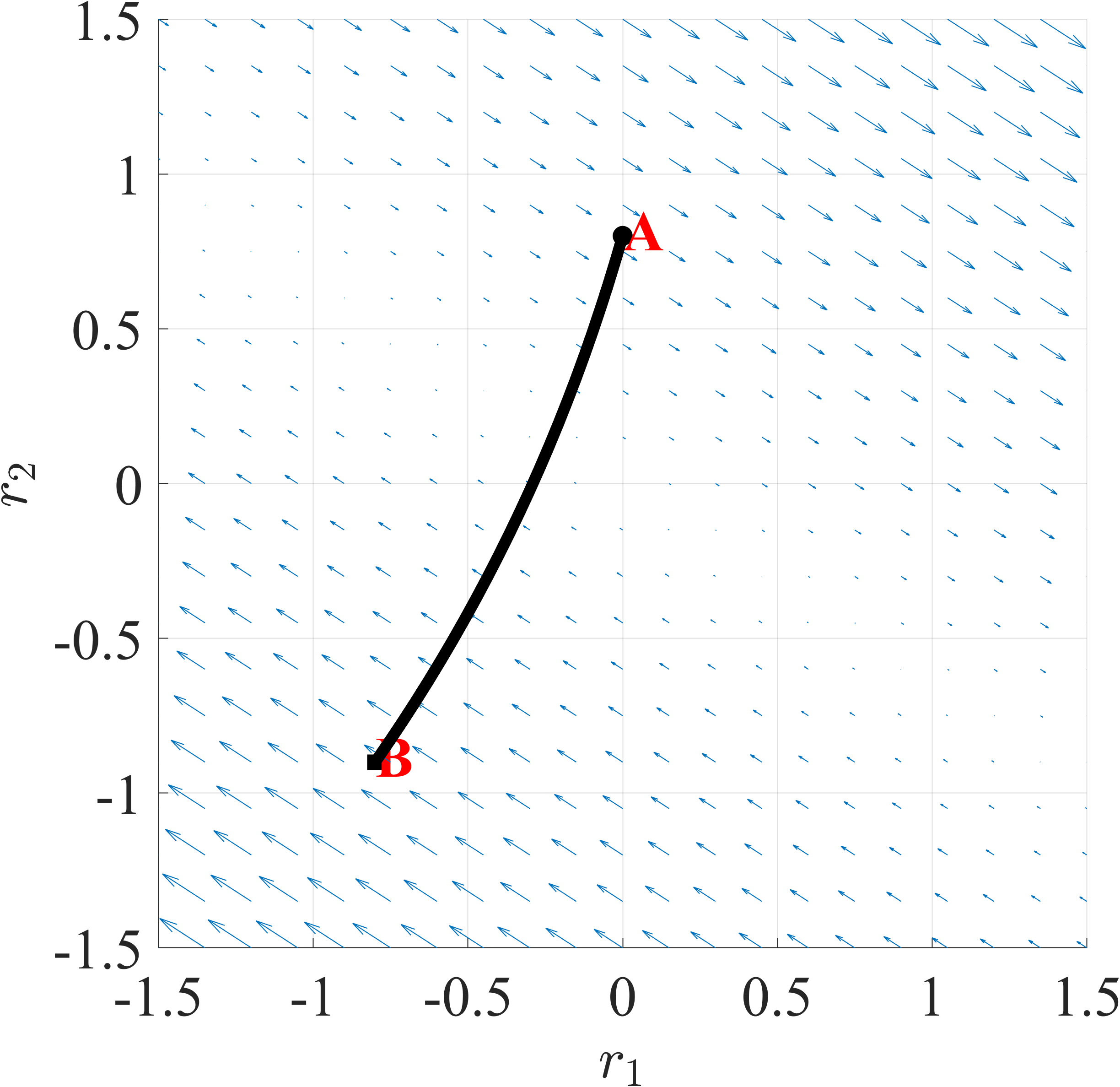}}
	\caption{Examples of optimal trajectories between fixed end points
		\revnew{(indicated in red)}
		in varying wind fields (indicated in blue) for the Zermelo problem.}
	\label{fig-zermelo-ex1}
\end{figure}

\figf{fig-zermelo-ex1} {provides examples of}
$\vec{\posState}^*$ trajectories for fixed initial 
and final points in various wind fields {of the form}
$\vec{w}(\vec{\posState}) = \frac{a_3}{a_1^2 + a_2^2}( (-a_1a_2 \posState_1 + a_2^2 \posState_2), 
(-a_1^2 \posState_1 + a_1a_2 \posState_2)  ),$ where scalars $a_1, a_2 \neq 0, 
a_3 \in \clint{0}{0.25}$ are arbitrarily chosen for each trial. The constant $a_3$ 
is indicative of the highest wind speed in normalized units.

\subsection{Minimum Threat Exposure Problem} 
\label{sec-min-threat}

\def\threatConstant{\lambda}
\def\threat{c}

Consider the motion of a vehicle where the objective is to minimize
its exposure to a spatially varying positive scalar field that we call the
\mydef{threat field} $\threat:\real[2] \rightarrow \real_{>0}.$
As in \scn{sec-min-time}, the vehicle's motion is modeled by 
simple planar particle kinematics:
\begin{align*}
	\dot{\posState}_1(t) &= V \cos\yaw(t),
	& \dot{\posState}_2(t) &= V \sin\yaw(t),
\end{align*}
For prespecified initial and 
final points $\vec{\posState}_0$ and $\vec{\posState}_1,$ 
we would like to find the time of travel $t^*$ and the 
desired heading profile $\yaw^*(t)$ over the entire interval 
$\clint{0}{t^*}$ such that the boundary conditions $\vec{\posState}(0) = 
\vec{\posState}_0$ and $\vec{\posState}(t^*) = \vec{\posState}_1$ are satisfied
and such that the total exposure to the threat
\begin{align}
	J[u] &\defeq \int_{0}^{t^*} (\threat(\vec{\posState}{(t)}) 
	+ \threatConstant) ~\df t
	\label{eq-threat-cost}
\end{align} 
is minimized. Here $\vec{\posState}$ is the vehicle's trajectory driven 
by a control $\yaw$ and $\threatConstant > 0$ is a scaling constant.
The first-order necessary conditions for this problem are somewhat similar to those
of the Zermelo problem in \scn{sec-min-time}. Namely, the minimum exposure trajectory
$\vec{\posState}^* = (\posState_1^*, \posState_2^*)$ and heading angle $\yaw^*$
must satisfy
\begin{align}
	\dot{\posState^*}_1(t) &= V \cos\yaw^*(t),
	\qquad \dot{\posState^*}_2(t) = V \sin\yaw^*(t), 
	\label{eq-minthreat-dynamics1} \\
	\dyaw^*(t) &= \frac{V }{\threat(\vec{\posState}^*(t)) + \threatConstant}
	\left(\cos\yaw^*(t) \parder{\threat}{\posState_2} (\vec{\posState}^*(t)) - 
	\sin\yaw^*(t) \parder{\threat}{\posState_1}(\vec{\posState}^*(t)) \right).
	\label{eq-minthreat-dynamics2}
\end{align}
As in \scn{sec-min-time}, the Hamiltonian $\hamiltonian$ 
remains zero along any optimal trajectory, namely,
\begin{align}
	\hamiltonian(\vec{\posState}^*, \yaw^*, \vec{\coState}^*) &\defeq 
	\threat(\vec{\posState}^*(t)) + \lambda + 
	V (\coState_1^*(t) \cos\yaw^*(t)  + 
	\coState_2^*(t) \sin\yaw^*(t)) = 0
	\label{eq-hamiltonian-minthreat}
\end{align}
for all $t \in \clint{0}{t^*}.$

To instantiate \prb{prob-general}, we consider on minimum exposure trajectories 
described by the state $\xState = (\vec{\posState},\yaw)$ and the 
dynamics~\eqnsnt{eq-minthreat-dynamics1}{eq-minthreat-dynamics2}. Let 
$\{\vec{\posState}^1, \vec{\posState}^2, \ldots, \vec{\posState}^N \}$ be a set of 
prespecified locations (e.g., grid points); we then 
identify $\wParamsVec \defeq (\threat(\vec{\posState}^1), \threat(\vec{\posState}^2), \ldots,  
\threat(\vec{\posState}^N) )$ as the parameter indicating threat intensities
at these grid point locations.
Finally, the {output} is $\yOutput = (\xState, \vec{\coState}).$

\begin{figure}[t]
	\centering
	\subfigure{\includegraphics[width=\thisfigwidth]{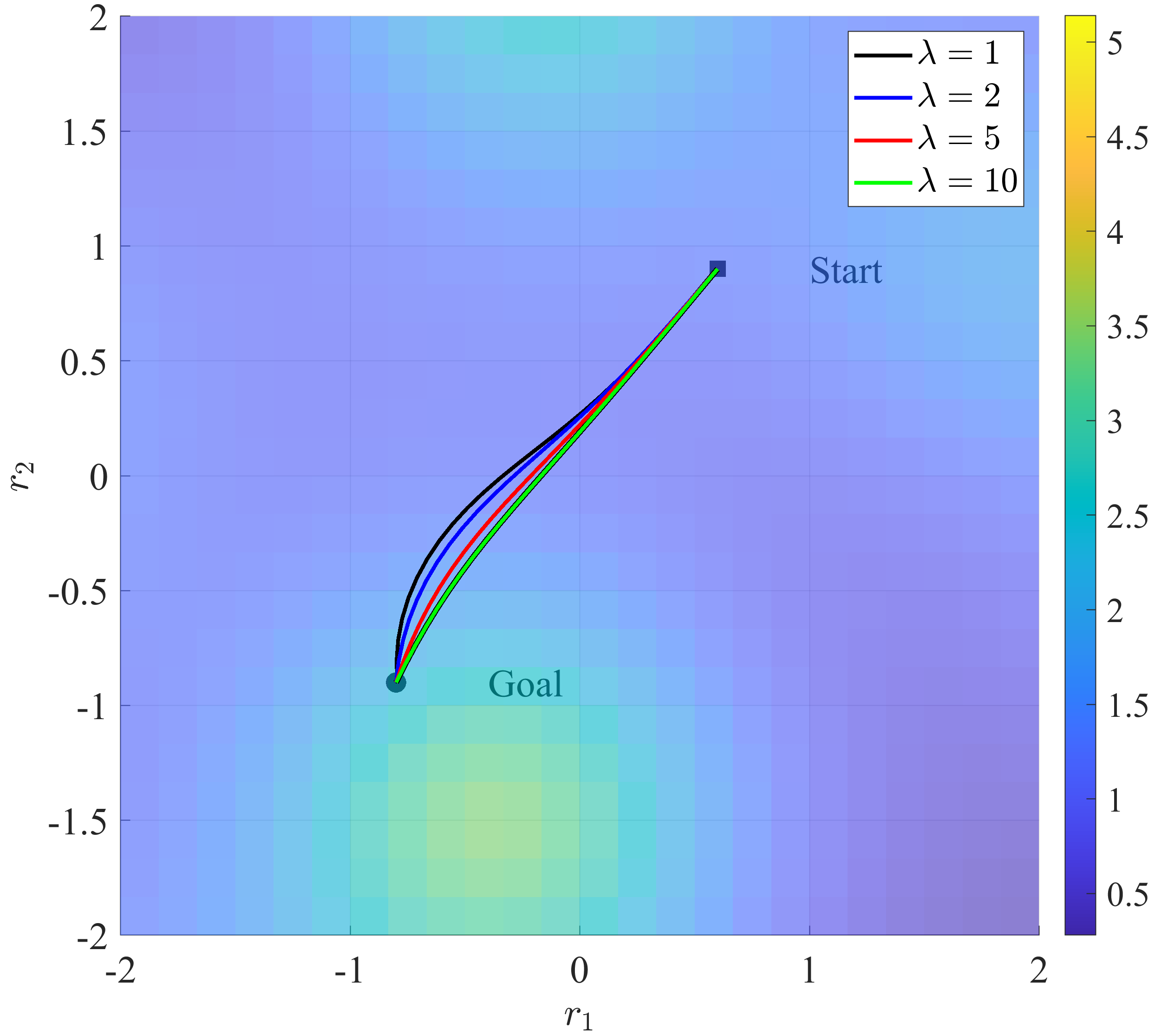}}
	\hspace{\thisfigspace}
	\subfigure{\includegraphics[width=\thisfigwidth]{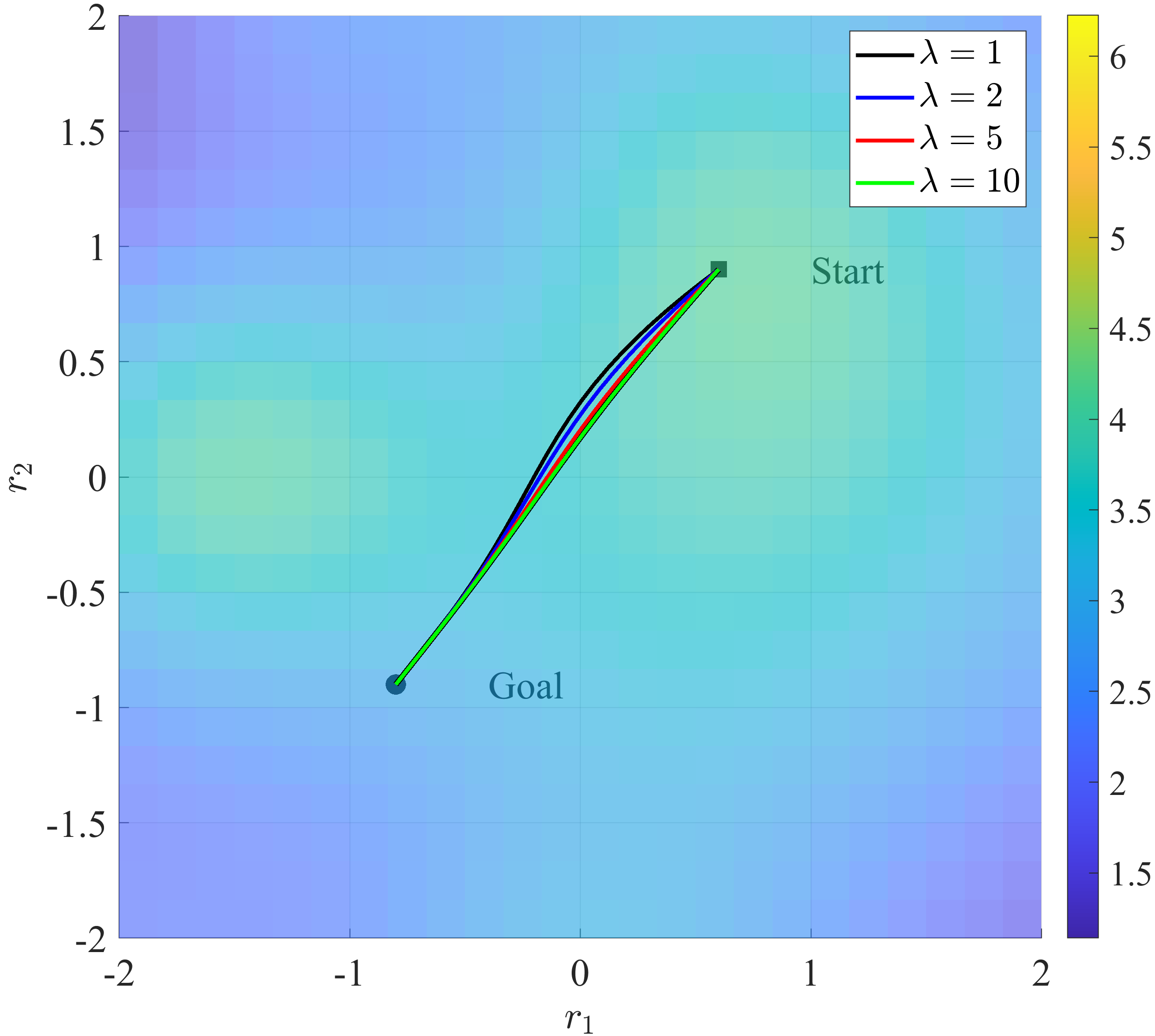}}
	\hspace{\thisfigspace}
	\subfigure{\includegraphics[width=\thisfigwidth]{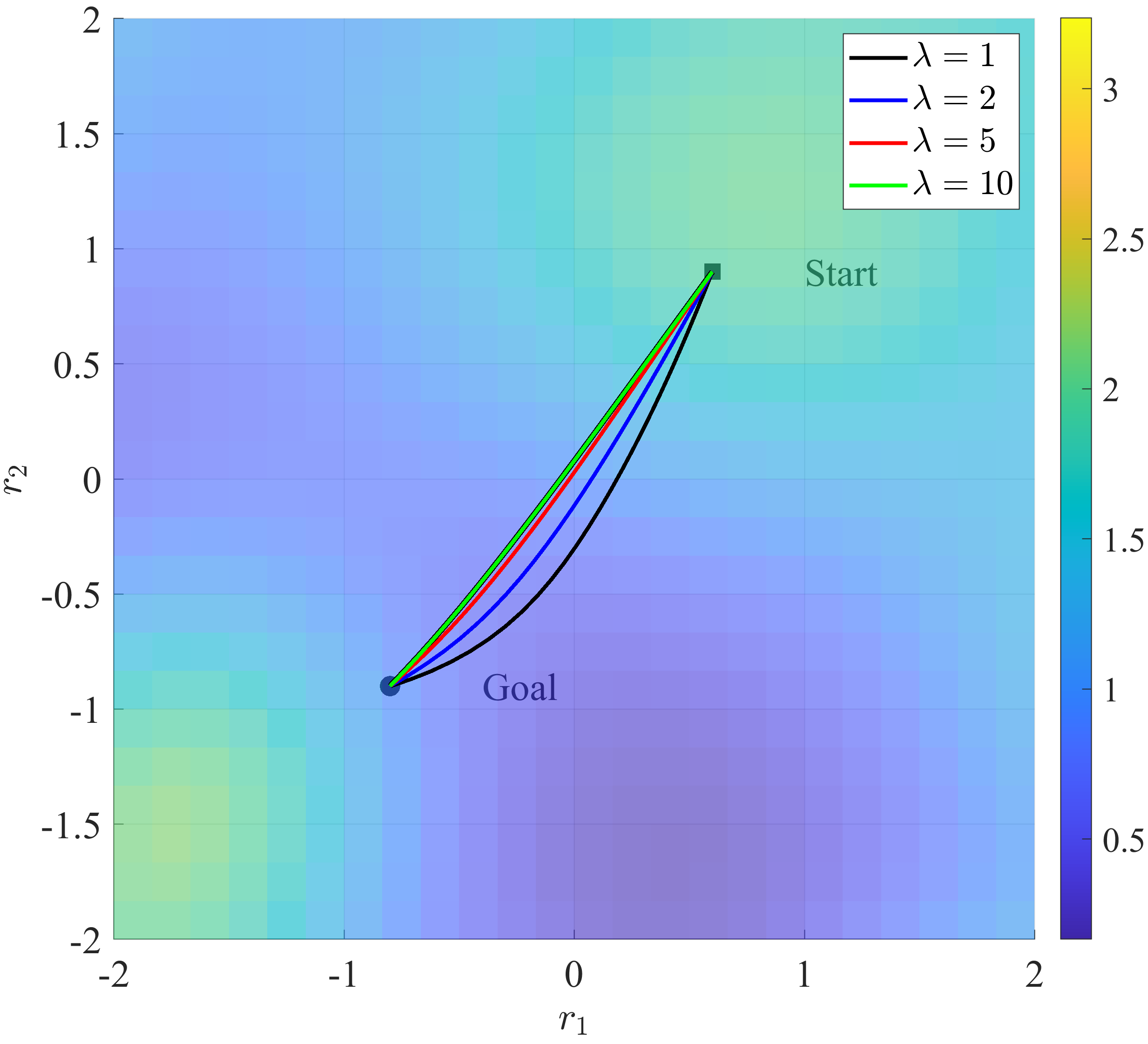}}
	\hspace{\thisfigspace}
	\subfigure{\includegraphics[width=\thisfigwidth]{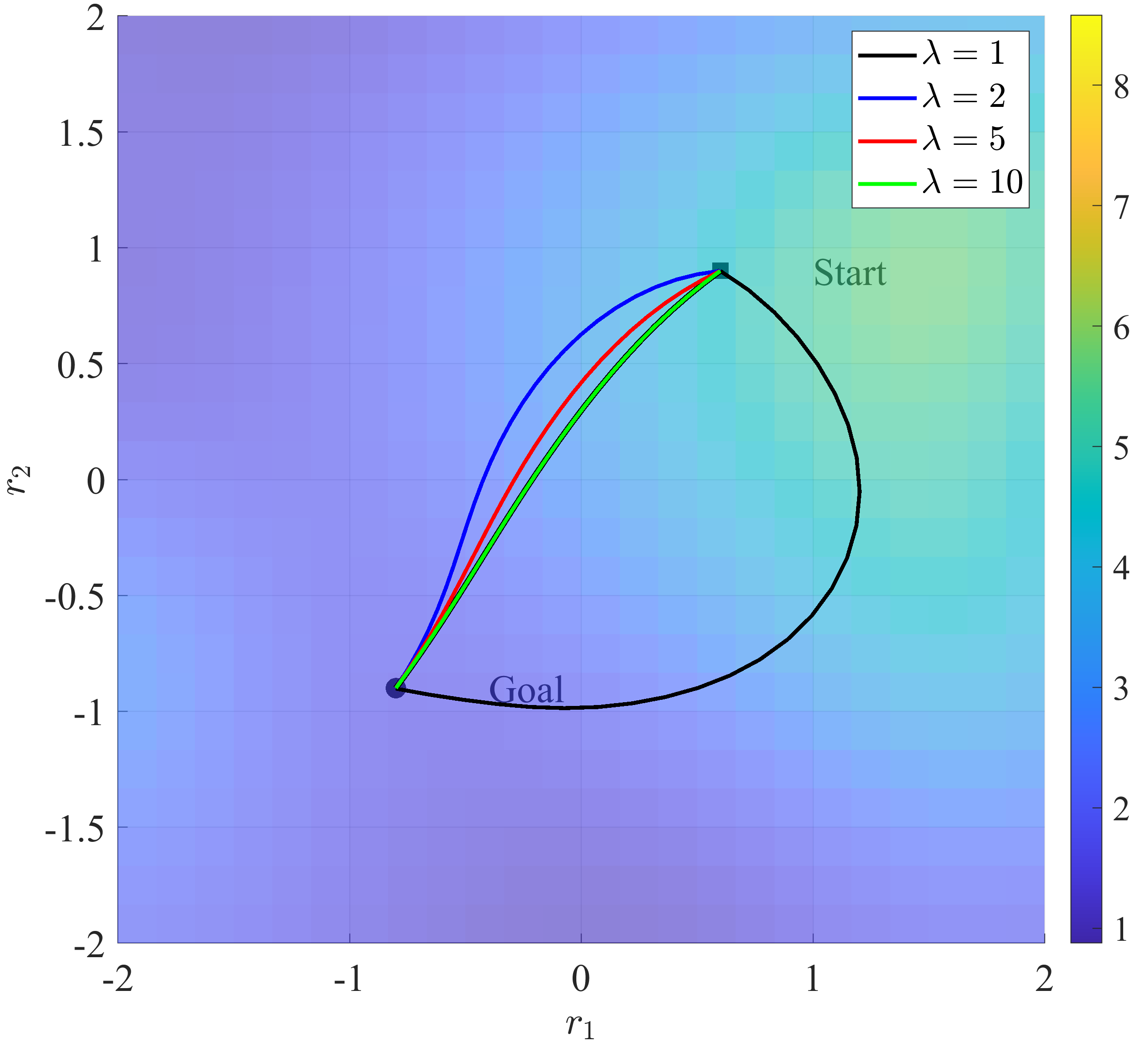}}
	\caption{Example of \mydef{model} trajectory ($\lambda =1$) and \mydef{observed} 
	trajectories ($\lambda =2,5,10$) between fixed end points for the minimum threat 
	exposure problem.}
	\label{fig-minthreat-sample}
\end{figure}

\figf{fig-minthreat-sample} {provides examples of}
$\vec{\posState}^*$ trajectories for fixed initial 
and final points in various threat fields {of the form}
$\threat(\vec{\posState}) = 1 + \vec{\Phi}^\intercal(\pos) 
\threatState.$ Here $\vec{\Phi}$, 
is a spatial basis function vector, e.g., radial basis functions,
and
the constant coefficient vector $\threatState$ is chosen arbitrarily.

For this problem, unlike \scn{sec-min-time}, we \emph{do not} assume that
{the model trajectories are identical to the observed trajectories.} We do assume
that the \mydef{model} and \mydef{observed} trajectories are optimal, but 
the optimality objective functions
may slightly differ. Specifically, the parameter $\threatConstant$ in the cost
functional \eqnnt{eq-threat-cost} may be different for the model and observed trajectories.

\revnew{For both these problems, the discretization scheme used for the OTD
	can be chosen as needed, i.e., the GAIM training process to follow 
	does not impose any specifications for discretization. As described 
	in the next section, the generated data mirror the discretization
	pattern of the training examples. As a result, the number and arrangement
	of features in the generated data are consistent with those in the training
	set, ensuring compatibility without the need for a prespecified discretization
	scheme.}

\section{Generative Model Architectures}
\label{sec-gaims}

A generative model is a map $\gen: \latentSpace \rightarrow \real[\nDimData]$
that maps a vector $\latentVector$ from a sample space $\latentSpace,$ called
the \mydef{latent space,} to a vector $\datum \in \real[\nDimData].$ The ideal
generative model learns this transformation from a training dataset $\dataset$
such that the statistical distribution of $\datasetGen = \gen(\latentSpace)$ is
the same as $\probab{\dataset}.$ The distribution of latent vectors in
$\latentSpace$ is prespecified, e.g., a uniform distribution \revnew{or a standard
normal distribution}. Informally, then,
the generative model maps random vectors from $\latentSpace$ to outputs that
resemble the training dataset without interpolation or extrapolation. In this
paper, we study several GAIMs for solving \prb{prob-general} as described in the
remainder of this section. \tbl{tbl-gaim-summary} provides a brief 
\revnew{and informal} summary of these GAIMs and their salient properties
that we observed. \revnew{Quantitative performance are described in detail
in \scn{sec-results}, where the exact meanings of the performance qualifiers
and dataset sizes mentioned in \tbl{tbl-gaim-summary} will become clear.}

\begin{table}[h]
	{\centering
	\caption{Summary of the GAIMs studied and main observations for 
		large and small sizes $\nData$ of training data.}
	\label{tbl-gaim-summary}
	\begin{tabular}{p{0.25\columnwidth} p{0.09\columnwidth}  p{0.09\columnwidth}  
	p{0.09\columnwidth} p{0.09\columnwidth}  p{0.09\columnwidth} p{0.09\columnwidth} }
		\toprule
		& \multicolumn{4}{c}{Minimum time (Zermelo) problem} & 
		\multicolumn{2}{c}{Minimum threat problem} \\ 
		& S-GAN & Z-GAN & S-VAE & Z-VAE & S-VAE & Split-VAE \\ \midrule
		Performance with large $\nData$ & Poor & Moderate\textsuperscript{*} 
		& Moderate & Excellent & Moderate & Excellent 
		\\ 
		Performance with small $\nData$ & -- & -- & Poor & Moderate & Poor & Good
		\\
		\bottomrule		
	\end{tabular}}
	{\footnotesize
		
		\textsuperscript{*}%
		Good performance in satisfying governing equations, 
		but poor statistical similarity due to 
		mode collapse.
	}
\end{table}

\subsection{Generative Adversarial Network Models} \label{sec-gan}

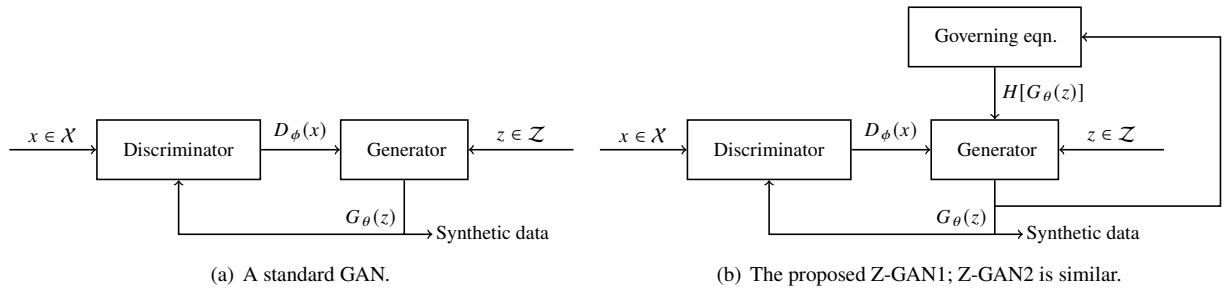
\begin{figure}
	\centering
	\subfigure[A standard GAN.]{\scalebox{0.75}{%
\begin{tikzpicture}
	
	\node [input, name=rinput] (rinput) {};
	\node [block, thick, right of = rinput, 
	node distance = 3cm] (disc-block){\quad Discriminator ~~~\quad} {};
	\draw [thick, ->] (rinput) -- node[anchor = south]{$\datum \in \dataset$} (disc-block);
	
	\node [block, thick, right of = disc-block, 
	node distance = 4cm] (gen-block){\quad Generator ~~\quad} {};
	\draw [thick, ->] (disc-block) -- node[anchor = south]{$\dsc(\datum)$} (gen-block);
	
	\node [input, name=latent-input, right of = gen-block,
	node distance = 3cm] (latent-input) {};
	\draw [thick, ->] (latent-input) -- node[anchor = south]{$\latentVector \in \latentSpace$} (gen-block);
	
	\node [tmp, below of=gen-block, node distance = 1.5cm] (tmp2){};
	\node [tmp, right of=tmp2, node distance = 0.5cm] (tmp3){};
	\draw [thick, ->] (tmp2) -- node[anchor = west, inner xsep = 0.3cm]
	{\revnew{Synthetic data}} (tmp3);
	
	\node [tmp, below of=disc-block, node distance = 1.5cm] (tmp1){};
	\draw [thick, ->] (gen-block) |- node[anchor = south east]{$\gen(\latentVector)$} 
	(tmp1) -|  (disc-block);
	
\end{tikzpicture}
}
	\label{fig-gan-concept}}
	\subfigure[The proposed Z-GAN1; Z-GAN2 is similar.]{\scalebox{0.75}{%
	\begin{tikzpicture}
		
		\node [input, name=rinput] (rinput) {};
		\node [block, thick, right of = rinput, 
		node distance = 3cm] (disc-block){\quad Discriminator ~~~\quad} {};
		\draw [thick, ->] (rinput) -- node[anchor = south]{$\datum \in \dataset$} (disc-block);
		
		\node [block, thick, right of = disc-block, 
		node distance = 4cm] (gen-block){\quad Generator ~~\quad} {};
		\draw [thick, ->] (disc-block) -- node[anchor = south]{$\dsc(\datum)$} (gen-block);
		
		\node [input, name=latent-input, right of = gen-block,
		node distance = 3cm] (latent-input) {};
		\draw [thick, ->] (latent-input) -- 
		node[anchor = south]{$\latentVector \in \latentSpace$} (gen-block);

		\node [tmp, below of = gen-block, node distance = 1.5cm] (tmp1){};
		\node [tmp, below of = gen-block, node distance = 1cm] (tmp2){};
		\node [tmp, right of = tmp2, node distance = 4cm] (tmp3){};
		\draw [thick, ->] (gen-block) |- node[anchor = south east]{$\gen(\latentVector)$} 
		(tmp1) -|  (disc-block);

		\node [tmp, right of=tmp1, node distance = 0.5cm] (tmp4){};
		\draw [thick, ->] (tmp1) -- node[anchor = west, inner xsep = 0.3cm]
		{\revnew{Synthetic data}} (tmp4);

		\node [block, thick, above of = gen-block,
		node distance = 2cm] (gov-block){\quad Governing eqn. ~~\quad} {};
		\draw [thick, ->] (tmp2) -- (tmp3) |- (gov-block);
		\draw [thick, ->] (gov-block) -- node[anchor = west]
		{$\hamiltonian[\gen(\latentVector)]$} (gen-block);
		
	\end{tikzpicture}
}
	\label{fig-zgan-concept}}
	\caption{Illustrations of GAN architectures.}
\end{figure}

The standard GAN consists of two ANNs called the
\mydef{generator} $\gen: \latentSpace \rightarrow \real[\nDimData]$ 
with parameters~$\nnParamGen,$
and the \mydef{discriminator}~$\dsc: \real[\nDimData] \rightarrow \clint{0}{1},$ 
with parameters~$\nnParamDsc.$
These ANNs are trained simultaneously over a zero-sum game
whose value function is~\cite{goodfellow2014generative}:
\begin{align}
		\min_{\nnParamGen} \max_{\nnParamDsc} \E{\log \dsc(\datum)}{\datum \in \dataset}
		+ \E{\log(1 - \dsc(\gen(\latentVector)) ) }{\latentVector \in \latentSpace}.
		\label{eq-goodfellow-value}
\end{align}
$\dsc$ is a supervised classifier that outputs a probability $\dsc(\datum)$
that $\datum \sim \probab{\dataset},$ i.e., that $\datum$ is ``real.''
$\gen$ maps a random vector
$\latentVector$ from a latent space $\latentSpace$ 
to a vector $\gen(\latentVector) \in \real[\nDimData].$ $\dsc$ learns to minimize
the misclassification loss, i.e., to correctly identify the generator's output as ``fake''.
It is trained on data from $\dataset$ labeled ``real''
and data from the generator's output labeled ``fake.'' $\gen$ learns to maximize the
discriminator's loss. $\gen$ and $\dsc$ train iteratively in a feedback loop as 
\fig{fig-gan-concept} illustrates.
After training $\gen$ and $\dsc,$ the desired dataset $\datasetGen$
can be produced as $\datasetGen = \{\gen(\latentVector_i)\}_{i = 1}^{\nGen},$
where $\latentVector_i$ are random samples drawn from $\probab{\latentSpace}.$ 
Because the evaluation
of $\gen(\latentVector_i)$ is an easy computation, we may choose 
$\nGen$ as large as needed. \revhl{Note that the discriminator is needed
for training the generator, but not for synthetic data generation.}

We develop three GAN models to solve the instantiation of the 
data generation problem discussed in \scn{sec-problem-formulation}.
We call these models the \mydef{standard-GAN} (S-GAN) and \mydef{Zermelo-GAN,} 
of which there are two variants Z-GAN1 and Z-GAN2.
The S-GAN applies the standard architecture described above to this problem,
i.e., it trains on data from $\dataset$ but altogether ignores the 
equations~\eqnser{eq-zermelo-dynamics}{eq-costate}.
The two Z-GANs do incorporate some of these equations: 
specifically Z-GAN1 incorporates the Hamiltonian equation~\eqnnt{eq-hamiltonian-mintime}, 
and Z-GAN2 incorporates \eqns{eq-hamiltonian-mintime}{eq-costate-control}. 
Further details of these architectures, illustrated in \figs{fig-gan-concept}{fig-zgan-concept},
are provided next.

The S-GAN value function is similar to~\eqnnt{eq-goodfellow-value}, except
that we replace the binary cross entropy term by a mean squared error (MSE) term as follows:
\begin{align}
	\hspace*{-2ex}
	\max_{\nnParamGen} \min_{\nnParamDsc}  \E{ (\dsc(\datum) - 1)^2 }{\datum \in \dataset} + 
	\E{ (\dsc(\gen(\latentVector)))^2 }{\latentVector \in \latentSpace}.
	\label{eq-value}
\end{align}
We propose similar MSE value functions for the two Z-GANs, but with 
additional terms related to the governing equations. For Z-GAN1 and Z-GAN2, 
these value functions are of the form
\begin{align}
	\max_{\nnParamGen} \min_{\nnParamDsc}  &
	~\E{ (\dsc(\datum) - 1)^2}{\datum \in \dataset} +
	\E{ (\dsc(\gen(\latentVector)))^2 + 
		\Gamma(\gen(\latentVector))
	}{\latentVector \in \latentSpace}.
	\label{eq-new-value}
\end{align}
For Z-GAN1 we consider $\Gamma(\gen(\latentVector)) \defeq 
\scaleConstant_1 \|\hamiltonian[\gen(\latentVector)]\|^2,$ 
where $\scaleConstant_1$ is a normalizing factor. For Z-GAN2,
$\Gamma(\gen(\latentVector)) \defeq \scaleConstant_1 \|\hamiltonian[\gen(\latentVector)]\|^2 + $ 
$\scaleConstant_2 \|\tan\yaw[\gen(\latentVector)] - \frac{\coState_2[\gen(\latentVector)]}
{\coState_1[\gen(\latentVector)]}\|^2.$
\revnew{The difference between the two loss functions of Z-GAN1 and Z-GAN2
is terms of \emph{which} governing equations are incorporated in the GAN
training. For Z-GAN1, we incorporate the Hamiltonian invariance equation, only,
whereas for Z-GAN2, we additionally incorporate the optimal control expression
resulting from variational necessary conditions. The purpose of introducing these two
separate GAN models is to study how model training and performance is affected
by the various details in the governing equations.}
In a minor abuse of notation, 
we reuse the symbols $\hamiltonian, \yaw,$ 
$\coState_1,$ and~$\coState_2$ to indicate the Hamiltonian,
the control input, and the costates, respectively
along a generated trajectory. Specifically, $\hamiltonian[\gen(\latentVector)]$
denotes the Hamiltonian calculated at instants $t_1, \ldots, t_\nSample$ along the
trajectory associated with the output $\gen(\latentVector),$ while
$\yaw[\gen(\latentVector)], \coState_1[\gen(\latentVector)],$
and $\coState_2[\gen(\latentVector)]$
denote the inputs and costates along that trajectory at those instants.

The discriminator in all three GANs learns to minimize the MSE loss
\begin{align}
	\loss_{D}(\nnParamDsc) &\defeq \E{ (\dsc(\datum) - 1)^2 }{\datum \in \dataset} + 
	\E{ (\dsc(\gen(\latentVector)))^2 }{\latentVector \in \latentSpace}.
	\label{eq-discriminator-loss}
\end{align}
The S-GAN generator learns to minimize the MSE loss
\begin{align}
	\loss_{SG}(\nnParamGen) &\defeq \E{(\dsc(\gen(\latentVector)) - 1)^2}{\latentVector \in 
	\latentSpace}.
	\label{eq-sgan-loss}
\end{align}
We formulate the following MSE losses that the Z-GAN1 and Z-GAN2 generators 
learn to minimize:
\begin{align}
	\loss_{ZG1}(\nnParamGen) &\defeq \E{(\dsc(\gen(\latentVector)) - 
	1)                                       ^2 + 
		\scaleConstant_1 \|\hamiltonian[\gen(\latentVector)]\|^2}{\latentVector \in \latentSpace}, \\
	\loss_{ZG2}(\nnParamGen) &\defeq \mathbb{E}_{\latentVector \in \latentSpace}
	\left[(\dsc(\gen(\latentVector)) - 1)^2 + 
	\scaleConstant_1 \|\hamiltonian[\gen(\latentVector)]\|^2 + 
	\scaleConstant_2 \|\tan\yaw[\gen(\latentVector)] - 
	\myfracB{\coState_2[\gen(\latentVector)]}
	{\coState_1[\gen(\latentVector)]}\|^2 \right].
\end{align}
The proposed loss functions 
$\loss_{Z\gen1}$ and $\loss_{Z\gen2}$ penalize the $\gen$ output's violation 
of the equations governing optimal trajectories in the Zermelo problem.
By contrast, the S-GAN loss relies only on the $\dsc$ output, which
in turn, trains only on the OTD $\dataset$ but not the
governing equations. Informally, whereas the S-GAN learns to generate 
trajectories that ``look like'' those in the OTD, the Z-GANs also
``understand'' some underlying fundamental properties of these trajectories.

\revnew{The MSE loss function plays a role similar to the more commonly used
	Binary Cross Entropy (BCE) loss in GAN training. For example, the BCE loss
	equivalent to \eqnnt{eq-discriminator-loss} would be $\E{\log
		\dsc(\datum)}{\datum \in \dataset} + \E{\log(1 - \dsc(\gen(\latentVector)) )
	}{\latentVector \in \latentSpace}.$ We were unable to find model hyperparameters
	for convergence of the BCE loss, and therefore we used the MSE loss.}

At first glance, it may seem that the discriminator is superfluous in 
the two Z-GANs, especially because we assume that the model of governing 
equations~\eqnsernt{eq-zermelo-dynamics}{eq-costate} is perfect. However, 
relying on the governing equations alone can easily lead to what is known as 
\mydef{mode collapse.} This is a phenomenon where $\gen$ locally minimizes
its loss but maps the latent space to a small (non-diverse) set of \revnew{outputs~\cite{LI2021107646}.} \revnew{Mode collapse is the 
consequence of convergence of the NN parameter optimization to a local minimum.}
As an extreme example, $\gen$ in Z-GAN1 may learn \emph{exactly one output }
$\gen(\latentVector),$ for all $\latentVector \in \latentSpace,$ 
such that $\hamiltonian[\gen(\latentVector)] = 0$ is satisfied.
The $\dsc$-dependent terms in $\loss_{Z\gen1}$ and $\loss_{Z\gen2}$ are 
intended to avoid mode collapse. {The $\dsc$-dependent
	terms are especially important when the observed and model trajectories
	are \emph{not} identical. In this case, the OTD (on which $\dsc$ trains) 
	provides information about the system operation that differs from the 
	state space model of the system.}

\begin{algorithm}[t]
	\begin{algorithmic}[1]
		\FOR{ $i = 1, \ldots, \nEpoch$}
		\STATE Initialize training epoch $j \defeq 0$
		\WHILE{$j\batchSize \leq \nData$}
		\STATE Select minibatch $\batch \subset \dataset,$ with $|\batch| = \batchSize$
		\STATE Sample a batch $\latentSpaceBatch \sim \probab{\latentSpace}$ 
		of latent vectors with $|\latentSpaceBatch| = \batchSize$
		\STATE Update discriminator ANN parameters as
		$$ \nnParamDsc \defeq \arg\min_{\nnParamDsc} \frac{1}{\batchSize} \sum_{\datum \in \batch} 
		(\dsc(\datum) - 1)^2 + 
		\frac{1}{\batchSize} \sum_{\latentVector \in \latentSpaceBatch} 
		(\dsc(\gen(\latentVector)))^2 $$
		\STATE Sample another batch $\latentSpaceBatch \sim \probab{\latentSpace}$  
		with $|\latentSpaceBatch| = \batchSize$
		\STATE Update generator ANN parameters as
		$$ \nnParamGen \defeq \arg \min_{\nnParamGen} \frac{1}{\batchSize} \sum_{\latentVector \in 
		\latentSpaceBatch} 
		(\dsc(\gen(\latentVector)) - 1)^2 + (\hamiltonian[\gen(\latentVector)])^2
		$$
		\STATE Increment iteration counter $j \defeq j + 1$
		\ENDWHILE
		\ENDFOR
		\caption{Iterative training of the generator and discriminator, 
			illustrated for Z-GAN1.}
		\label{alg-main}
	\end{algorithmic}
\end{algorithm}

All three GANs are trained using the iterative process illustrated in
\algoref{alg-main}. At each iteration, a batch of $\batchSize$ data points 
is extracted from the dataset $\dataset$ and a batch of
$\batchSize$ random samples are drawn from the latent distribution
$\probab{\latentSpace}.$ 
First, the
discriminator ANN parameters are updated by minimizing $\loss_D(\nnParamDsc)$ approximated
over the batches, while the generator parameters $\nnParamGen$ remain fixed. 
Next, a new batch of random samples is taken from the latent space.
With fixed~$\nnParamDsc,$ the generator parameters $\nnParamGen$ are then updated by 
minimizing its loss approximated over the new latent vector batch.
The iterations continue until all $\nData$ data points in $\dataset$ are used, 
which completes one training epoch. Training continues further over a user-specified
number of epochs $\nEpoch.$ \algoref{alg-main} shows the batch loss function for
Z-GAN1. Generator loss functions for S-GAN and Z-GAN2 are similarly constructed.

\subsection{Variational Autoencoder Models} \label{sec-vae}

A \mydef{variational autoencoder} (VAE) consists of two NNs called the
\mydef{encoder} $\encoder$ and \mydef{decoder} $\decoder$, respectively. 
The overlapping notation $\dec$ for the decoder
and the generator in \scn{sec-gan} is intentional because both of these ANNs serve
the purpose of mapping vectors from a latent space to desired outputs. 
A detailed explanation of the VAE is out of scope here; we refer the reader
interested to \cite{kingma2019introduction-vae} for a comprehensive and mathematically
rigorous description. A brief overview of the VAE follows.

The output space
of the encoder, which is also the input space of the decoder is the 
{latent space} $\latentSpace.$ The input space of the encoder,
which is also the output space of the decoder is the same as that of the data,
i.e., $\real[\nDimData].$
To synthesize the desired dataset $\datasetGen,$
the decoder maps samples drawn from a standard normal distribution over 
the $\latentSpace$ to its output space.
The encoder learns a mapping from points 
$\datum \in \dataset$ to distributions in the latent space such that the distribution 
of $\latentVector \sim \encoder(\datum)$ conditioned on $\datum$
is approximately a standard normal distribution,
in the sense of low KL divergence.

\begin{algorithm}
	\begin{algorithmic}[1]
		\STATE Initialize Encoder-Decoder Parameters: $\phi$, $\theta$
		\FOR{ $i = 1, \ldots, \nEpoch$}
		\STATE Initialize training epoch $j \defeq 0$
		\WHILE{$j\batchSize \leq \nData$}
		\STATE Select minibatch $\batch \subset \dataset,$ with $|\batch| = \batchSize$
		\STATE Encode: $\batch \rightarrow \enc(\batch)$ 
		\STATE Decode: $\enc(\batch) \rightarrow \dec(\enc(\batch))$
		\STATE Update ANN parameters as
		\begin{align*}
			\nnParamDec,\nnParamEnc &= \arg\min_{\nnParamDec,\nnParamEnc}
			\loss\msub{SVAE}(\nnParamDec,\nnParamEnc)
		\end{align*}

		\STATE Increment iteration counter $j \defeq j + 1$
		\ENDWHILE
		\ENDFOR
		\caption{Iterative training of the encoder and decoder
			illustrated for Z-VAE.}
		\label{alg-main_vae}
	\end{algorithmic}
\end{algorithm}

More precisely, let $\nnParamEnc, \nnParamDec$ be
the parameters of the encoder and decoder ANNs, respectively.
We denote by $\pdDec$ the \mydef{likelihood}, 
i.e., the conditional distribution of the decoder's outputs $\datum$ 
given samples $\latentVector$ from the latent space. The objective of
statistical similarity between $\dataset$ and $\datasetGen,$ decoder
parameters are sought to maximize the log-likelihood.
Next, we denote by $\pdEnc$ 
the conditional distribution of $\latentVector$ given $\datum.$
We can formulate this distribution as a normal distribution,
i.e., $\pdEnc \sim 
\mathcal{N}( \mu(\datum; \nnParamEnc), \Sigma(\datum; \nnParamEnc)),$
where $\mu$ and $\Sigma$ are the mean and covariance to be learned by
the encoder during training. The encoder and decoder are 
trained simultaneously by minimizing the loss
\begin{align}
	\loss\msub{VAE}(\nnParamEnc, \nnParamDec) &\defeq 
	- \E{\log \pdDec }{z \sim \pdEnc} 
	+
	\klDiv{\vphantom{1^{2}} 
		\mathcal{N} (\mu(\datum; \nnParamEnc), \Sigma(\datum; \nnParamEnc))~||~ 
		\mathcal{N}(0, I) }.
	\label{eq-vae-loss}
\end{align}
The first term in $\loss\msub{VAE}$ is a \emph{reconstruction loss,} which penalizes
outputs statistically dissimilar from the training data. The second term
in $\loss\msub{VAE}$ is a \emph{similarity loss,} which penalizes the 
difference of the learned latent space distribution to the decoder's sampling distribution
(standard normal). For brevity in the subsequent discussion, we denote this similarity
loss by $\lossSim(\mu, \Sigma) \defeq \klDiv{\vphantom{1^{2}} 
	\mathcal{N} (\mu(\datum; \nnParamEnc), \Sigma(\datum; \nnParamEnc))~||~ 
	\mathcal{N}(0, I) }.$
We develop two VAE models - the standard-VAE (S-VAE) and Zermelo-VAE, namely Z-VAE. Similar to the 
GAN approach, the S-VAE trains only on data from $\dataset$.
The Z-VAE enforces the Hamiltonian constraint~\eqref{eq-hamiltonian-mintime} on the generated 
outputs.

We consider the following loss function for the S-VAE:
\begin{equation} \label{eq-svae-loss}
	\loss\msub{SVAE}(\nnParamDec,\nnParamEnc) \defeq 
	\mathbb{E}_{\datum \in \dataset} \left[ (\datum - \dec(\enc(\datum)))^2 + 
	\alpha_{1} \lossSim(\mu, \Sigma) \right],
\end{equation}
which implements \eqnnt{eq-vae-loss}. Here $\alpha_1 > 0$ is a constant.
For the Z-VAE we consider the loss function
\begin{equation}\label{eq-zvae-loss}
	\loss\msub{ZVAE}(\nnParamDec,\nnParamEnc) \defeq 
	\mathbb{E}_{\datum \in \dataset} \left[  (\datum - \dec(\enc(\datum))  )^2 
	+ \alpha_{1} \lossSim(\mu, \Sigma) + \alpha_2 
		\| \hamiltonian [\vphantom{1^2}\dec(\enc(\datum))]  \|^2 \right],
\end{equation}
where $\alpha_2 > 0$ is a constant. As before,
we reuse the symbol $\hamiltonian$ to indicate the Hamiltonian 
along a generated trajectory, i.e.,
the term $\hamiltonian [\vphantom{1^2}\dec(\enc(\datum)) ]$ in \eqnnt{eq-zvae-loss}
denotes the Hamiltonian calculated at points $t_1, \ldots, t_\nSample$ along the
trajectory associated with the output $\dec(\enc(\datum)).$
This term in the loss $\loss\msub{ZVAE}$ penalizes violations in the decoder output
of the zero Hamiltonian necessary condition in \eqnnt{eq-hamiltonian-mintime}). 
Note that the S-VAE loss function $\loss\msub{SVAE}$ does not consider the 
necessary conditions of optimality at all.

The two VAEs are trained per \algoref{alg-main_vae}.
At each iteration, a batch of $\batchSize$ data points is extracted from the 
dataset~$\dataset$, and a batch of $\batchSize$ samples is drawn 
from the latent space. The latent space samples are 
passed through the decoder. The decoder and encoder ANN parameters 
$\nnParamDec$ and $\nnParamEnc,$ respectively are
updated by minimizing the loss approximated over the batches. 
Next, a new batch of random samples is taken from the latent space. 
The iterations continue until all $\nData$ data points in
the dataset $\dataset$ are used, which completes one training epoch. Training
continues further over a user-specified number of epochs $\nEpoch.$
\algoref{alg-main_vae} shows the batch loss function for Z-VAE.

\subsection{Split Variational Autoencoder Model} \label{sec-splitvae}

\revhl{We were unable to use the Z-VAE idea of adding a Hamiltonian violation term to
the loss function,} in \eqnnt{eq-zvae-loss}, \revhl{to develop a similar VAE for the
minimum threat problem.} \revhl{This issue arises because the OTD in the minimum threat
problem is ``noisy,'' i.e., the trajectories in the OTD do not exactly satisfy
the model governing equations} in \scn{sec-min-threat}. To remedy this issue, we
develop \revrpl{another}{a new} VAE-based model called the \mydef{Split-VAE} as
follows. We augment the training dataset $\dataset$ with an additional synthetic
dataset, $\datasetSup$, which we refer to as "noiseless". This "noiseless"
dataset consists of model trajectories.  Thus, the cumulative training dataset
becomes $\mathcal{X}\msub{e} = (\dataset,\datasetSup)$, where $\dataset$
consists of observed(i.e.,noisy) trajectories.

The proposed Split-VAE has a conditioned latent space such that each subspace of
the latent space captures different representations of the OTD. We train the
Split-VAE on the cumulative dataset, $\mathcal{X}\msub{e}$. The latent space is
partitioned such that two components $\latentVectorA$ and $\latentVectorB$ of
the latent vector $\latentVector = (\latentVectorA, \latentVectorB)$, where
$\latentVectorA$ is dedicated to noisy and $\latentVectorB$ to both noisy and
noiseless input trajectories. We formulate the two conditional distributions as
normal distributions of the form
\begin{align*}
	\mathbb{P}_{\nnParam}(\latentVectorA|\datum\in\dataset)
	&\sim \mathcal{N}(\mu_1(\datum\in\dataset; \nnParamEnc),
	\Sigma_1(\datum\in\dataset;	\nnParamEnc)), \\
	\mathbb{P}_{\nnParam}(\latentVectorB|\datum\in\mathcal{X}\msub{e})
	&\sim \mathcal{N}( \mu_2(\datum\in\mathcal{X}\msub{e}; \nnParamEnc),
	\Sigma_2(\datum\in\mathcal{X}\msub{e}; \nnParamEnc)).
\end{align*}
The motivating idea is
that the noiseless model trajectories, which satisfy the governing equations,
are \emph{abundant}. The noisy observed trajectories are relatively few, and it
is easier to map the shared features between the noisy and the noiseless
trajectories in the latent space.
We train the Split-VAE to minimize the  loss function
\begin{align}\label{eq-splitvae-loss}
	\loss\msub{split} &=
	\E{(\datum|_{\datum\in\dataset} - \dec(\enc(\datum)) )^2  + 
		\alpha_1 \lossSim(\mu_1, \Sigma_1) + 
		\alpha_2 \lossSim(\mu_2, \Sigma_2)\indicator(\datum))) 
	}
	{\datum \in \mathcal{X}\msub{e}}.
\end{align}
The indicator function $\indicator(\datum)$ indicates
whether the training input $\datum$ belongs to $\dataset$
or if it is a model (noiseless) trajectory.
\begin{align*}
	\indicator(\datum) = \begin{cases}
		0 & \mbox{if } \datum \in \dataset, \\
		1 & \mbox{otherwise.}
	\end{cases}
\end{align*}
Finally, $\alpha_1,\alpha_2$ are user-specified constants. The loss term
$(\datum \big|_{\datum \in \dataset} - \dec(\enc(\datum)))^2$ ensures that the
decoder generates samples that align with the manifold of the noisy dataset.
Meanwhile, the KL divergence terms in the loss function guide the VAE to capture
shared features in $\latentVectorB$ while isolating features unique to the noisy
dataset in $\latentVectorA$. This approach effectively reduces the total
features to be learned associated with the noisy dataset, leading to more
efficient training and improved outcomes.
Using this loss function, the iterative training process for 
the Split-VAE is similar to that of the S-VAE and Z-VAE shown
in \algoref{alg-main_vae}.

 \revnew{Further insight into the Split-VAE model architecture is as follows.
	From a Bayesian perspective, this decomposition of the latent space leads to
	posterior regularization\cite{bouchacourt2018multi}. In standard VAEs, the
	evidence lower bound (ELBO) minimizes \eqref{eq-svae-loss}, which penalizes
	deviations of the approximate posterior from the standard normal distribution.
	With limited training data, the posterior may overfit the training examples,
	yielding poor generalization. In SplitVAE, the variational objective explicitly
	partitions the latent space into two components, as shown in
	\eqref{eq-splitvae-loss}, thereby imposing a structured factorization on the
	posterior. This architectural separation introduces an inductive bias that
	aligns with the nature of the data: the latent variable $\latentVectorB$, which
	captures features common to both the noisy and noiseless datasets, is inferred
	from a larger pool of training examples.}

\revnew{The noiseless data drawn by solving the governing equations are
	contribute to a more reliable estimation of $\latentVectorB$, improving both
	posterior regularization and prior matching. In contrast, the latent variable
	$\latentVectorA$, which encodes the residual variability unique to the noisy
	data (e.g., stochastic effects, unmodeled dynamics), is inferred solely from the
	limited noisy dataset. However, because $\latentVectorA$ is tasked only with
	modeling domain-specific deviations, its dimensionality can be kept small and
	its scope narrowly defined, reducing the risk of overfitting. This selective
	encoding leads to improved generalization, as the model leverages the abundant,
	low-variance information from the model data to stabilize learning, while
	preserving the capacity to represent noise-induced variability when needed. The
	latent space capacity is allocated more efficiently, and the generalization gap
	is reduced as a consequence.}

\revnew{Thus, the SplitVAE mitigates the overfitting risks associated with
	limited noisy data while leveraging prior knowledge from model-based clean data
	to stabilize training and enhance sample quality.}

\section{Results and Discussion}
\label{sec-results}

We implemented all the GAIMs described in \scn{sec-gaims} using PyTorch \cite{paszke2019pytorch}, 
which is a library of Python-based software tools for implementing NN various 
architectures. \revnew{Functional sample code of our implementation is available
at the following links:}
\begin{itemize}
	\item Code for training the proposed models (GitHub link): \url{https://shorturl.at/1Cz2z}
	\item Datasets (Google Drive link): \url{https://shorturl.at/2Ejlp} 
\end{itemize}

Training datasets were synthesized \matlab-based numerical solutions of the variational 
necessary conditions of optimality equations for the Zermelo navigation problem
(\scn{sec-min-time}) and the minimum threat exposure problem (\scn{sec-min-threat}).
The number of sample time instants per datum were set to $\nSample = 25.$ 
Details regarding the OTD, network architecture, and performance indices 
for the implemented GAIMs are discussed next. 
All of the hyperparameter values chosen for the different GAIMs were established 
after numerical experiments with different hyperparameter combinations.

\subsection{S-GAN and Z-GAN Implementation}
For all three GANs, a total of $\nData = 3000$ optimal trajectories were generated.
As explained in \ref{sec-min-time}, the output in this problem is 
 $\yOutput = (\xState, \vec{\coState})$
and the parameter $\wParamsVec \defeq (w_1(\vec{\posState}^1), 
w_1(\vec{\posState}^2), \ldots, 
w_2(\vec{\posState}^1), \ldots, w_2(\vec{\posState}^N) )$ consists of wind 
velocities. In our implementation we choose $\nSample = N = 25,$ which leads
to $\nDimData = 175.$ We choose the latent space $\latentSpace$ as the hypercube 
$\clint{-1}{1}^{20}$ and $\probab{\latentSpace}$ as a uniform distribution over
this hypercube. The generator $\gen$ and discriminator $\dsc$ in all GANs
were implemented as multilayer perceptrons with eight hidden layers, leaky 
rectified linear unit (ReLU) function activation functions~\cite{Banerjee2019}, 
and with dropout layers of probability 0.2.
The dimensions of each layer are provided in \tbl{tbl-gan-dimensions}. 
The $\dsc$ output layer was chosen to be sigmoidal.
For training via \algoref{alg-main}, the batch size was chosen as 
$\batchSize = 64.$ For the S-GAN generator
and for all the Z-GAN generators and discriminators, learning rates were set at
as 0.01, whereas the S-GAN discriminator learning rate was set at 0.001.

Note that the discriminator input layer is $50$ rather than $\nDimData = 175,$
for the following reasons. After several unsuccessful\footnote{\revnew{We
		consider a training attempt ``successful'' if the loss function converges to a
		small value near zero, and ``unsuccessful' otherwise.}} attempts at training
$\dsc,$ we reduced the complexity of the $\dsc$ classification problem by
reducing the dimension of $\yOutput$ by redefining $\yOutput = \vec{\posState}.$
\revnew{However, the generator still produces output trajectories of $\nDimData
	= 175,$ features. The generator $\gen$ incorporates as constraints the governing
	equations which enforce the correct physical relationships and correlations
	across all features. As a result, even though the discriminator $\dsc$ only sees
	a reduced subset of features, $\gen$ learns to maintain consistency across the
	entire trajectory. This setup effectively balances reduced feature
	dimensionality for $\dsc$ with physics-informed constraints for $\gen$.}

Nevertheless, training the discriminator on fewer features of the data leads to
inferior performance of the GAN. In our study all GAN models performed
significantly worse compared to the VAE models, as discussed next.

\def\thisfigwidth{0.23\columnwidth}
\def\thisfigspace{0.01\columnwidth}
\begin{figure}
	\centering
	\subfigure{\includegraphics[width=\thisfigwidth]{\figpath/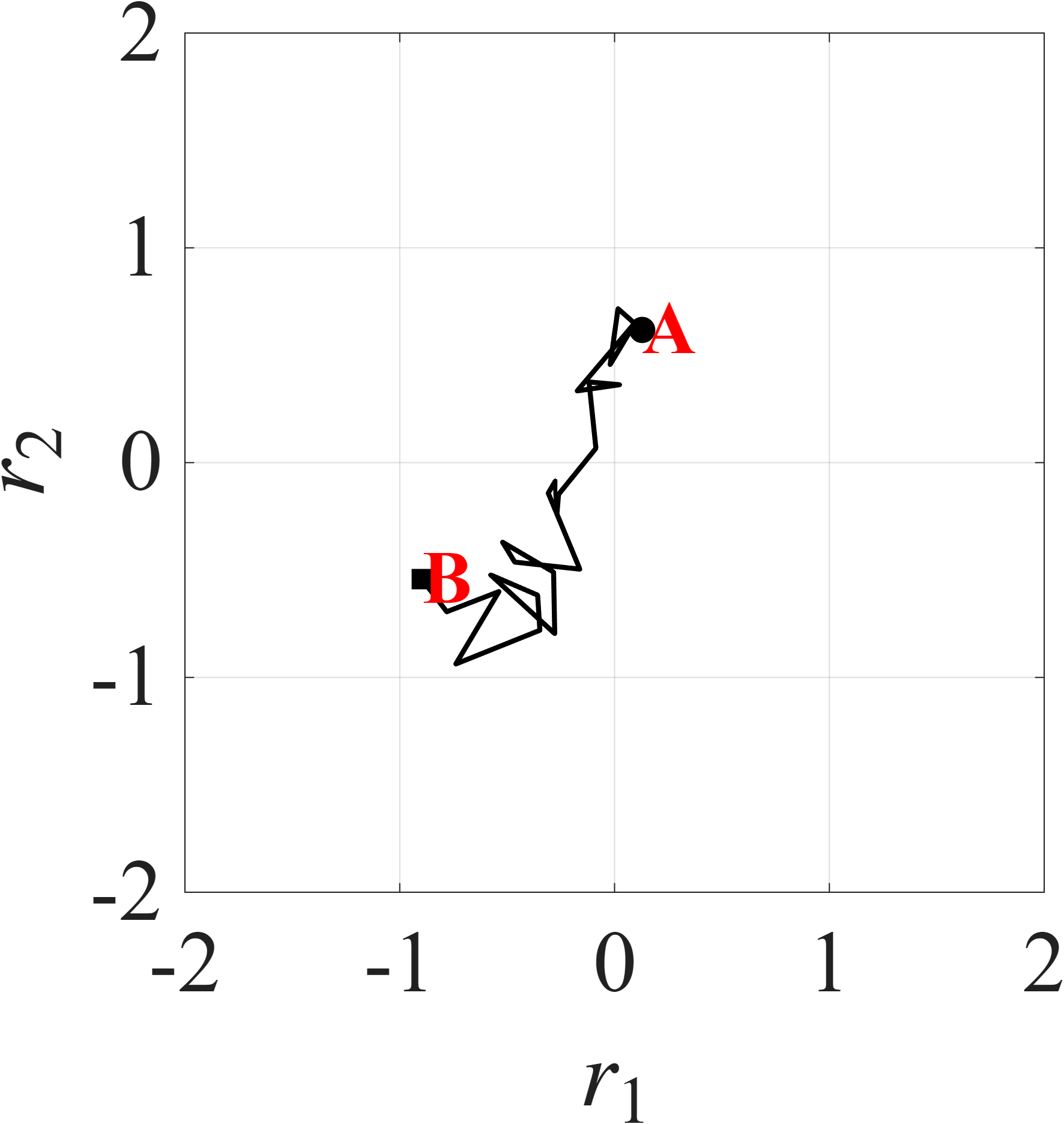}}
	\hspace{\thisfigspace}
	\subfigure{\includegraphics[width=\thisfigwidth]{\figpath/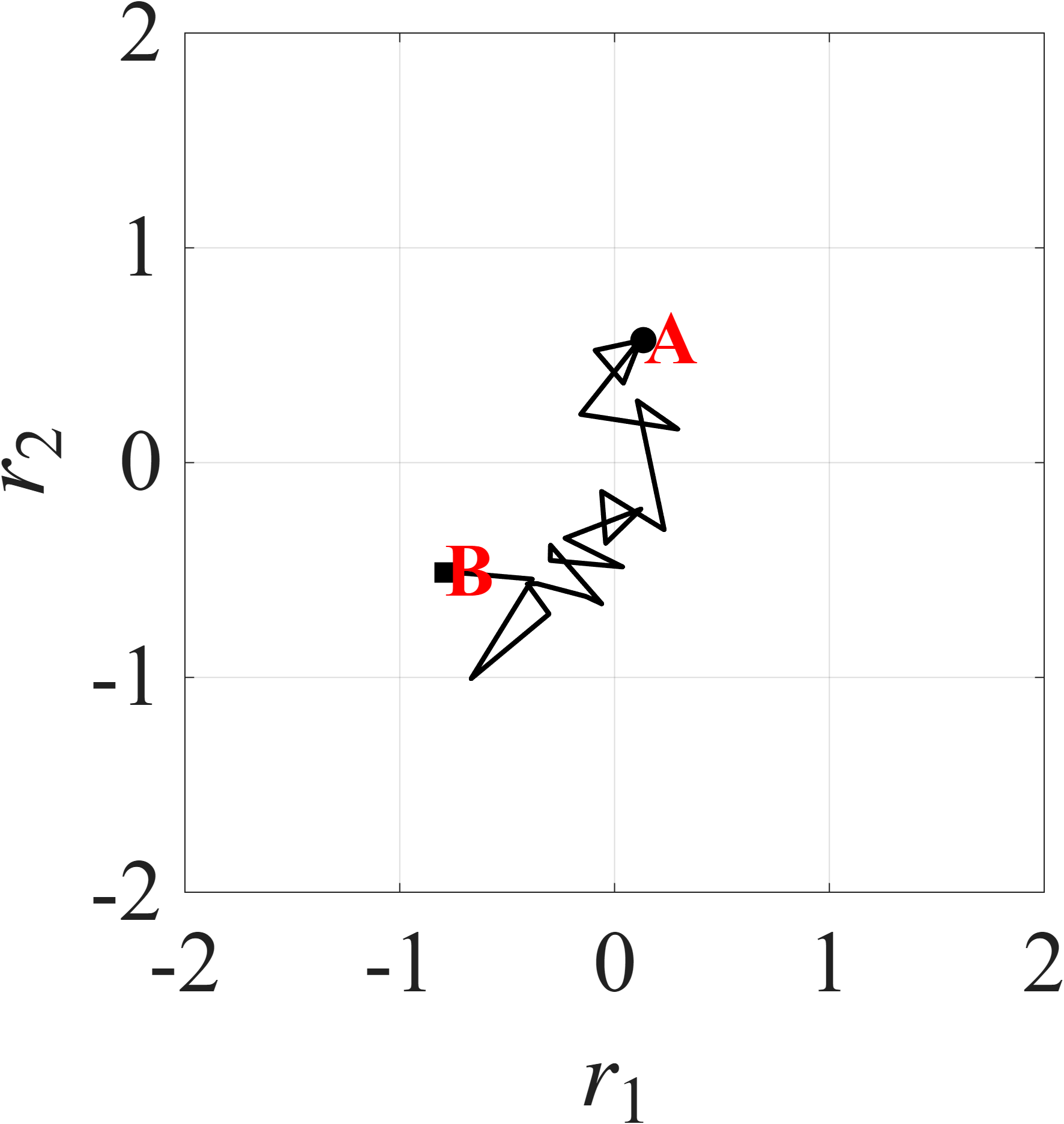}}
	\hspace{\thisfigspace}
	\subfigure{\includegraphics[width=\thisfigwidth]{\figpath/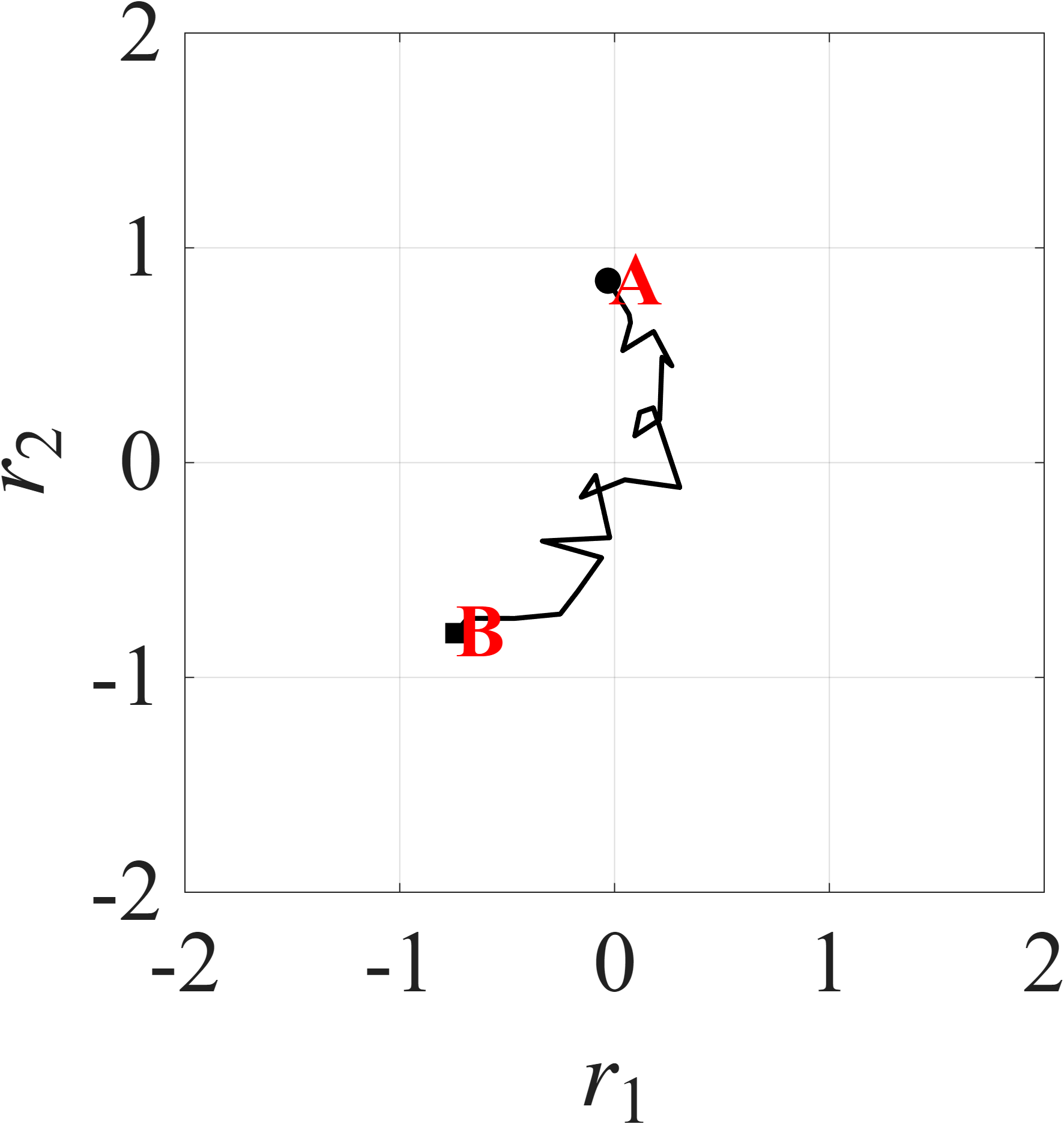}}
	\hspace{\thisfigspace}
	\subfigure{\includegraphics[width=\thisfigwidth]{\figpath/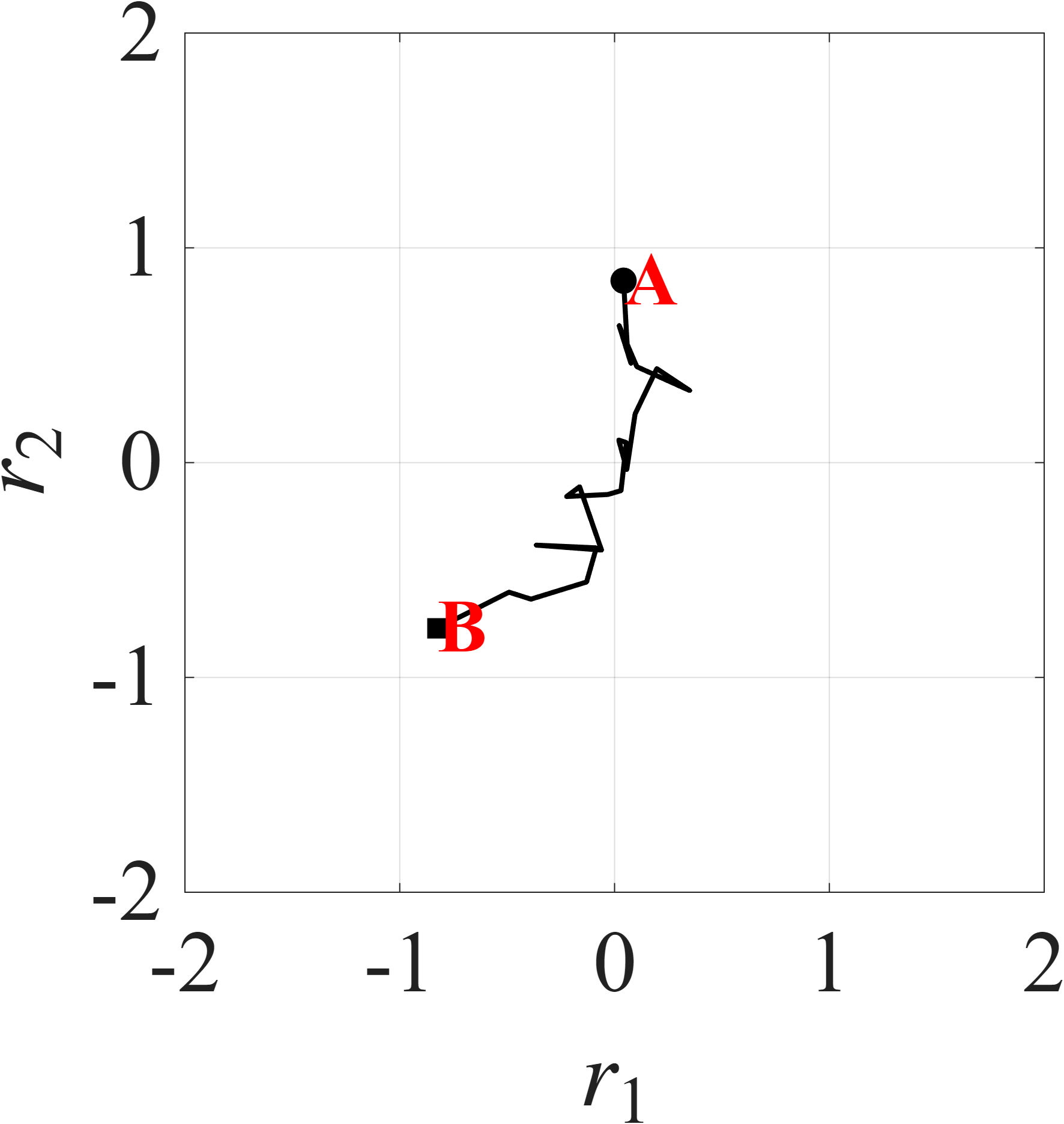}}
	\caption{Sample outputs of the S-GAN generator.}
	\label{fig-SGAN-mintime}
\end{figure}

The quality of the generated dataset $\datasetGen$ was assessed by two 
complementary methods: (1) a direct comparison of the first four statistical
moments of $\datasetGen$ to those of $\dataset,$ and (2) calculation
of performance indices related to deviations from the necessary conditions 
of optimality \eqnsernt{eq-hamiltonian-mintime}{eq-costate}
stated in \scn{sec-min-time}. Specifically, for each 
generated output $\datum \defeq \gen(\latentVector)$
for some sample $\latentVector \sim \probab{\latentSpace},$ we calculate: 
\begin{align}
	\label{eq-sigma-gan}
	\simMet_{1} &\defeq  \| \hamiltonian[\gen(\latentVector)] \|^2, \qquad
	%
	%
	\simMet_{2} \defeq \| \tan\yaw [\gen(\latentVector)] - 
	\frac{\coState_2[\gen(\latentVector)]} {\coState_1 [\gen(\latentVector)]} \|^2, \nonumber \\
	%
	\simMet_{3} &\defeq \| \coState_1 [\gen(\latentVector)] + 
	\textstyle{\frac{\cos\yaw [\gen(\latentVector)]} {\costateden[\gen(\latentVector)]}} \|^2 +
	\| \coState_2 [\gen(\latentVector)]  + 
	\textstyle{\frac{\sin\yaw[\gen(\latentVector)] }{\costateden[\gen(\latentVector)] }}  \|^2.
\end{align}
Note that 
Z-GAN1 learns to satisfy~\eqnnt{eq-hamiltonian-mintime}, Z-GAN2
learns to satisfy~\eqnsnt{eq-hamiltonian-mintime}{eq-costate-control},
but~\eqnnt{eq-costate} is ``new knowledge'' to both GANs. Beside these
two quantitative methods of evaluating $\datasetGen,$ 
a visual assessment of generated output samples is also helpful.

\def\thisfigwidth{0.23\columnwidth}
\def\thisfigspace{0.01\columnwidth}
\begin{figure}
	\centering
	\subfigure{\includegraphics[width=\thisfigwidth]{\figpath/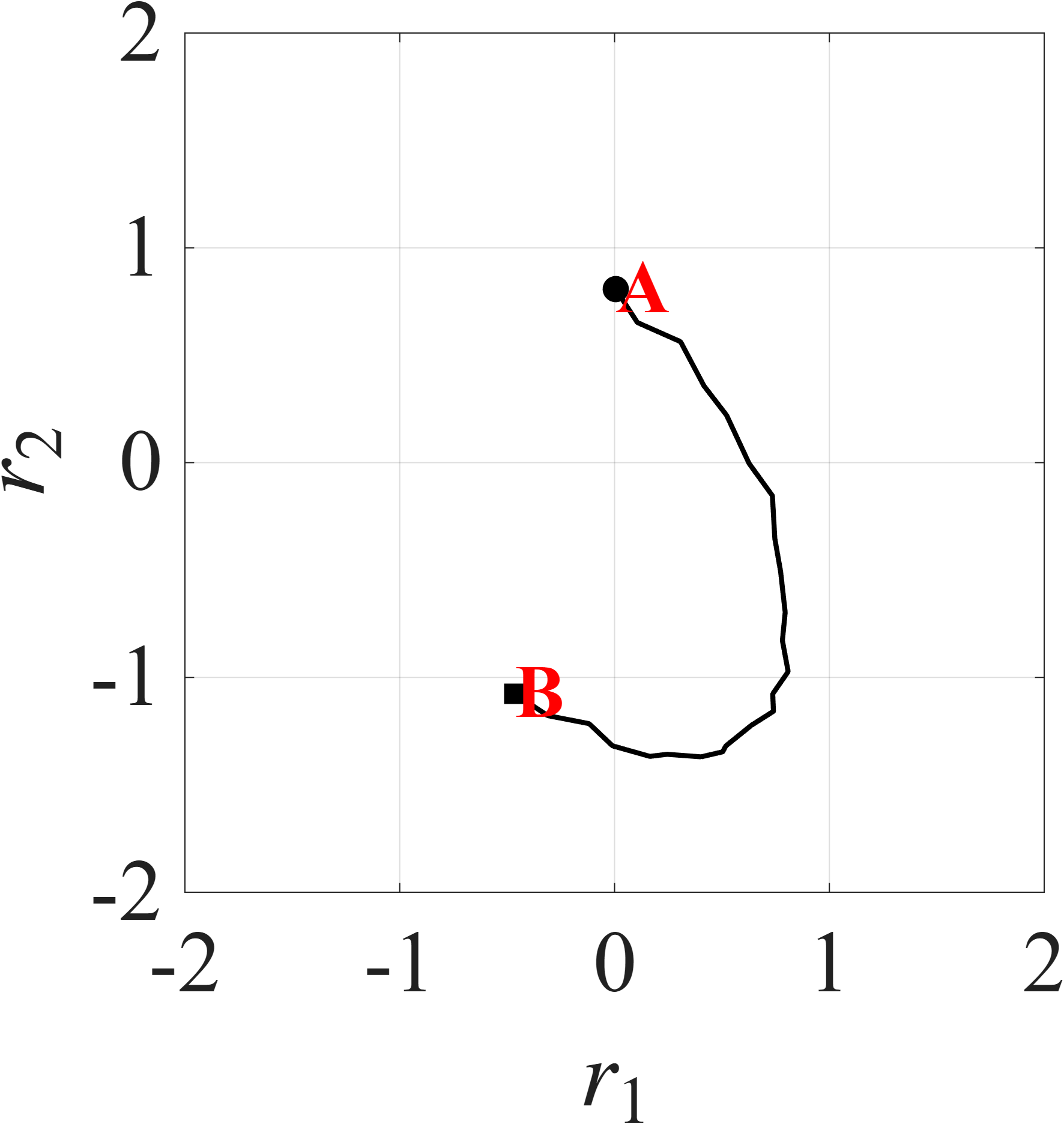}}
	\hspace{\thisfigspace}
	\subfigure{\includegraphics[width=\thisfigwidth]{\figpath/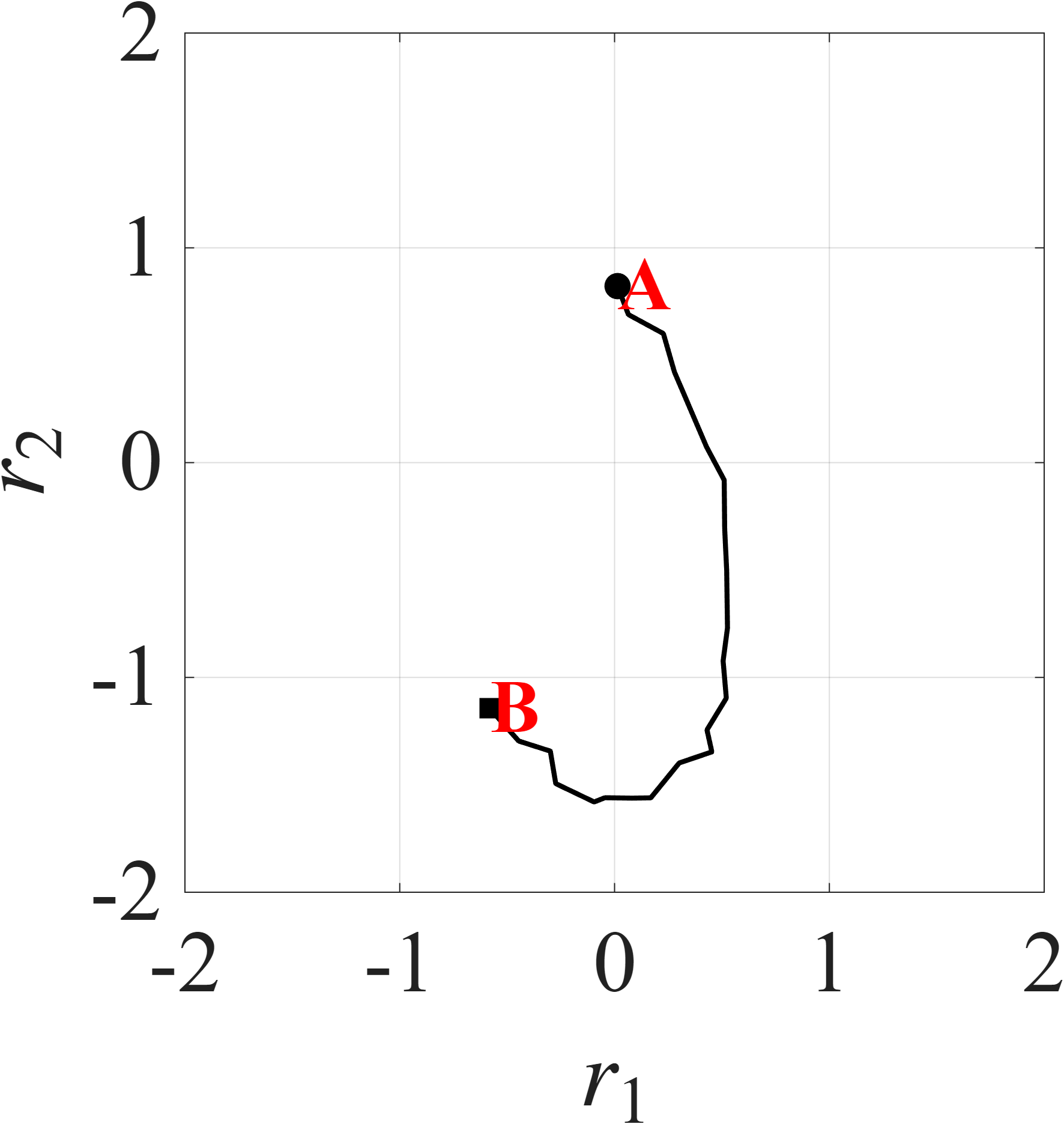}}
	\hspace{\thisfigspace}
	\subfigure{\includegraphics[width=\thisfigwidth]{\figpath/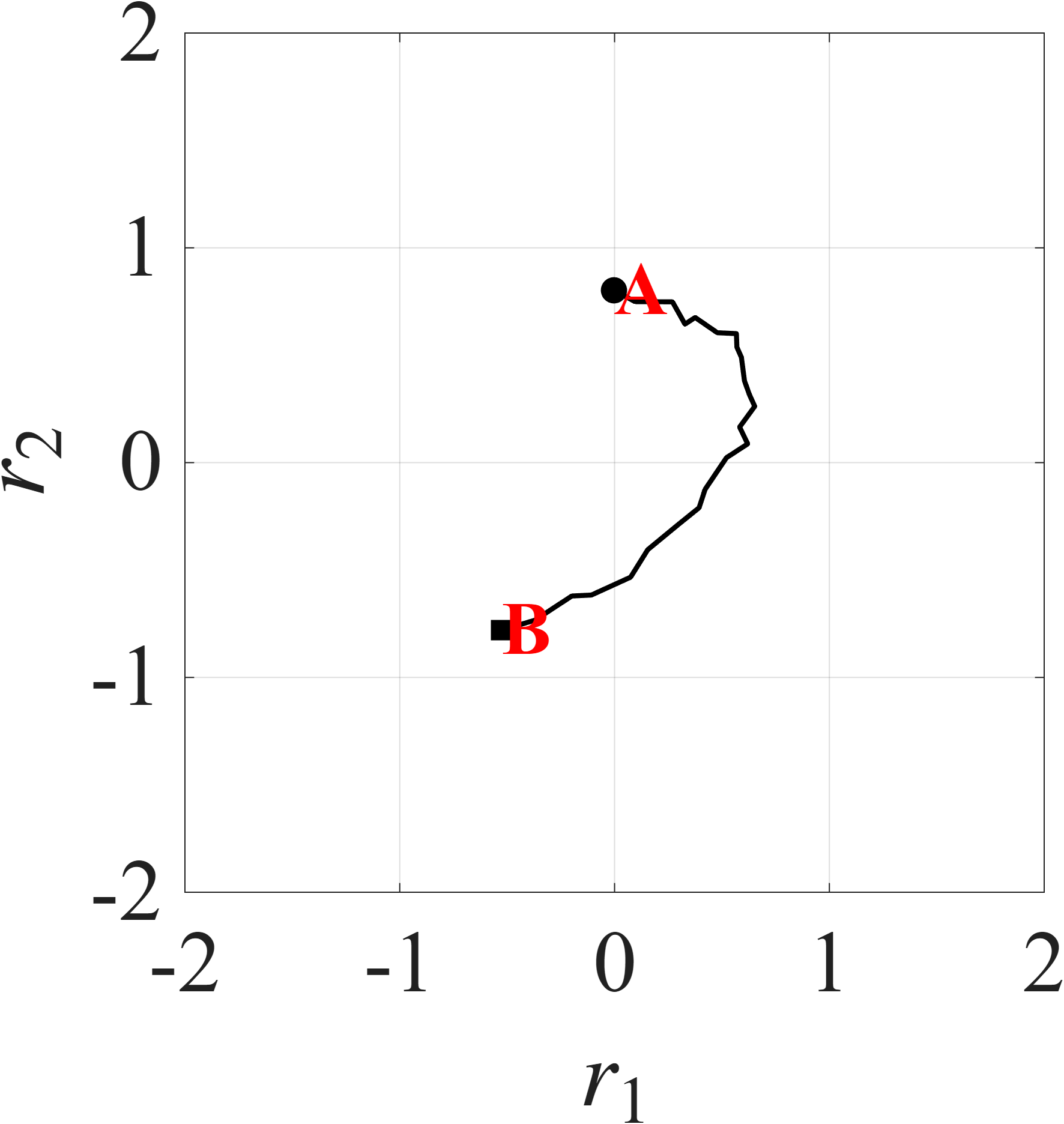}}
	\hspace{\thisfigspace}
	\subfigure{\includegraphics[width=\thisfigwidth]{\figpath/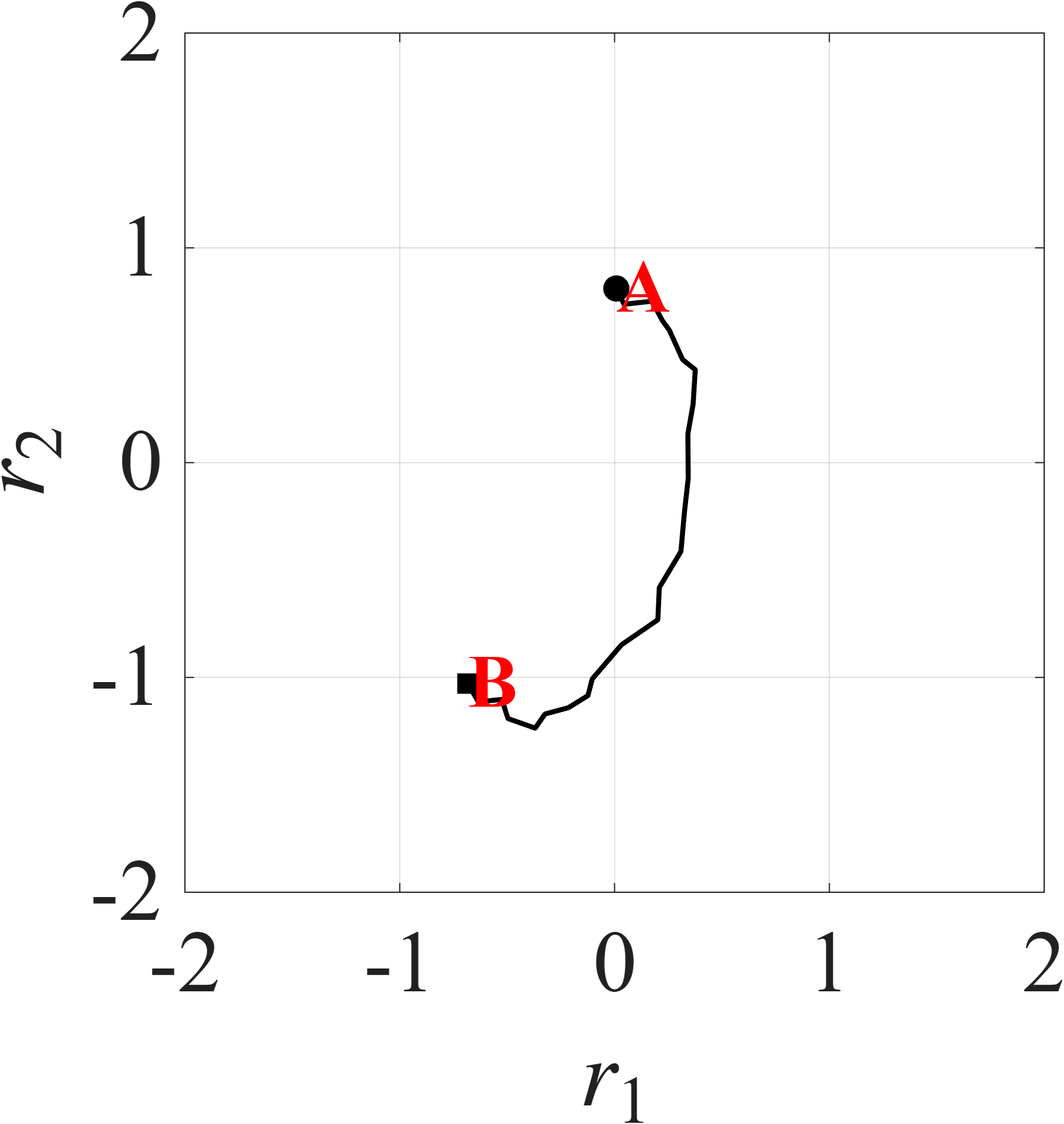}}
    \caption{Sample outputs of the Z-GAN1 generator.}
    \label{fig-zgan1-sample}
\end{figure}

\def\thisfigspace{0.01\columnwidth}
\begin{figure}
	\centering
	\subfigure{\includegraphics[width=\thisfigwidth]{\figpath/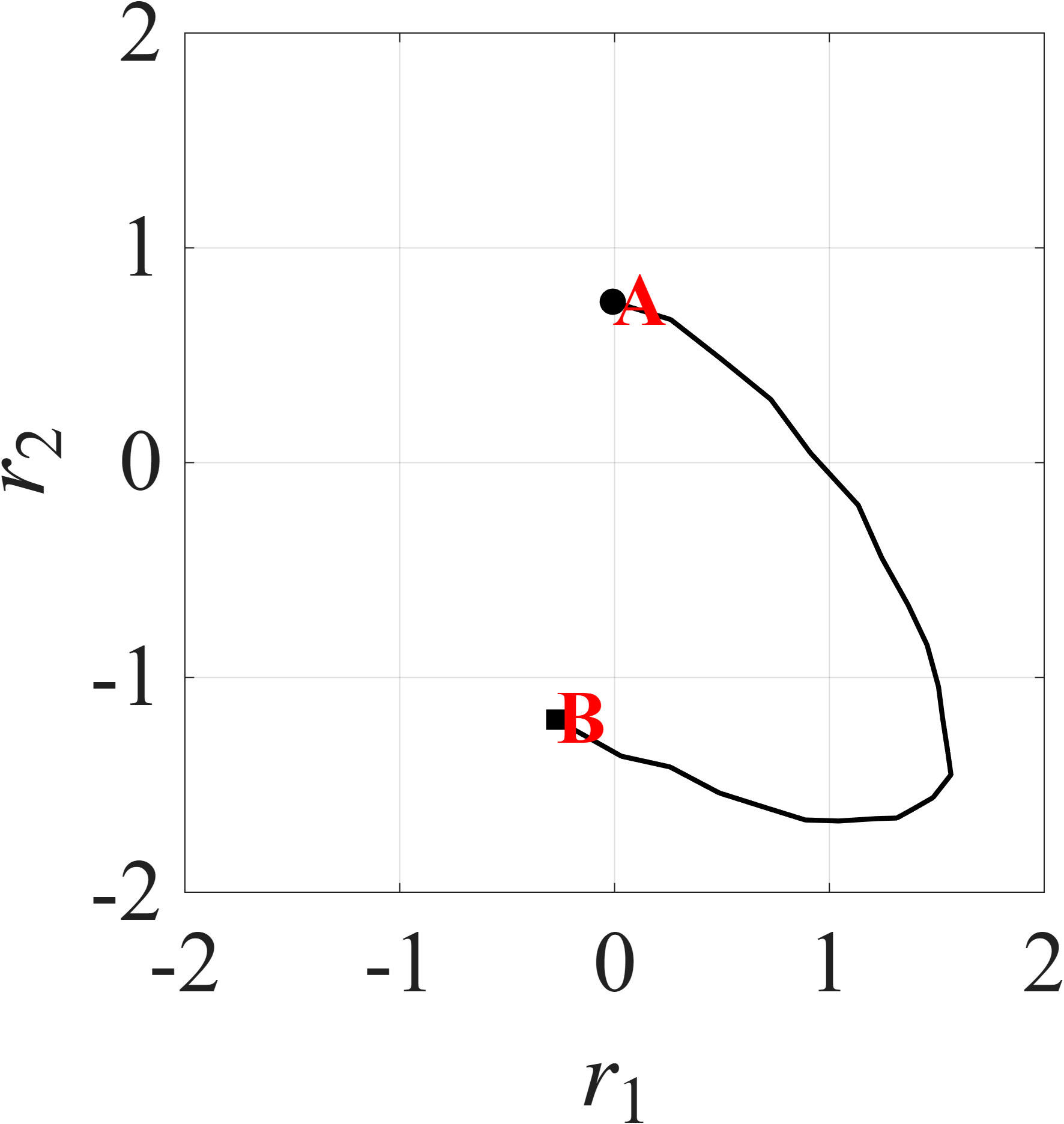}}
	\hspace{\thisfigspace}
	\subfigure{\includegraphics[width=\thisfigwidth]{\figpath/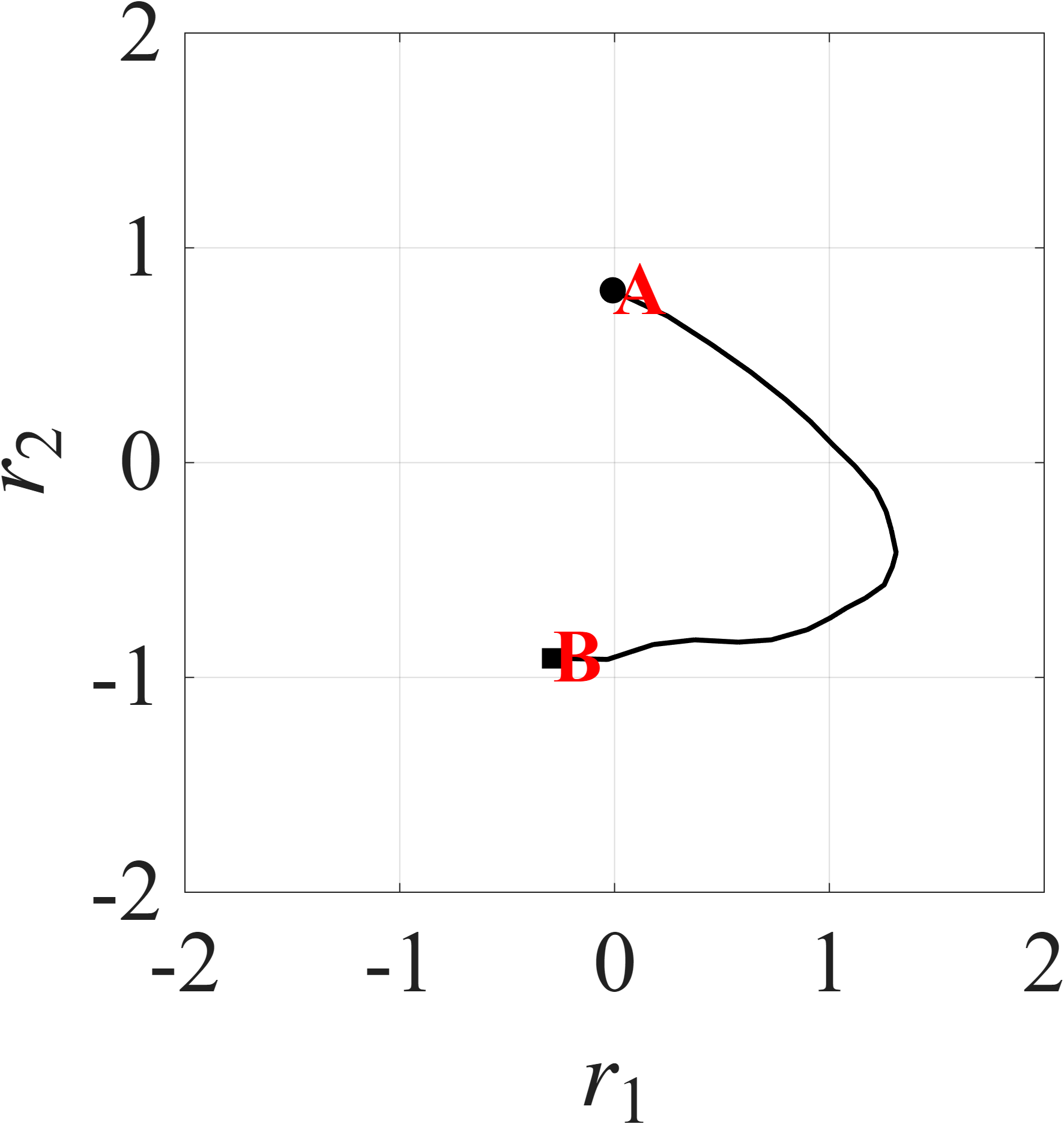}}
	\hspace{\thisfigspace}
	\subfigure{\includegraphics[width=\thisfigwidth]{\figpath/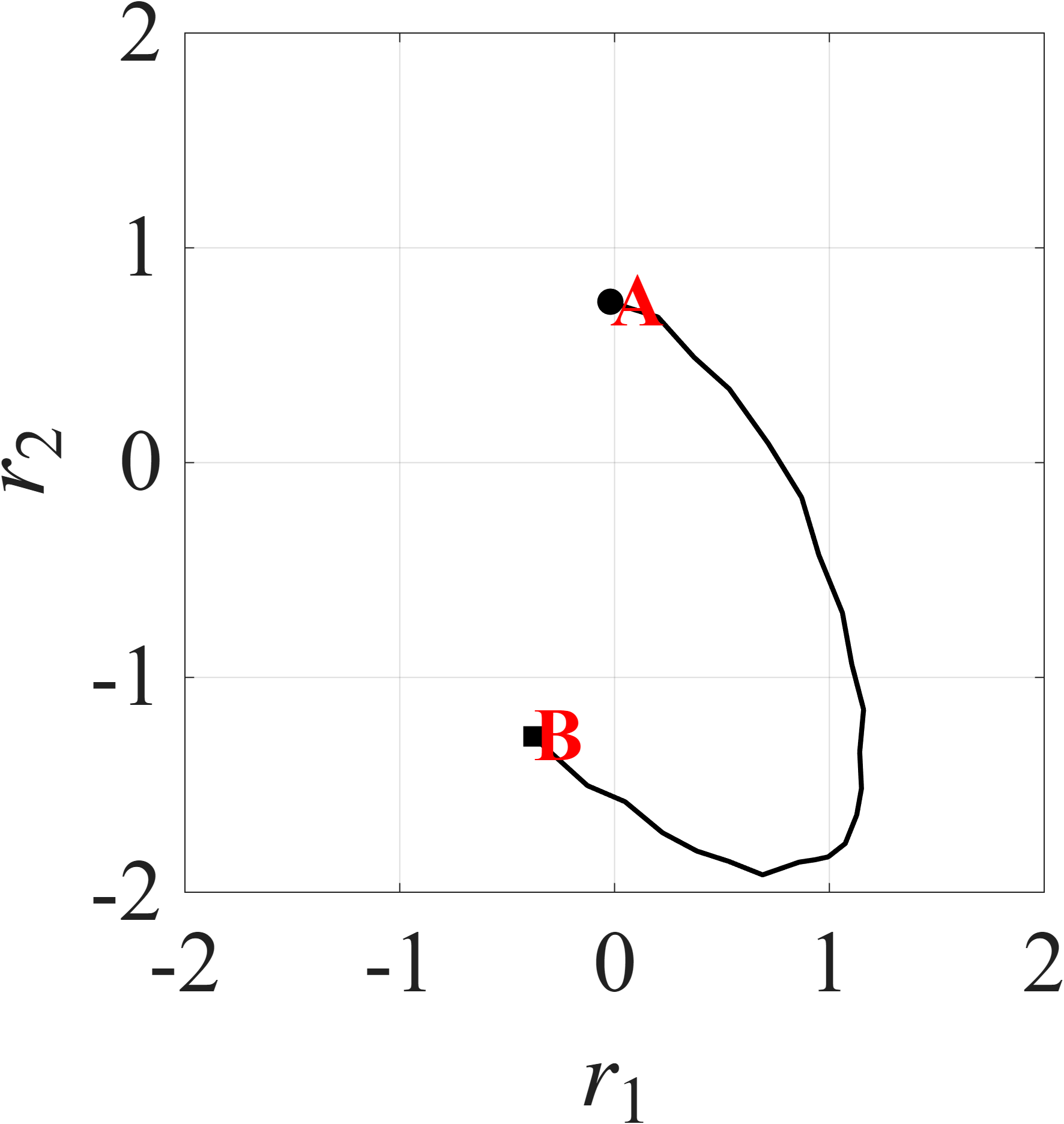}}
	\hspace{\thisfigspace}
	\subfigure{\includegraphics[width=\thisfigwidth]{\figpath/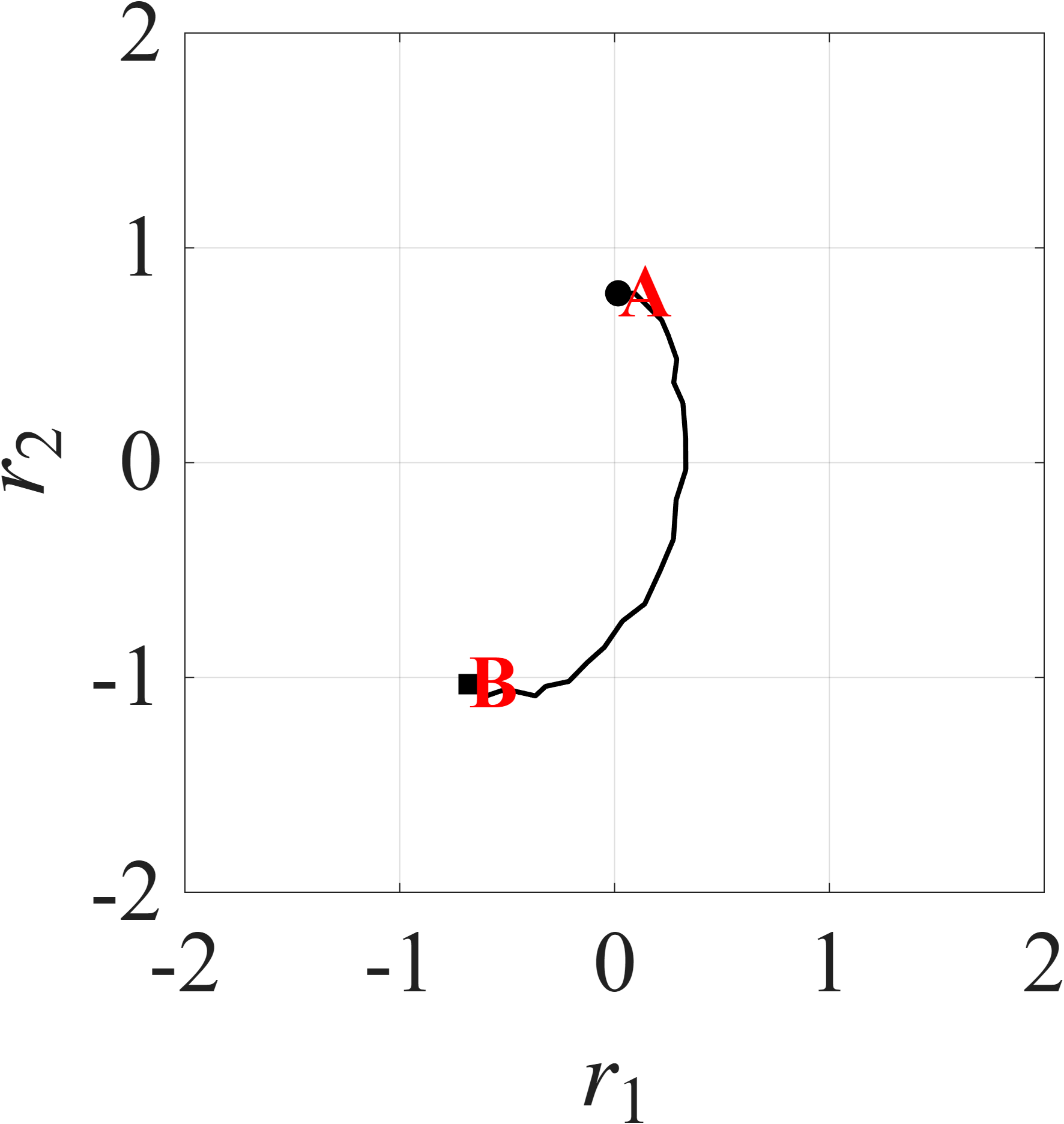}}
	\caption{Sample outputs of the Z-GAN2 generator.}
	\label{fig-zgan2-sample}
\end{figure}

\paragraph{Visual Assessment:}
\figser{fig-SGAN-mintime}{fig-zgan2-sample} shows the position~$\vec{\posState}$ 
components of generator output samples from
each of the three GAN models. Note that the S-GAN outputs do not resemble the OTD, 
whereas the Z-GAN outputs visually resemble the OTD samples.

\paragraph{Statistical Similarity:}
To measure statistical similarity, \tbl{tbl-moments-gan} shows statistical
moments (up to four significant digits) of the first three principal components
of the OTD $\dataset$ in comparison to those of the datasets $\datasetGen$
generated by the S-GAN and Z-GAN2 models. \revnew{Quantities nearest to the
	training dataset moments are indicated in bold font.}
\revrpl{We can notice that the S-GAN and
Z-GAN2 dataset moments differ widely from those of the OTD along the first and
third principal component. The principal moment along the second component is
similar to that of the OTD.}{Notice that the S-GAN and Z-GAN2 show large
differences compared to the OTD.} Note also that the GAN outputs are
clustered together which is indicative of mode collapse. \figf{fig-pca-gan}
provides a scatter plot visualization of these observations, where mode collapse
is evident in the dense clustering of the generated outputs (red and green
dots).

\def\thisfigwidth{0.65\columnwidth}
\def\thisfigspace{0.00\columnwidth}
\begin{figure}
	\centering
	{\includegraphics[width=\thisfigwidth]{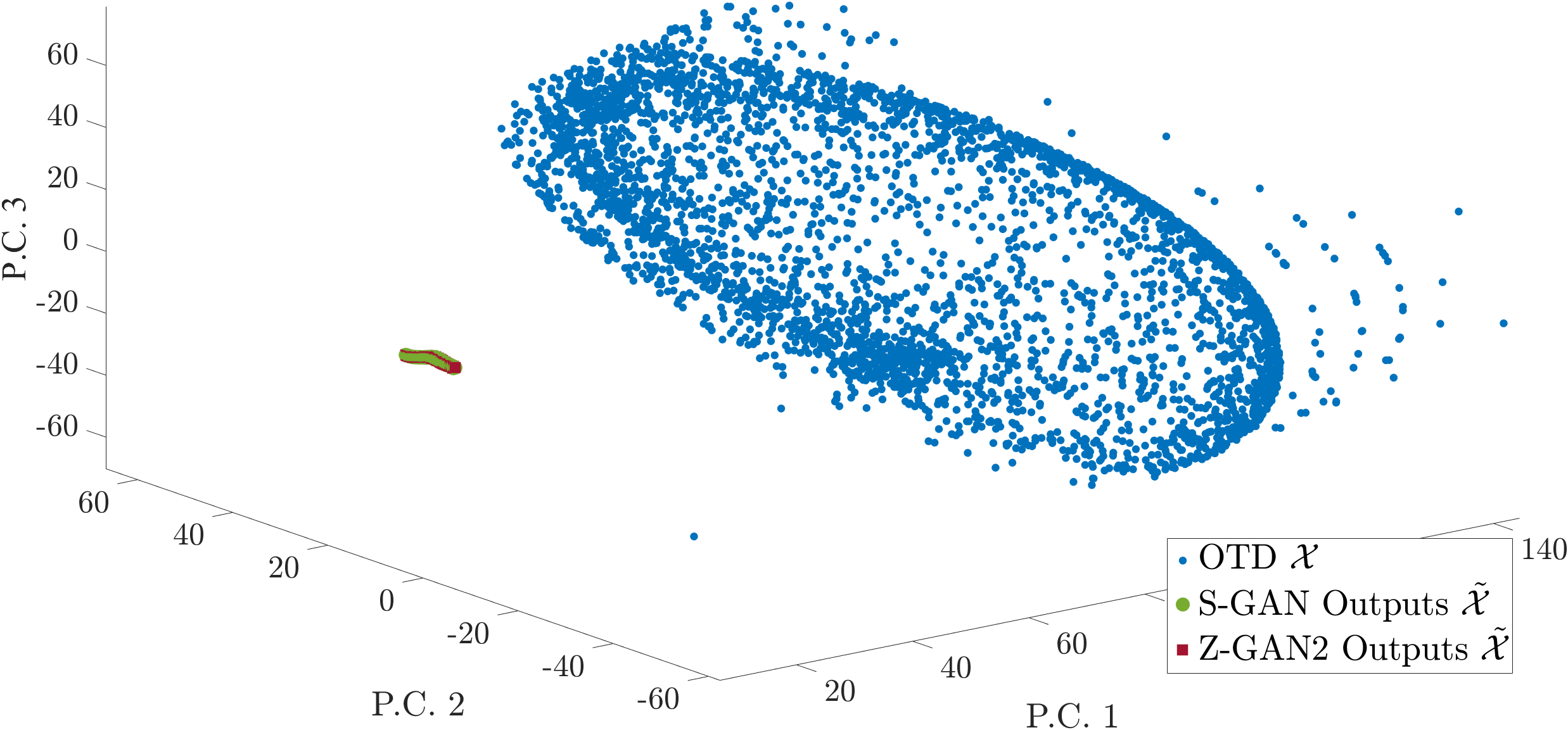}}
	\hspace{\thisfigspace}
	\caption{Scatter plot of first three principal components (P.C.) of data points in the OTD 
	and generated datasets for S-GAN and Z-GAN2 with $\nData = 3000$ and $\nGen = 1000$. Mode 
	collapse is evident.}
	\label{fig-pca-gan}
\end{figure}

\begin{table}[htbp]
	\centering
	\caption{Statistical moments of GAN-generated datasets for the 
		Zermelo navigation problem with $\nGen = 1000$.}
	\label{tbl-moments-gan}
	\renewcommand{\arraystretch}{1.3}
	\resizebox{\textwidth}{!}{%
		\begin{tabular}{l ccc ccc ccc ccc}
			\toprule
			\textbf{} & \multicolumn{3}{c}{\textbf{Mean}} &
			\multicolumn{3}{c}{\textbf{Variance}} &
			\multicolumn{3}{c}{\textbf{Skewness}} &
			\multicolumn{3}{c}{\textbf{Kurtosis}} \\
			\cmidrule(lr){2-4} \cmidrule(lr){5-7} \cmidrule(lr){8-10} \cmidrule(lr){11-13}
			$\dataset$ & 93.83 & 2.900 & 2.410 & 
			161.9 & 1407 & 1212 & 
			-0.1800 & -0.0400 & 0.3100 & 
			2.700 & 1.780 & 1.850 \\
			
			$\datasetGen$ (S-GAN) & -94.85 & \textbf{1.027} & \textbf{0.0736} & 
			\textbf{76.28} & \textbf{136.1} & \textbf{18.97} & 
			0.8060 & \textbf{0.0764} & \textbf{-0.04820} & 
			\textbf{2.543} & \textbf{1.615} & \textbf{2.218} \\
			
			$\datasetGen$ (Z-GAN2) & \textbf{8.011} & 0.1570 & -6.100\engE{-3} & 
			0.4960 & 3.373 & 0.1170 & 
			\textbf{-0.7310} & 0.3410 & -0.7610 & 
			1.991 & 2.011 & 4.286 \\
			\bottomrule
		\end{tabular}
	}
\end{table}

\paragraph{Performance Indices:}
\tbl{tbl-sigma-gan-new} shows the performance of the GAN models by indicating
the minimum, maximum, mean, and standard deviation on the performance indices in
\eqnnt{eq-sigma-gan}. \revnew{Lowest values are indicated in bold font.}
Note that the Z-GAN2 performance measure on
all statistical measures is better than Z-GAN1 and S-GAN, with the exception of
$\simMet_{1}$ for Z-GAN1, which shows the best performance. Both Z-GAN2 and
Z-GAN1 outperform S-GAN on all the defined performance indices. 

%

\begin{table}
	\centering
	\caption{Performance indices for the three GAN models with $\nData = 3000$ and 
		$\nGen = 1000$ for the Zermelo  problem.}
	\label{tbl-sigma-gan-new}
	\resizebox{\textwidth}{!}{%
		\begin{tabular}{l cccc cccc cccc}
			\toprule
			\textbf{} & \multicolumn{4}{c|}{$\simMet_{1}$} & \multicolumn{4}{c|}{$\simMet_{2}$} & 
			\multicolumn{4}{c}{$\simMet_{3}$} \\
			
			\textbf{} & Mean & Std.dev. & Max. & Min. & Mean & Std.dev. & Max. & Min. & Mean & 
			Std.dev. & Max. & Min. \\
			\midrule
			S-GAN  & 81.89 & 8.147 & 100.5 & 69.59& 3.387 & 0.3392 &4.169 & 2.917 & 303.2 & 
			4743 & $1.494\engE{5}$ & 25.09\\
			\midrule
			Z-GAN 1 & \textbf{1.279} & \textbf{0.6417} & 4.967 & 0.4336 & 1.526 & 0.1267 & 1.917 & 1.207 & 
			112.3 & 1933 & $5.960\engE{4}$& 3.619\\
			\midrule
			Z-GAN 2 & 2.399 & 1.232 & \textbf{4.713} & \textbf{0.3536} & $\textbf{0.0050}$ & 
			$\textbf{0.0025}$ & $\textbf{0.0136}$ & $\textbf{0.0017}$ & 
			\textbf{0.1095} & $\textbf{0.0594}$ &\textbf{0.2305} & $\textbf{0.0168}$\\
			\bottomrule
		\end{tabular}
	}
\end{table}

\paragraph{Other Characteristics:}
We tested the proposed Z-GANs with OTDs consisting of trajectory data sampled 
at a higher rate, i.e., we increased $\nSample$ from 25 to $50$ and then to $100$. 
No significant difference in performance was observed.

The discriminator is a classifier, and for classifier training
it is common to use a binary cross-entropy (BCE) loss function.
Our choice of an MSE loss instead of BCE is driven by observations of the 
generator's performance. We implemented different versions of the GANs with BCE and MSE
losses. The S-GAN performance did not significantly change. For the Z-GANs, 
using the BCE loss instead of MSE caused mode collapse that we could not resolve.

%



\subsection{S-VAE and Z-VAE Implementation}

For the two VAEs, the encoder and the decoder NNs were implemented as multilayer
perceptrons with a latent space size of 32. The S-VAE was implemented with six
hidden layers, and the Z-VAE with five hidden layers. The rectified linear unit
(ReLU) function~\cite{Banerjee2019} was chosen as the activation function for
both VAEs. The batch size was chosen as $M_b = 32$, and the learning rates was
set to 0.001.

%

\paragraph{Visual Assessment:}
\figfs{fig-svae-mintime-500}{fig-zvae-mintime-500} show the position $r$
sample outputs from the two VAE models. The visible deviation between the S-VAE
outputs and the real trajectories is more pronounced compared to that of the
Z-VAE. Additionally, one might observe several ``kinks'' in the output samples of
S-VAE that are absent in Z-VAE. Also, the error in time of flight and the
physical shape is more pronounced in the S-VAE generated samples.

We also present the percentage deviations in both the physical trajectory shape
and the time of flight, denoted as~$\Delta r$ and~$\Delta t$ respectively. These
deviations are expressed as the percentage change relative to a real trajectory
for that specific initial conditions and parameter choice (i.e the wind field),
\revnew{i.e., lower these deviations, closer are the generated outputs to the true optimal.
Notice that the Z-VAE generated times of flight (which is the metric of optimality)
are closer to the true optimal.}

\paragraph{Statistical Similarity:}
\tbl{tbl-vae-mintime} provides statistical moments of the VAE-generated datasets
in comparison to the OTD for $\nData = 4000.$
Note that, \revnew{even with the large volume of training data
when one expects the S-VAE to match the physics-informed Z-VAE,} 
the moments along all principal axes
for the Z-VAE are closer to those of the OTD. 
This is further illustrated in the scatter plots in
\figf{fig-pca-vae-mintime}, where it is evident for $\nData = 4000$ 
that the Z-VAE outputs (red dots) are distributed similarly as the 
OTD (blue dots), whereas the S-VAE output
distribution (green dots) is dissimilar. \revnew{Note that with $\nData = 500,$
i.e., with low training data volume, the Z-VAE outputs are somewhat similarly
distributed as the S-VAE. This observation leads us to conclude that merely
adding a residual term of the governing equations to the loss, as is done for
$\loss\msub{ZVAE}$ in \eqnnt{eq-zvae-loss} may not suffice to train a GAIM
when training data volume is low. In the next subsection, we show that the 
Split-VAE performs better even with scarce training data.}

\begin{table}
	\centering
	\caption{Statistical moments of VAE-generated datasets
	for the Zermelo navigation problem with $\nGen = 1000$.}
	\label{tbl-vae-mintime}
	\renewcommand{\arraystretch}{1.3}
	\resizebox{\textwidth}{!}{%
		\begin{tabular}{l ccc ccc ccc ccc}
			\toprule
			\textbf{} & \multicolumn{3}{c}{\textbf{Mean}} &
			\multicolumn{3}{c}{\textbf{Variance}} &
			\multicolumn{3}{c}{\textbf{Skewness}} &
			\multicolumn{3}{c}{\textbf{Kurtosis}} \\
			\cmidrule(lr){2-4} \cmidrule(lr){5-7} \cmidrule(lr){8-10} \cmidrule(lr){11-13}
			$\dataset$ & 152.0 & 3.100 & 1.249 &
			267.1 & 1384 & 1295 &
			0.3382 & -0.1550 & 0.0748 &
			1.790 & 1.797 & 1.830 \\
			
			$\datasetGen$ (S-VAE) & -154.5 & -1.605 & -2.122 &
			238.8 & 1097 & 1058 &
			0.0166 & 0.0026 & \textbf{0.3067} &
			\textbf{1.767} & \textbf{1.943} & \textbf{2.090} \\
			
			$\datasetGen$ (Z-VAE) & \textbf{154.4} & \textbf{2.494} & \textbf{-1.067} &
			\textbf{246.1} & \textbf{1161} & \textbf{1124} &
			\textbf{0.0330} & \textbf{-0.0495} & 0.5337 &
			1.806 & 1.978 & 2.205 \\
			\bottomrule
		\end{tabular}
	}
\end{table}

\paragraph{Performance Indices:} The performance of the two VAEs is evaluated
using the indices defined in \eqnnt{eq-sigma-gan} across various values of
$\nData$. The Z-VAE demonstrates superior performance across most statistical
measures. The results are presented in \tbl{tbl-vae-4000-mintime-new}. For clarity,
the best-performing measures are highlighted in bold, indicating better
performance regardless of the volume of training data. As noted above,
the results are mixed with low training data volume ($\nData = 500$).

\def\thisfigwidth{0.23\columnwidth}
\def\thisfigspace{0.01\columnwidth}
\begin{figure}
	\centering
	\subfigure{\includegraphics[width=\thisfigwidth]{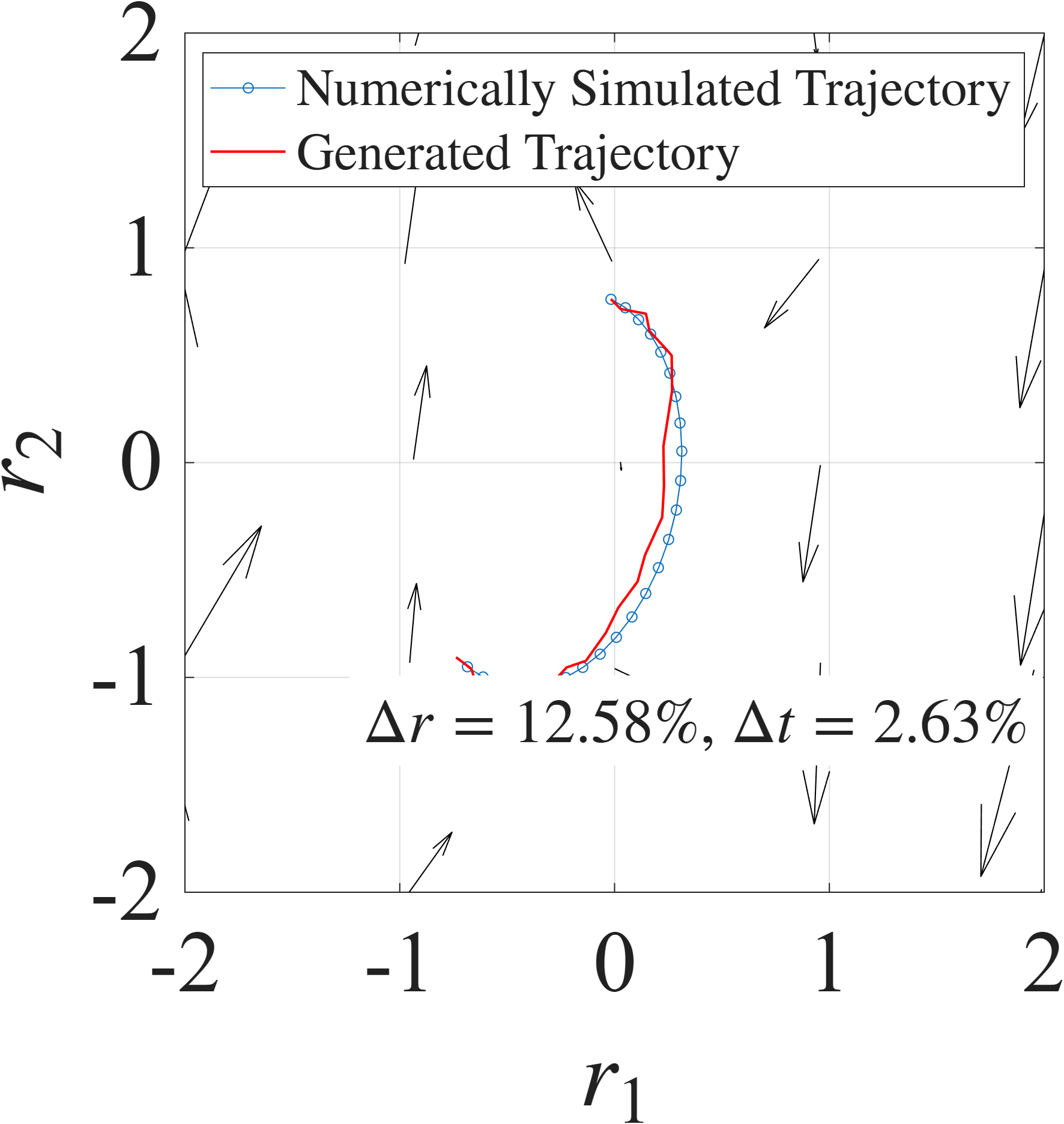}}
	\hspace{\thisfigspace}
	\subfigure{\includegraphics[width=\thisfigwidth]{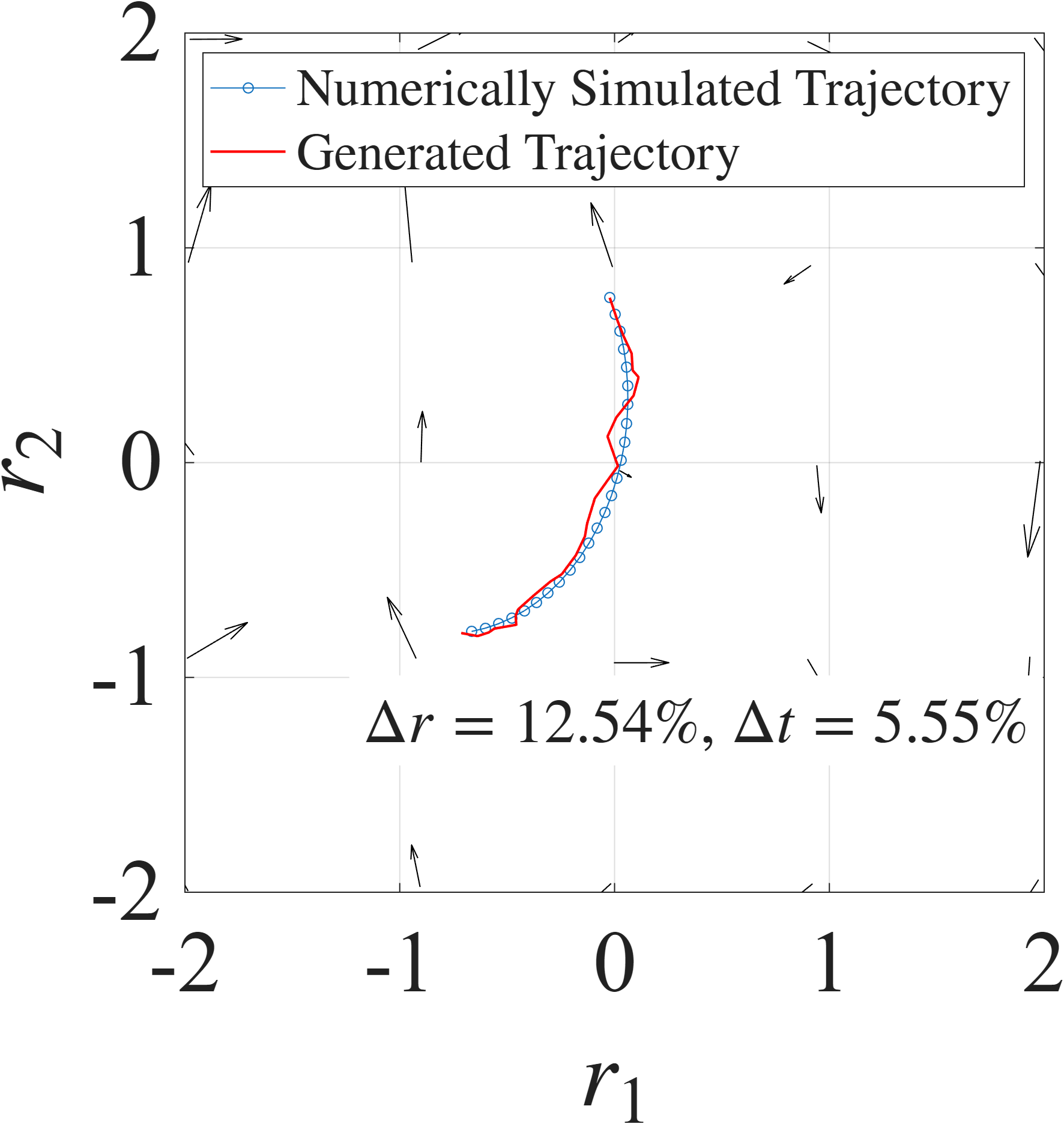}}
	\hspace{\thisfigspace}
	\subfigure{\includegraphics[width=\thisfigwidth]{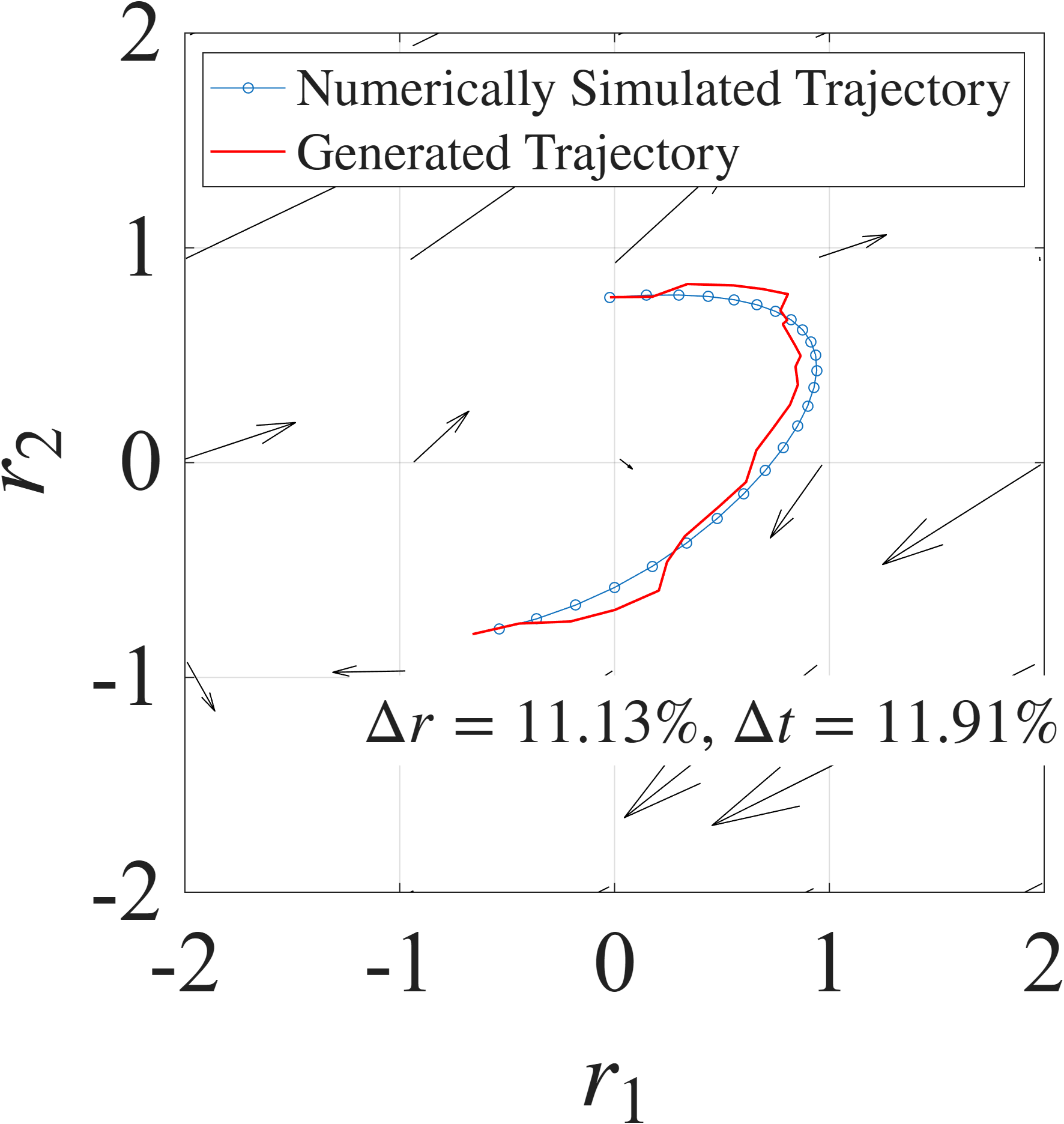}}
	\hspace{\thisfigspace}
	\subfigure{\includegraphics[width=\thisfigwidth]{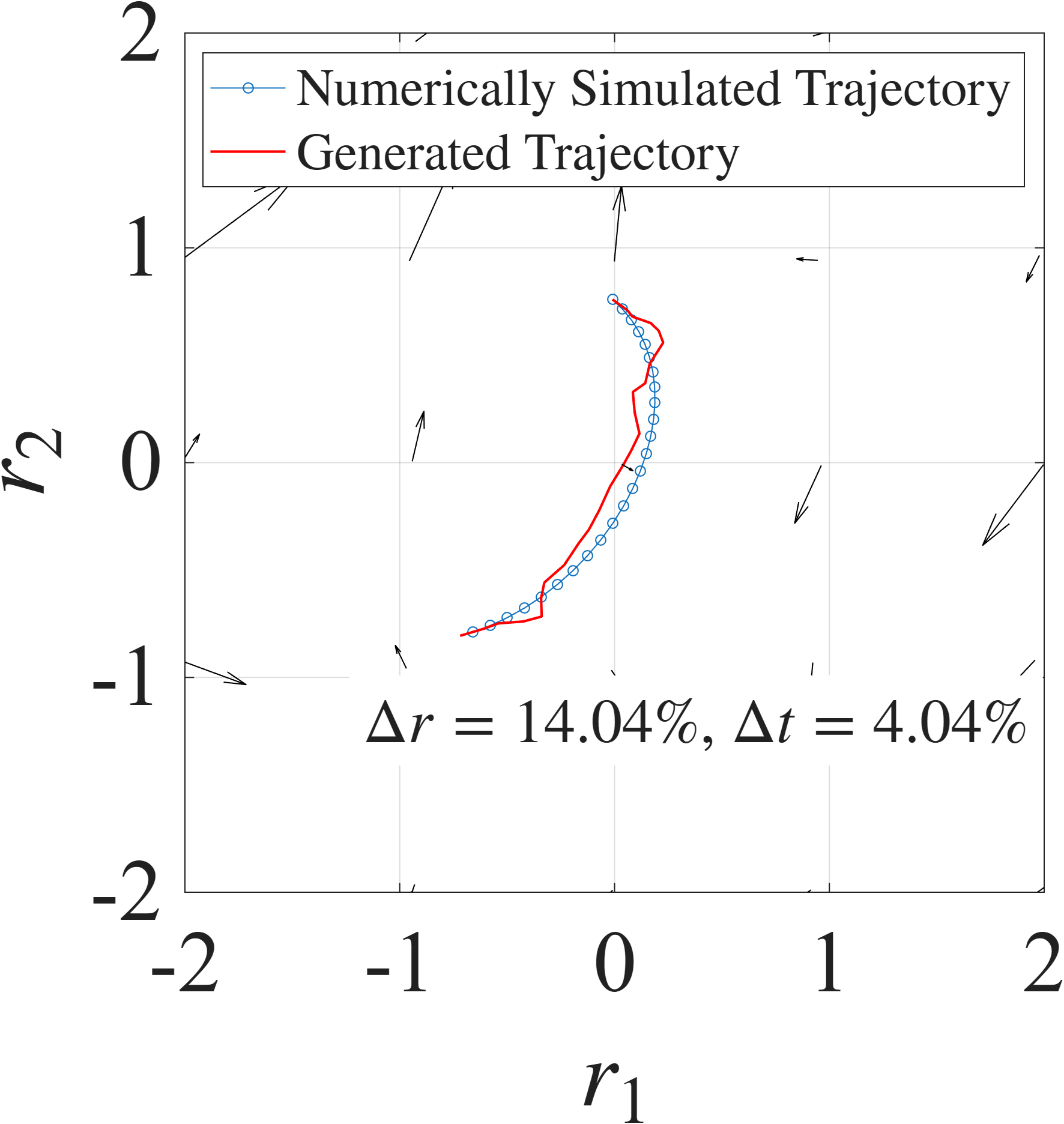}}
	\caption{Sample outputs of the S-VAE for $\nData = 500.$}
	\label{fig-svae-mintime-500}
\end{figure}
%

\begin{figure}
	\centering
	\subfigure{\includegraphics[width=\thisfigwidth]{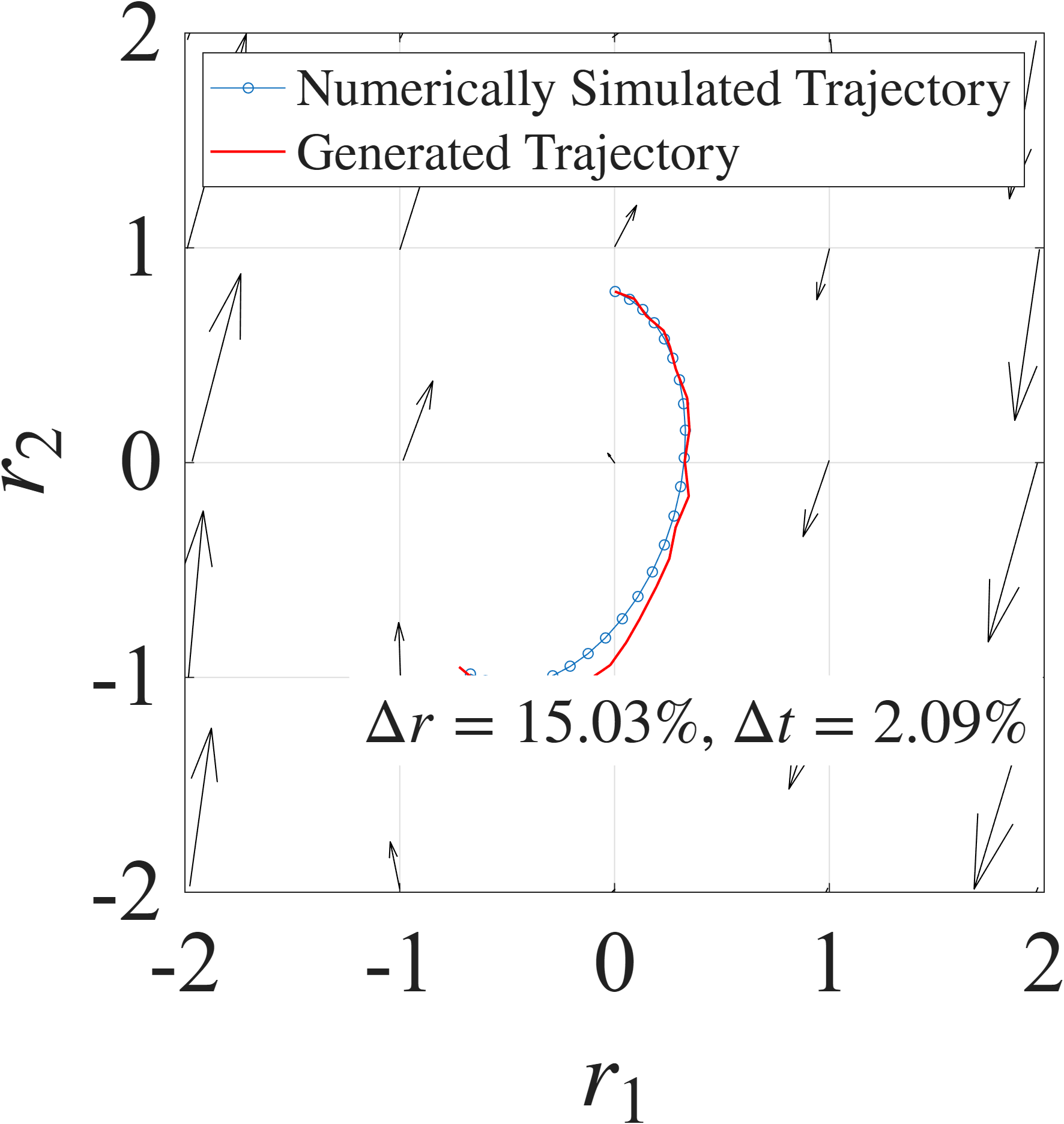}}
	\hspace{\thisfigspace}
	\subfigure{\includegraphics[width=\thisfigwidth]{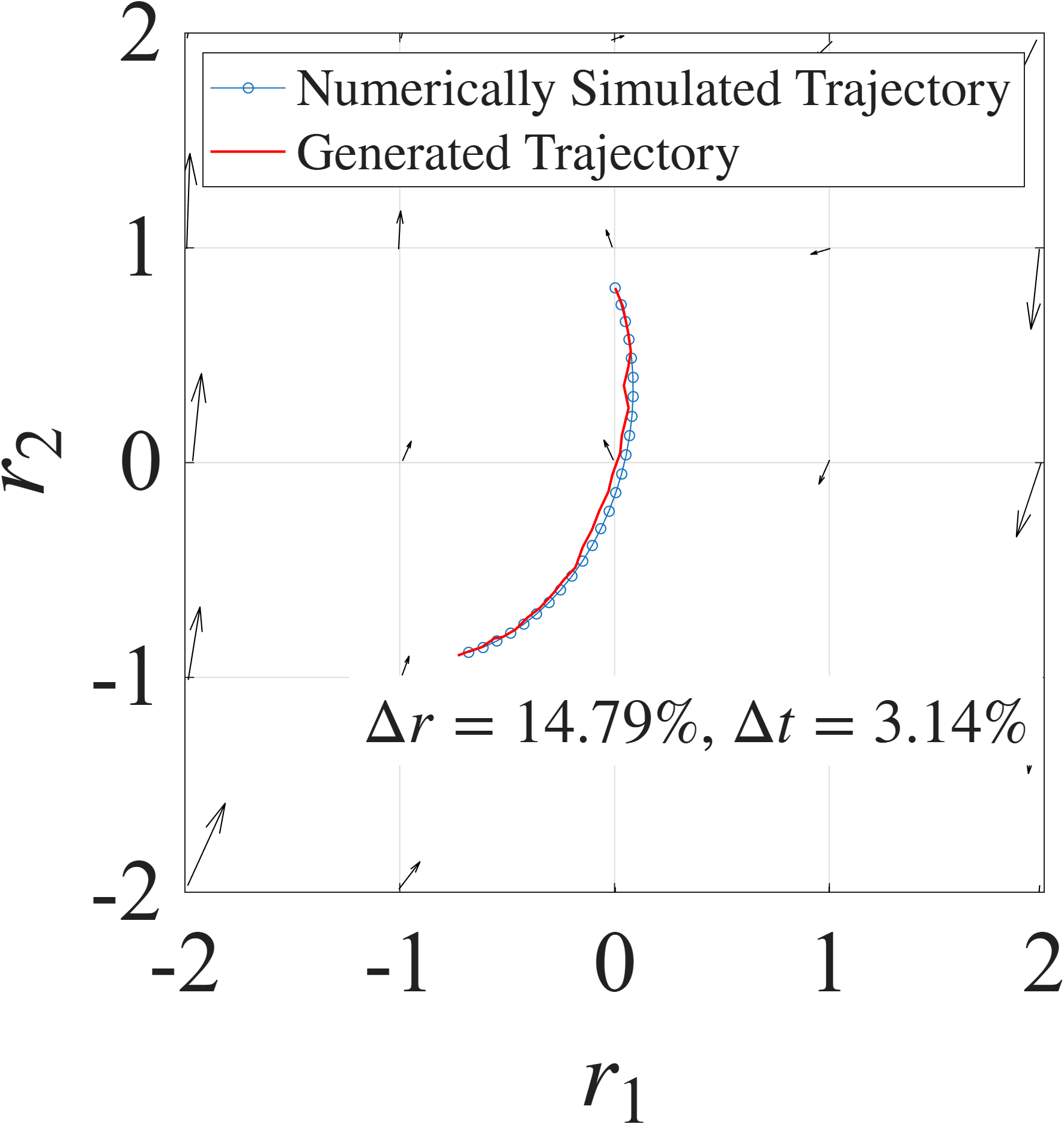}}
	\hspace{\thisfigspace}
	\subfigure{\includegraphics[width=\thisfigwidth]{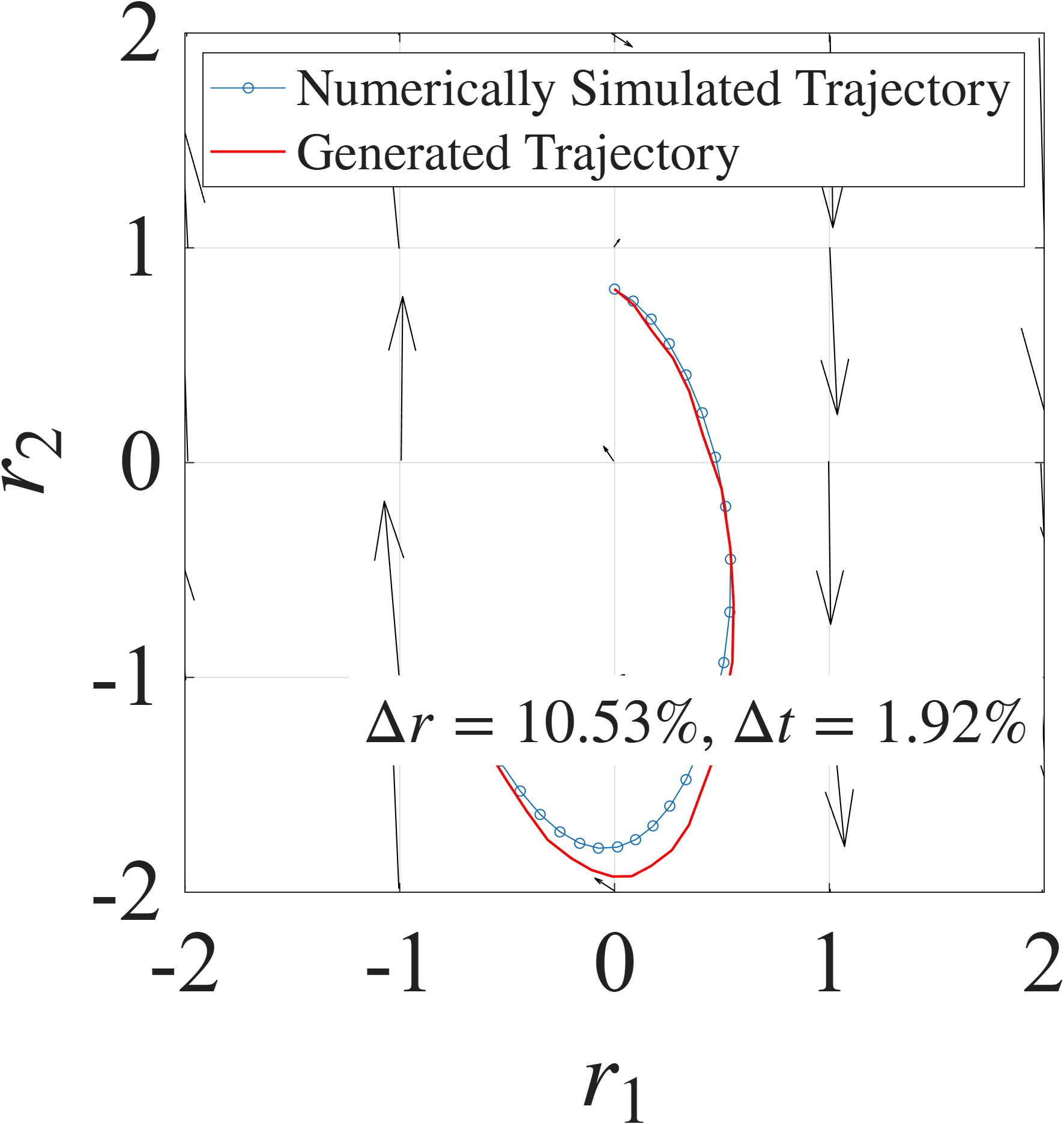}}
	\hspace{\thisfigspace}
	\subfigure{\includegraphics[width=\thisfigwidth]{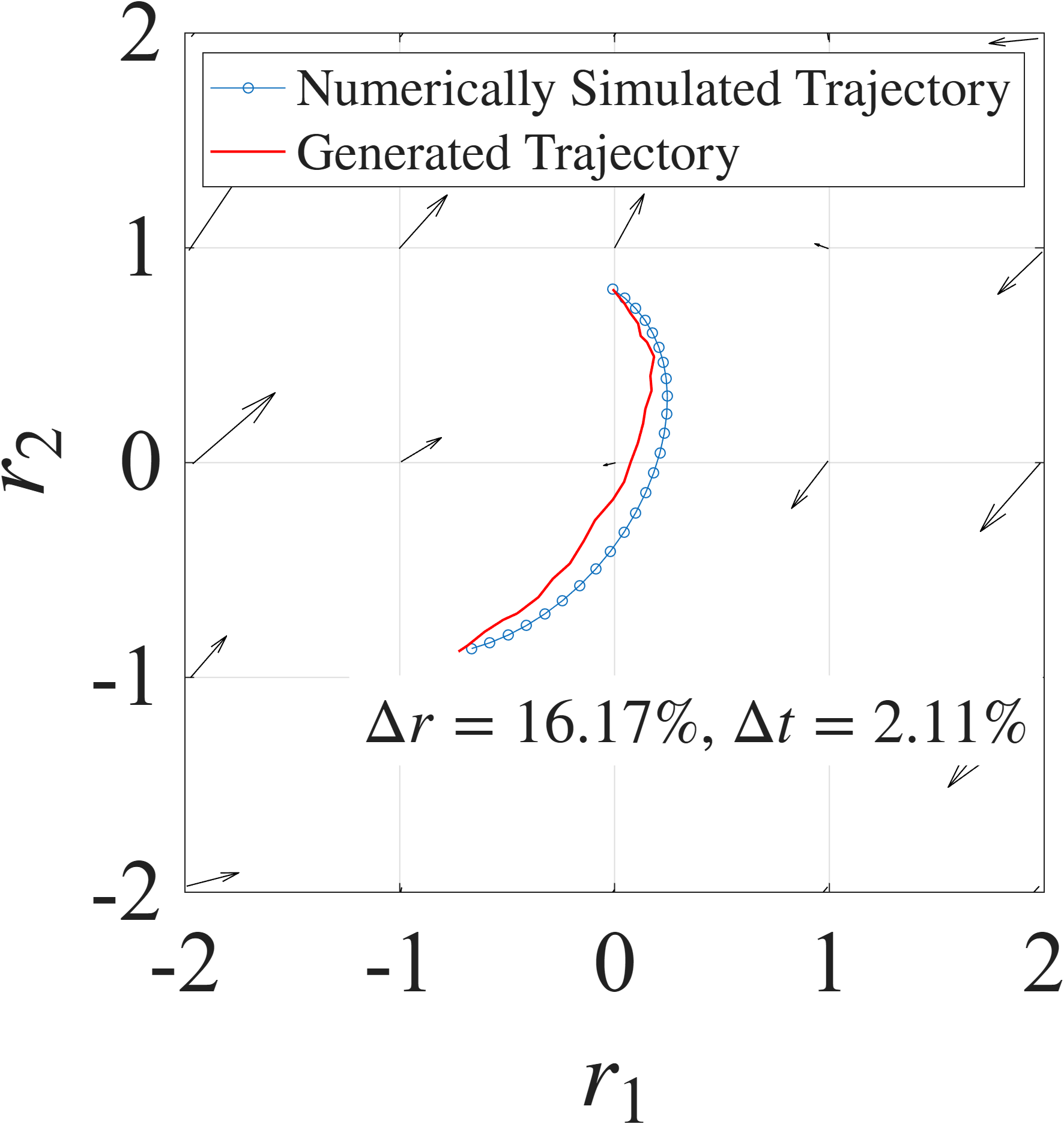}}
	\caption{Sample outputs of the Z-VAE for $\nData = 500.$}
	\label{fig-zvae-mintime-500}
\end{figure}
%

\def\thisfigwidth{0.65\columnwidth}
\def\thisfigspace{0.01\columnwidth}
\begin{figure}[ht]
	\centering
	\subfigure[\revnew{$\nData = 4000.$}]
	{\includegraphics[width=\thisfigwidth]{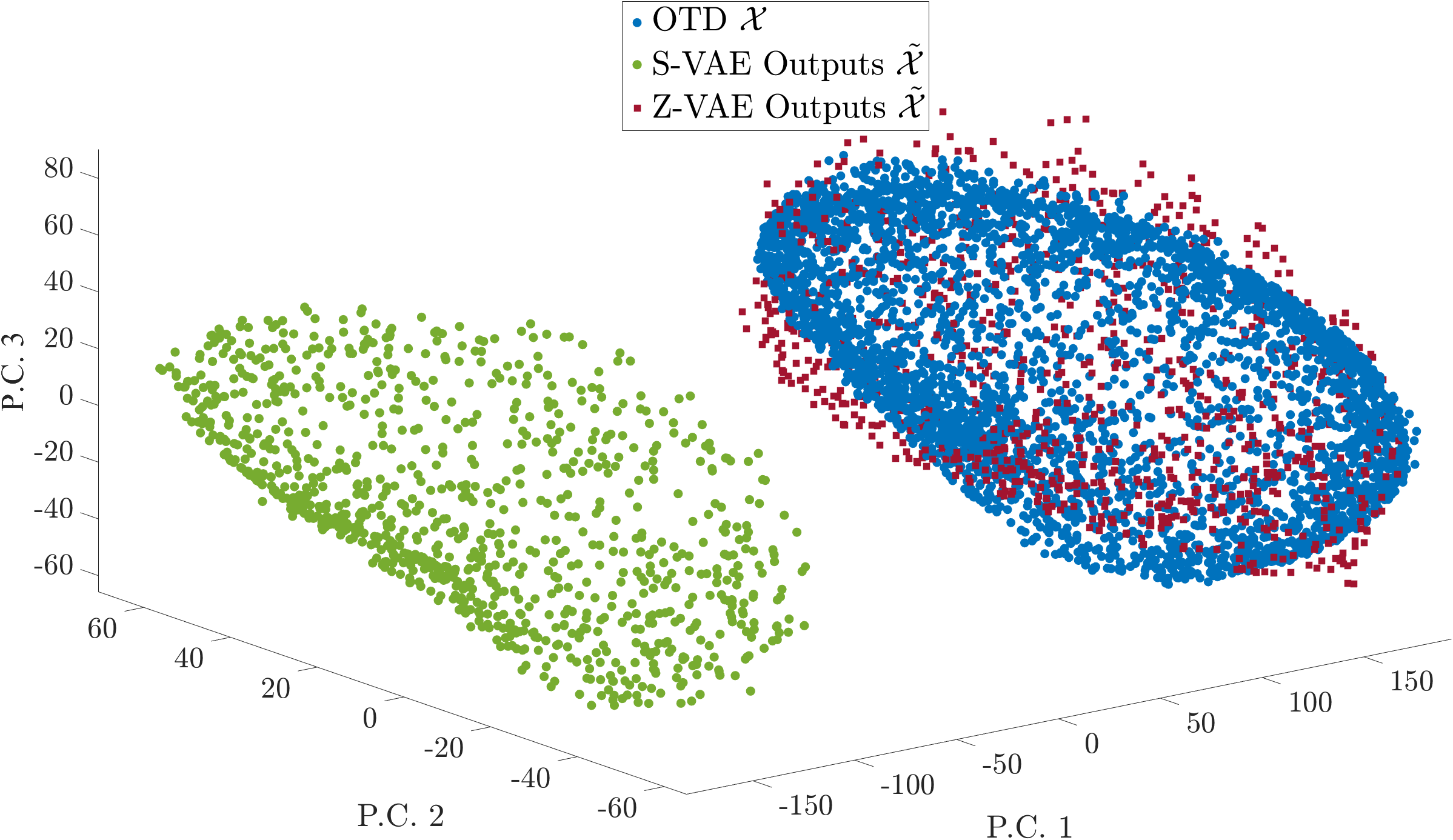}}
	\hspace{\thisfigspace}
	\subfigure[\revnew{$\nData = 500,$ i.e., not all points in the OTD pool are used for training.}]
	{\includegraphics[width=\thisfigwidth]{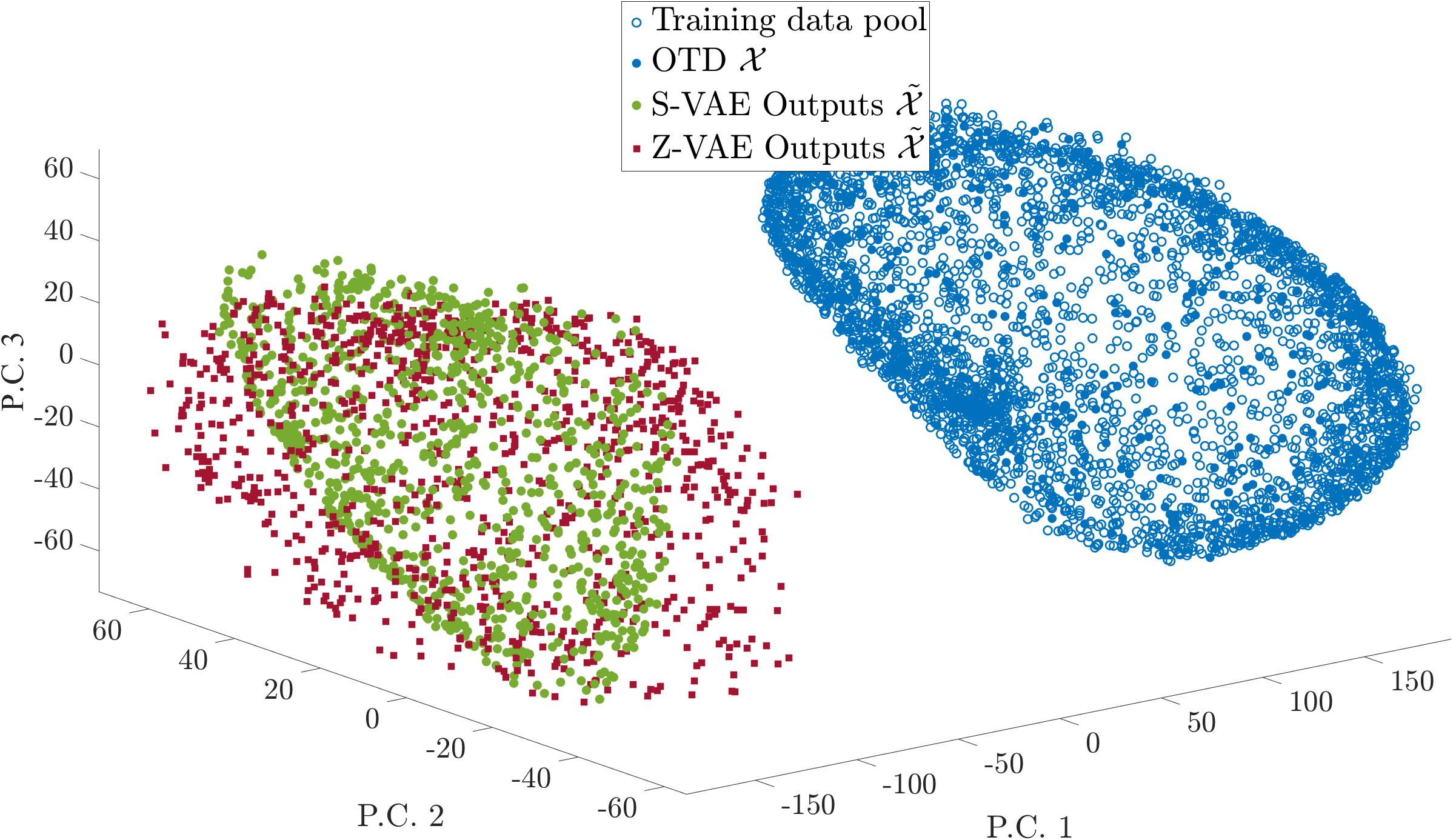}}
	\hspace{\thisfigspace}
	\caption{Scatter plots of first three principal components (P.C.) of data points in the OTD 
		and generated datasets for S-VAE and Z-VAE with $\nGen = 1000$ for the Zermelo problem.}
	\label{fig-pca-vae-mintime}
\end{figure}

\begin{table}
	\centering
	\caption{Performance indices of the two VAE models with two different values
		of $\nData$ and with $\nGen = 1000$ for the Zermelo problem.}
	\label{tbl-vae-4000-mintime-new}
	\resizebox{\textwidth}{!}{%
		\begin{tabular}{l cccc cccc cccc}
			\toprule
			& \multicolumn{4}{c}{$\simMet_{1}$} 
			& \multicolumn{4}{c}{$\simMet_{2}$} & \multicolumn{4}{c}{$\simMet_{3}$} \\
			
			& Mean & Std.dev. & Max. & Min. & Mean & Std.dev. & Max. & Min. & Mean & Std.dev. 
			& Max. & Min. \\
			\midrule
			& \multicolumn{12}{c}{$\boldsymbol{\nData = 4000}$} \\\midrule
			S-VAE  & 6.769 & 8.255 & 56.44 & 0.9937 & 4.223 & 2.204 & 17.986 & \textbf{1.240} & 40.56 & 409.3 & $1.029\engE{4}$ & 1.831 \\
			\midrule
			Z-VAE  & \textbf{1.760} & \textbf{5.789} & \textbf{55.29} &\textbf{0.2040} & \textbf{4.167} &\textbf{1.892} & \textbf{14.95} & 1.406 & 
			\textbf{16.51} & $\textbf{1.409\engE{2}}$ & $\textbf{3.053\engE{3}}$ &\textbf{0.3962} \\
			\midrule 
			& \multicolumn{12}{c}{$\boldsymbol{\nData = 500}$} \\\midrule
			S-VAE  & 11.98 & \textbf{10.25}  & 54.54& 4.609 & 4.106 & \textbf{1.971} & 16.73 & 0.1109 & \textbf{123.90} & 
			\textbf{406.13} & $\textbf{6.7197\engE{3}}$ &7.002 \\
			\midrule
			Z-VAE  & \textbf{7.893} & 12.40 & \textbf{52.39} & \textbf{0.6394} & \textbf{3.892} & 1.976 &\textbf{16.26} & $\textbf{0.5124\engE{-1}}$ & 132.8 & 
			901.6 & $1.699\engE{4}$ & \textbf{1.666} \\
			\bottomrule
		\end{tabular}
	}
	
\end{table}

\subsection{Minimum Threat Exposure Problem} 
For this study, we considered as \emph{observed} training data solutions of
\eqnser{eq-minthreat-dynamics1}{eq-minthreat-dynamics2} with various values of
the cost weight parameter $\lambda.$ Specifically, the OTD consisted of
solutions of \eqnser{eq-minthreat-dynamics1}{eq-minthreat-dynamics2} with
$\lambda = 2, 5,$ and $10.$ A training data pool of 1000 such trajectories was
synthesized.
For the \emph{model} trajectories and governing equations, we fixed $\lambda =
1.$ In this sense, the observed trajectories do not exactly satisfy the
governing equations.

The encoder and the decoder ANNs for the two VAE models described in
\scn{sec-min-threat} were implemented as multilayer perceptrons with a latent
space size of 32. The rectified linear unit (ReLU) function was chosen as the
activation function. The learning rates were chosen as 0.001 for both VAEs.
\tbl{tbl-vae-min-threat-dimensions} in the Appendix provides the dimensions of
each layer for the two VAEs.

A crucial observation made during training the Split-VAE model was the
importance of an optimal amount of model trajectory data. This was essential
because providing a larger number of \mydef{model} trajectory samples led the
Split-VAE to generate outputs statistically similar to \mydef{model} trajectory
data, while providing fewer of these samples resulted in poor training outcomes.
Thus, finding the optimal combination of \mydef{observed} and \mydef{model}
trajectory samples was key to successful training. To that end, we used $\nData
=200$ of \mydef{observed} and trajectories for training, along with an equal
number of model trajectories.

For training the S-VAE, we used only the  training dataset of \mydef{observed}
trajectories with $\nData = 200$ samples. The training process was similar to that
of a standard VAE.
\revnew{The observed trajectories contain disturbances arising from 
unmodeled dynamics or stochasticity not captured by the governing equations. Noiseless 
(or model-based) trajectories are synthetically generated by simulating the known model dynamics 
under specified initial and boundary conditions. Therefore, noiseless trajectory examples are abundant and can easily exceed the number of OTD data points.}

\revnew{Importantly, these model trajectories serve not as exact analogs but as
	approximations of real-world behavior, with deviations primarily due
	to noise or unmodeled effects. In the Split-VAE, this separation gives us the ability
	to take advantage of the shared structure across both data domains. By
	leveraging the model trajectories, we can provide a robust inductive prior that
	guides the learning of latent representations from the real-world data. This
	improves generalization and robustness, particularly in scenarios where
	real-world data is limited, or heavily corrupted.}

\revnew{The proposed split latent space architecture framework is scalable and
	tolerant to dataset imbalance. Furthermore, for cases of extreme imbalance, we
	have the option of weighting reconstruction or KL-divergence terms during
	training. Alternatively, we can apply data rebalancing techniques which will not
	distort the true model.}

%

To assess the similarity of the generated dataset $\dataset =
\{\datum_{i}^{g}\}_{i=1}^{\nGen}$, we evaluated the performance of the VAEs on
$\simMet_{1i}$ which we redefined as: \begin{align*} \simMet_{1i} & := \lVert
	H_{\lambda}[\datum_{i}] \rVert^2, \end{align*} This performance index tests the
deviation of $\datum_{i}^{g}$ from ~\eqnnt{eq-hamiltonian-minthreat}. It is
important to note that the Hamiltonian is a function of the parameter $\lambda$.
Therefore, we must select the appropriate value of $\lambda$ for each observed
trajectory for $\simMet_{1}$ calculation.

\paragraph{Visual Assessment:}
\figfs{fig-splitvae-200_2}{fig-svae-200_2} show sample outputs from the two
VAE models, plotted in the position variables \(r\). The color bar on the side
represents the intensity of the threat field. Note that several irregularities
are visible in the trajectories generated by the S-VAE compared to those
produced by the Split-VAE.

Additional sample results comparing the outputs of the S-VAE and Split-VAE for
different values of the constant \(\lambda\) are provided in
\figser{fig-splitvae-200_5}{fig-splitvae-200_10} in the Appendix. To evaluate these
outputs quantitatively, we employed a total variance-based performance index
\cite{pedersen2011total} to quantify the overall ``smoothness'' of the outputs
generated by both VAEs. These results are summarized in
\tbl{tbl-minthreat-total-variance}, which displays the total variance computed
for each of the 1000 generated samples from both VAEs across different $\lambda$
values. \tbl{tbl-minthreat-total-variance} provides information on the mean,
standard deviation, maximum and the minimum values of total variance 
of the generated datasets.

Smoothness in the generated outputs is an essential criterion as it reflects
adherence to the underlying equations of motion. A lower total variance measure
indicates greater smoothness. The results show that the Split-VAE outputs
exhibit significantly lower total variance compared to those generated by the
S-VAE. This finding corroborates our earlier observation based on the physical
shapes of the outputs, where the S-VAE-generated samples displayed more
pronounced irregularities.

\begin{table}
	\centering
	\caption{Total variance measure 
		with $\nData = 200$ and $\nGen = 1000$ for the minimum threat 
		exposure problem.}
	\label{tbl-minthreat-total-variance}
	\begin{tabular}{l rr  rr  rr}
		\toprule 
		& \multicolumn{2}{c}{$\lambda = 2$} 
		& \multicolumn{2}{c}{$\lambda = 5$} 
		& \multicolumn{2}{c}{$\lambda = 10$} \\
		& S-VAE & Split-VAE & S-VAE & Split-VAE & S-VAE & Split-VAE \\
		\midrule
		Mean
		& 2.188 & \textbf{0.4174} & 5.366 & \textbf{0.5595} & 3.051 & \textbf{0.6715} \\
		Std. dev.
		& 0.5421& \textbf{0.1078}& 0.8556& $\textbf{0.09776}$ & 0.8069 & \textbf{0.3101} \\
		Maximum
		& 4.415 & \textbf{1.018} & 8.636 & \textbf{1.132}& 6.922 & \textbf{4.213}\\
		Minimum
		&0.9670 & \textbf{0.1887} & 2.570 & \textbf{0.3082} & 1.385 & \textbf{0.2448}\\
		\bottomrule
	\end{tabular}
\end{table}

\def\thisfigwidth{0.23\columnwidth}
\def\thisfigspace{0.01\columnwidth}
\begin{figure}[t]
	\centering
	\subfigure{\includegraphics[width=\thisfigwidth]{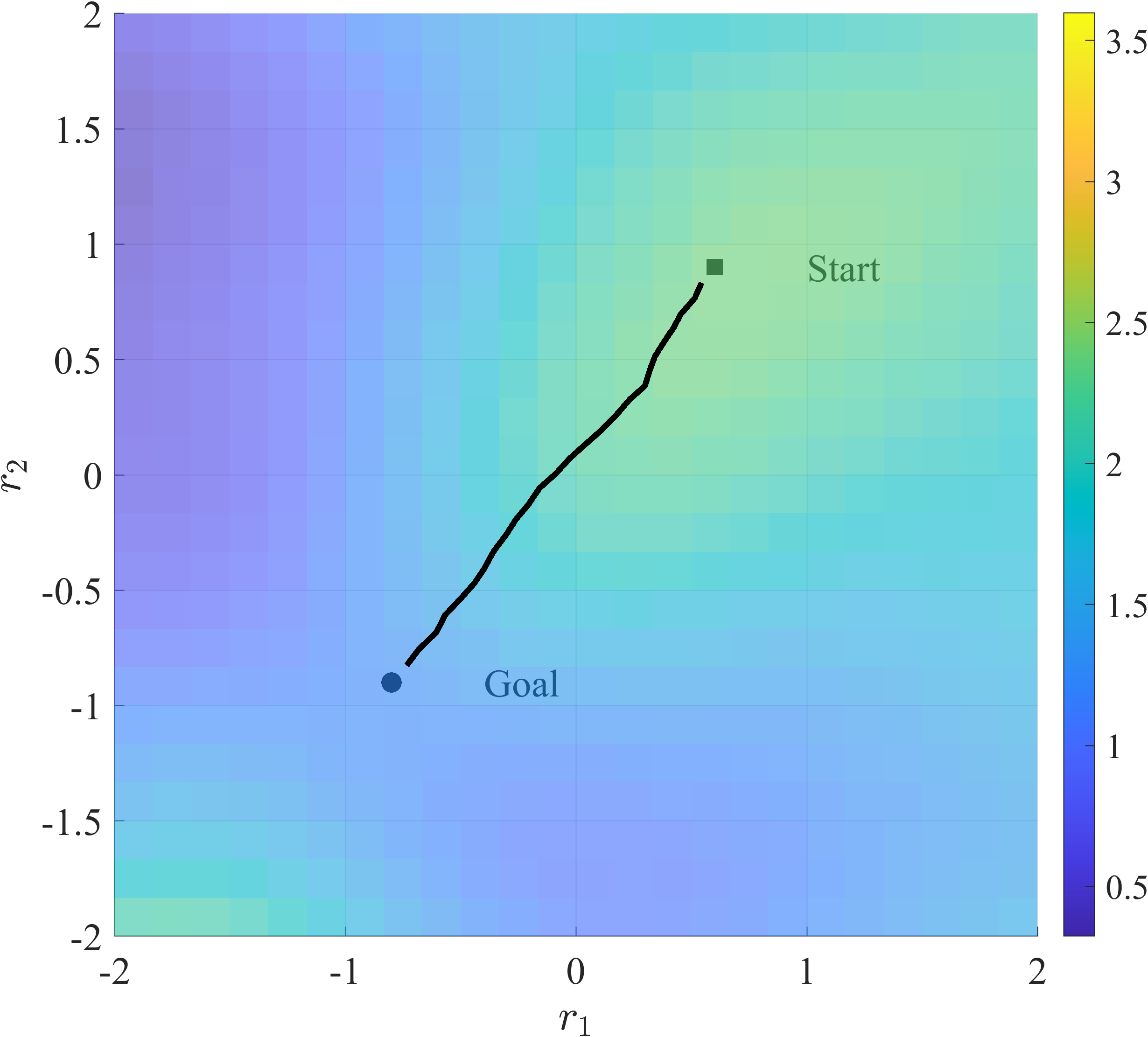}}
	\hspace{\thisfigspace}
	\subfigure{\includegraphics[width=\thisfigwidth]{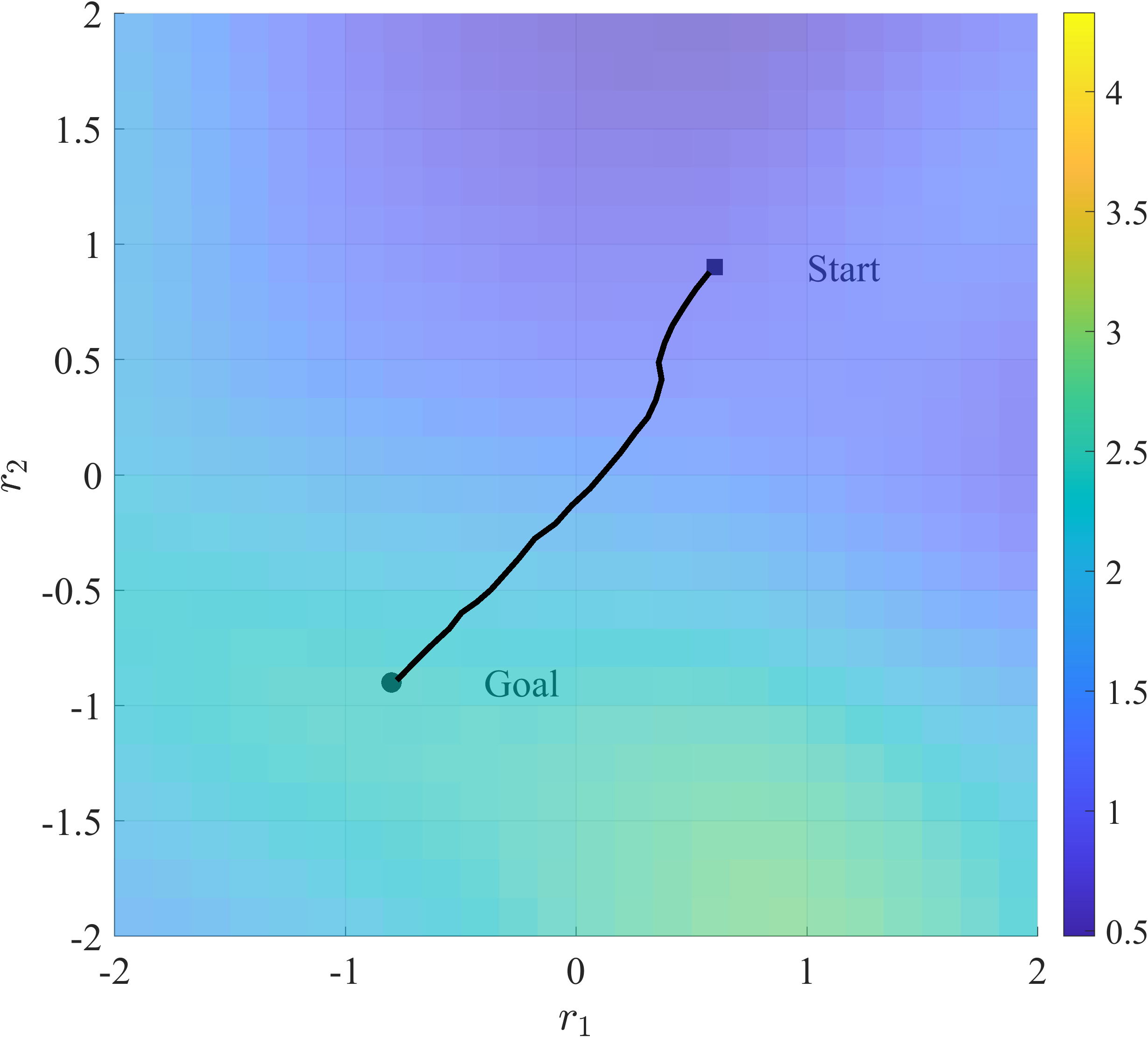}}
	\hspace{\thisfigspace}
	\subfigure{\includegraphics[width=\thisfigwidth]{\figpath/splitvae_minthreat_2_lambda_2}}
	\hspace{\thisfigspace}
	\subfigure{\includegraphics[width=\thisfigwidth]{\figpath/splitvae_minthreat_2_lambda_2}}
	\caption{Sample outputs of the Split-VAE for 200 training samples for $\lambda =2$.}
	\label{fig-splitvae-200_2}
\end{figure}

\def\thisfigwidth{0.23\columnwidth}
\def\thisfigspace{0.01\columnwidth}
\begin{figure}[h]
	\centering
	\subfigure{\includegraphics[width=\thisfigwidth]{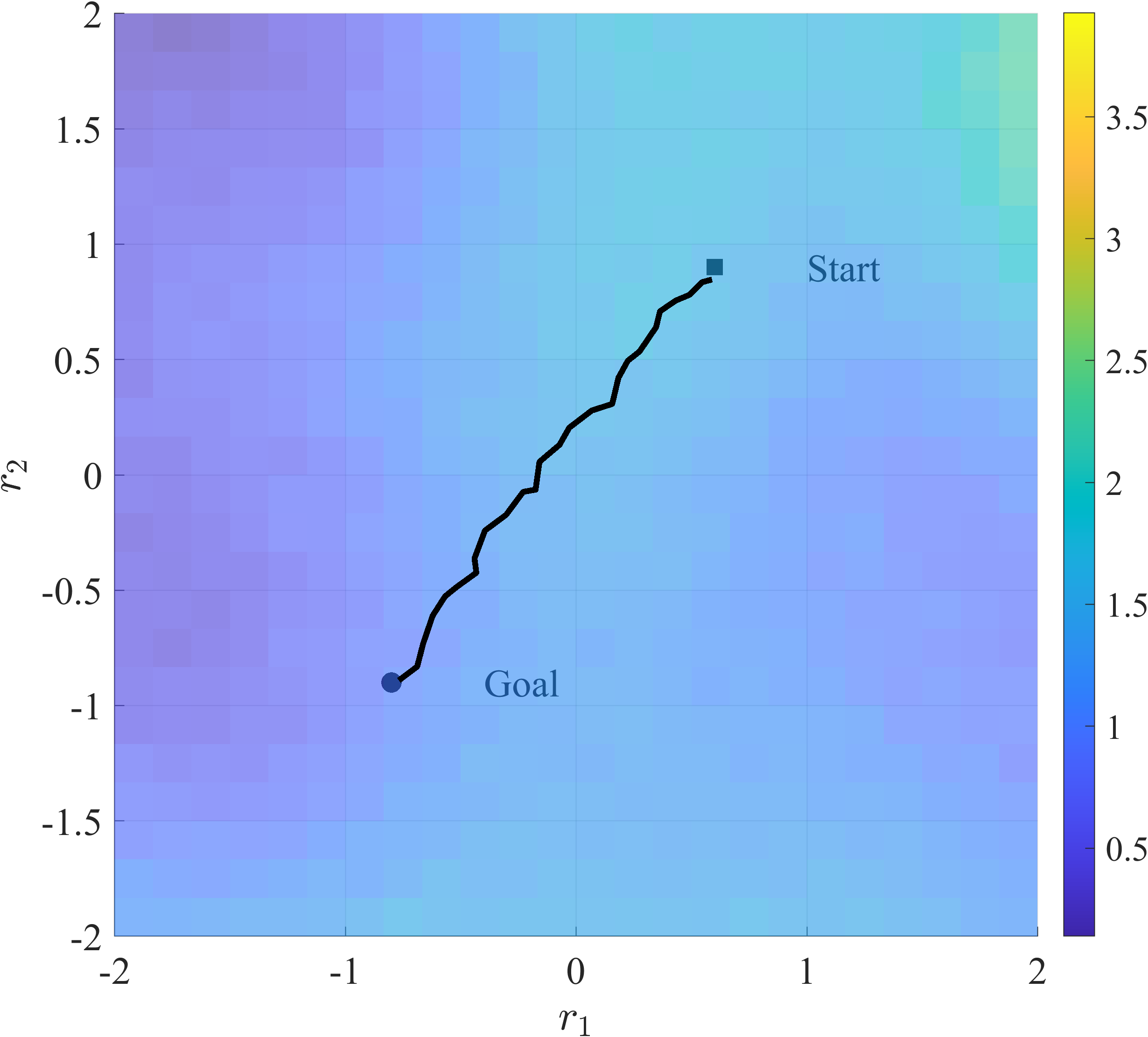}}
	\hspace{\thisfigspace}
	\subfigure{\includegraphics[width=\thisfigwidth]{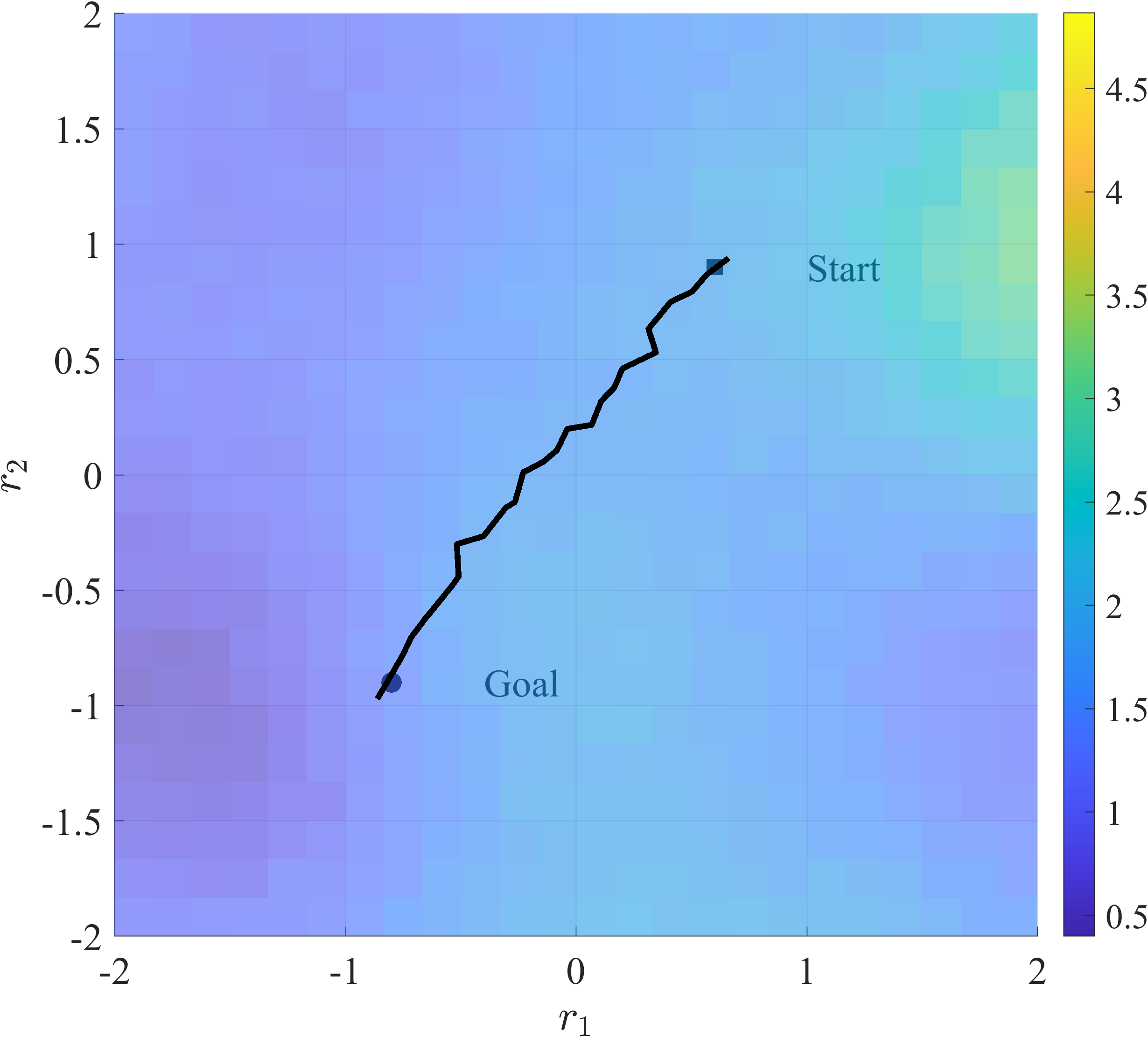}}
	\hspace{\thisfigspace}
	\subfigure{\includegraphics[width=\thisfigwidth]{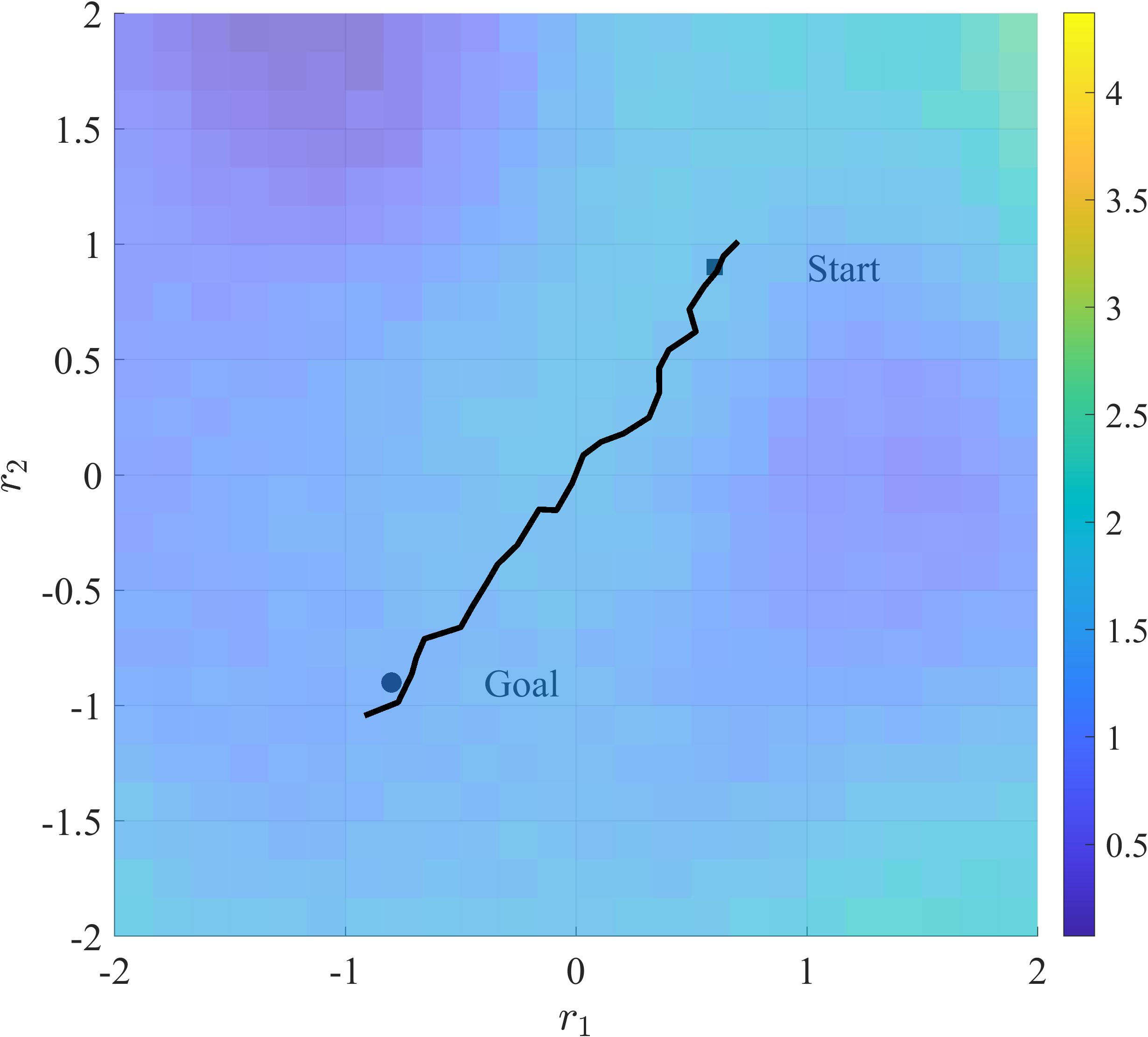}}
	\hspace{\thisfigspace}
	\subfigure{\includegraphics[width=\thisfigwidth]{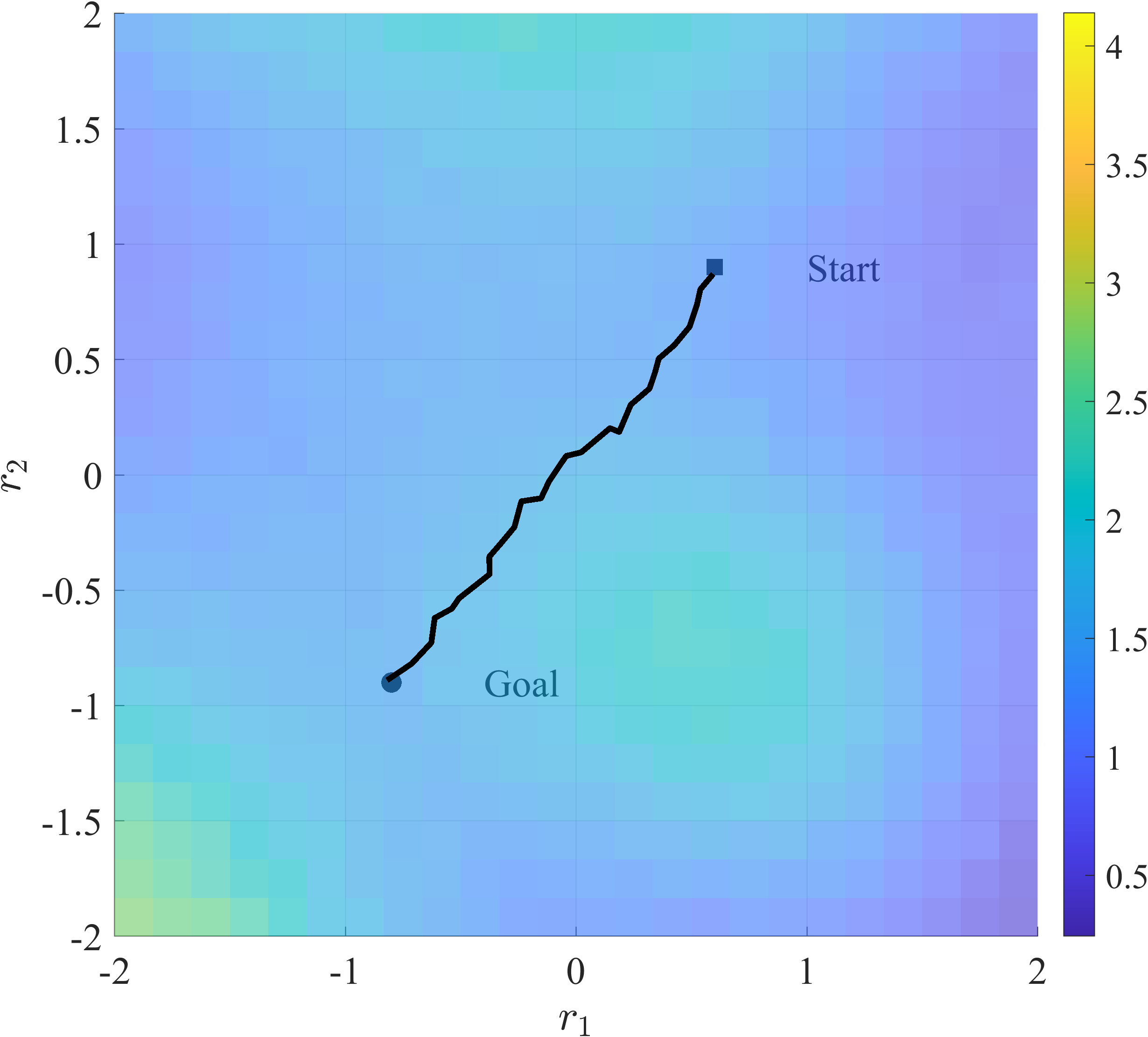}}
	\caption{Sample outputs of the S-VAE for 200 training samples for $\lambda =2$.}
	\label{fig-svae-200_2}
\end{figure}

\paragraph{Performance Indices:}
\tbl{tbl-minthreat-performance} provides a quantitative comparison of the results based on the 
performance metric $\simMet_{1}$. The table includes statistical measures such as the mean, 
variance, skewness, and kurtosis for each model's performance. 

A closer examination reveals that the Split-VAE consistently achieves more desirable values for the 
majority of these metrics, showcasing its superior ability to model the data. Specifically, the 
Split-VAE outperforms the S-VAE in terms of the minimum $\simMet_{1}$ value across all $\lambda$ 
values. This observation highlights the presence of high quality samples within the generated 
dataset.

\begin{table}
	\centering
	\caption{$\simMet_{1}$ performance 
		with $\nData = 200$ and $\nGen = 1000$ for the minimum threat 
		exposure problem.}
	\label{tbl-minthreat-performance}
	\begin{tabular}{l  rr  rr  rr}
		\toprule 
		& \multicolumn{2}{c}{$\lambda = 2$} 
		& \multicolumn{2}{c}{$\lambda = 5$} 
		& \multicolumn{2}{c}{$\lambda = 10$} \\
		& S-VAE & Split-VAE & S-VAE & Split-VAE & S-VAE & Split-VAE \\
		\midrule
		Mean
		& 1.911 & \textbf{1.415} & 7.364 & \textbf{6.303} & \textbf{5.891} & 6.466 \\
		Std. dev.
		& 2.595& \textbf{2.431}& \textbf{7.851} & 8.183 & \textbf{7.518} & 8.967 \\
		Maximum
		& \textbf{15.40} & 26.03 & \textbf{33.04} & 38.28 & 58.00 & \textbf{54.51}\\
		Minimum
		& 0.1090 & \textbf{0.0742} & 0.2516 & \textbf{0.1033} & 0.1783 & \textbf{0.0614}\\
		\bottomrule
	\end{tabular}
\end{table}

\paragraph{Statistical Similarity:}
From \tbl{tbl-vae-minthreat} the S-VAE and Split-VAE demonstrate successful
training by capturing the statistical properties of the three most dominant
features. \figf{fig-pca-vae-minthreat} provides a visualization of this result
for $\lambda = 5$, wherein the generated data aligns with the manifold of the
training dataset. The analysis was performed across all prescribed $\lambda$
values. The observations on the other $\lambda$ values were similar.

\begin{table}
	\centering
	\caption{Statistical moments of datasets generated by the VAE models for the minimum 
		threat problem for $\nGen = 1000$ and $\lambda = 5$.}
	\label{tbl-vae-minthreat}
	\renewcommand{\arraystretch}{1.3}
	\resizebox{\textwidth}{!}{%
		\begin{tabular}{l ccc ccc ccc ccc}
			\toprule
			\textbf{} & \multicolumn{3}{c}{\textbf{Mean}} & 
			\multicolumn{3}{c}{\textbf{Variance}} & 
			\multicolumn{3}{c}{\textbf{Skewness}} & 
			\multicolumn{3}{c}{\textbf{Kurtosis}} \\
			\cmidrule(lr){2-4} \cmidrule(lr){5-7} \cmidrule(lr){8-10} \cmidrule(lr){11-13}
			$\dataset$ & -92.12 & -2.294 & -0.3583 & 
			129.4 & 614.5 & 544.2 & 
			-0.0934 & 0.7989 & 0.0042 & 
			7.511 & 3.831 & 2.846 \\
			
			$\datasetGen$ (S-VAE) & -76.97 & \textbf{-0.2783} & \textbf{-0.3453} & 
			\textbf{48.00} & 168.9 & 117.6 & 
			0.1403 & \textbf{0.0462} & \textbf{0.0462} & 
			\textbf{3.060} & 2.427 & \textbf{2.656} \\
			
			$\datasetGen$ (Split-VAE) & \textbf{-86.73} & 0.4147 & 0.8289 & 
			43.74 & \textbf{374.5} & \textbf{292.9} & 
			\textbf{-0.2015} & -0.2899 & -0.0589 & 
			2.831 & \textbf{2.824} & 2.509 \\
			\bottomrule
		\end{tabular}
	}
\end{table}

\def\thisfigwidth{0.7\columnwidth}
\def\thisfigspace{0.01\columnwidth}
\begin{figure}
	\centering
	{\includegraphics[width=\thisfigwidth]{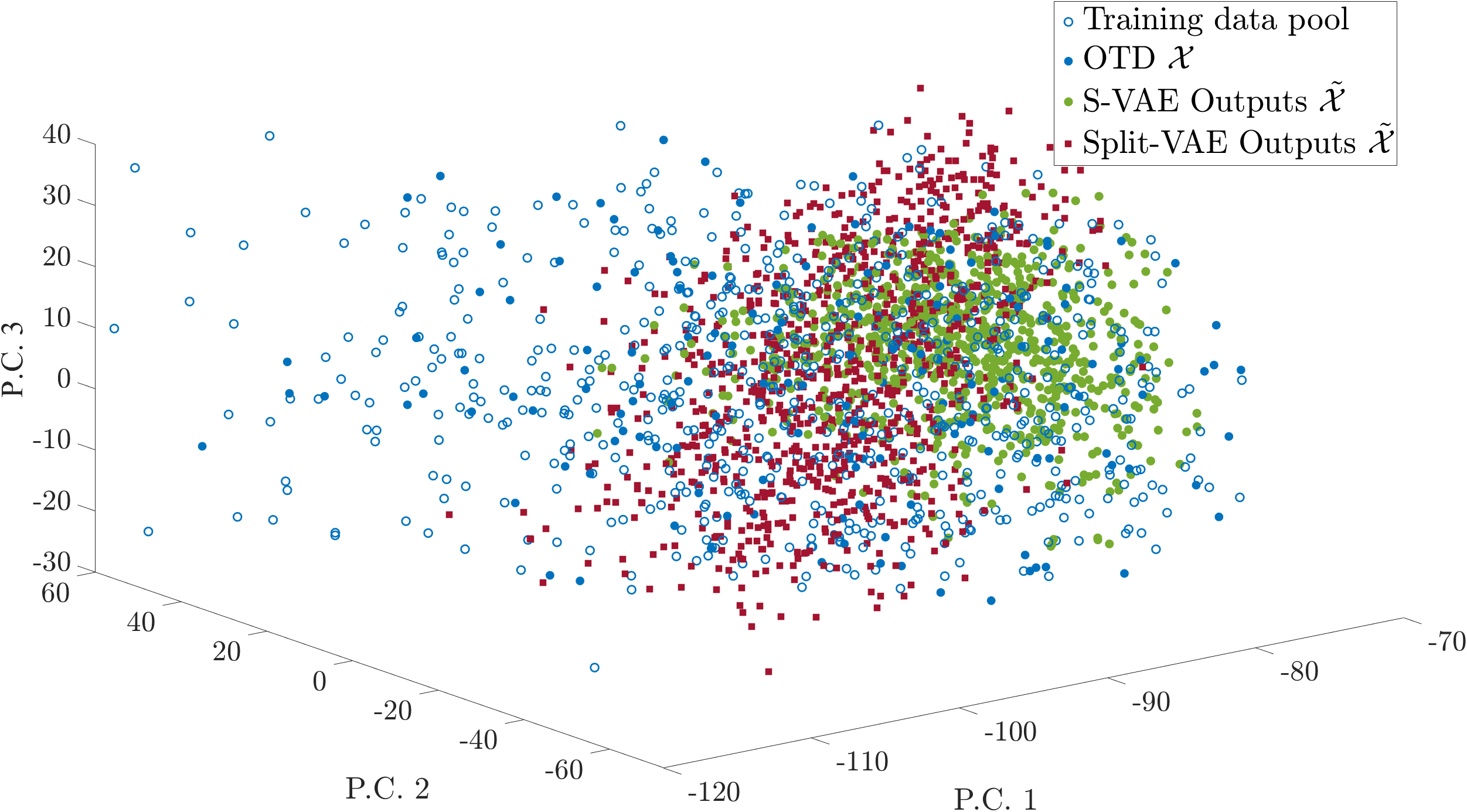}}
	\caption{Scatter plot of first three principal components (P.C.) of data points in the OTD and 
	generated datasets for S-VAE and Split-VAE with $\nData = 200$ and $\nGen = 1000$ for the 
	minimum threat problem.}
	\label{fig-pca-vae-minthreat}
\end{figure}

\subsection{\revnew{High-Dimensional Linear Time-Invariant (LTI) Systems}}
\revnew{To demonstrate the broader applicability of the Split-VAE architecture beyond
the minimum-time and minimum-threat problems, we consider the problem of synthesizing
trajectories of a family of linear time-invariant (LTI) dynamical systems. 
The governing equations are linear differential equations of the form}
\begin{equation}\label{lti-dyn-model}
	\revnew{\dot{\linstate} = A\linstate,}
\end{equation}  
\revnew{where $\linstate(t) \in \real[n]$ is the state and $A \in \real[n \times n].$ 
For these systems, we created OTDs by adding process noise, i.e.,
by solving equations of the form}
\begin{equation}\label{lti-dyn}
	\revnew{\dot{\linstate} = A\linstate + G\unoise}
\end{equation}  
\revnew{from various initial conditions. Here $G \in \real[n \times 1]$
is fixed, and $\unoise(t) \in \real$ is a noise process. Note that
the governing equation~\eqnnt{lti-dyn-model} involves no noise process
at all. Furthermore, to create OTDs $\dataset,$ we synthesized a noise
process such that $\unoise(t)$ is \emph{uniformly} distributed, unlike
standard control/estimation models where $\unoise(t)$ is assumed to
be normally distributed. The intention is to demonstrate that the Split-VAE
model can learn to synthesize data based on the distributions of
trajectories in the OTD, instead of making a priori assumptions about
the noise process. Furthermore, we considered high-dimensional state spaces,
namely, with $n= 10$ and $n= 100.$
}

\revnew{
We created three separate OTDs of size $\nData = 500$ each:
\begin{itemize}
	\item $\dataset_1$: $\linstate \in \mathbb{R}^{10}, 
	A \in \real[10 \times 10], G \in \real[10\times1]$
	\item $\dataset_2$: $\linstate  \in \mathbb{R}^{10}, 
	A \in \real[10 \times 10], G \in \real[10\times1]$
	\item $\dataset_3$: $\linstate  \in \mathbb{R}^{100}, 
	A \in \real[100 \times 100], G \in \real[100\times1]$
\end{itemize}
For each OTD, the $A$ and $G$ matrices were created randomly and fixed,
while ensuring that $A$ is Hurwitz, i.e., has all eigenvalues with
negative real parts.  The sequence length $T$ (refer to 
\scn{sec-problem-formulation}) was set to $1001,$ which leads to
$\nDimData = 10010$ for both $\dataset_1$ and $\dataset_2$, 
and $\nDimData = 100100$ for $\dataset_3$.
}

\revnew{For these systems, we trained the Split-VAE model and the S-VAE model
	for comparison. Recall that the S-VAE is trained only on data, and does not
	incorporate the governing equations.  Both models used a latent space of
	dimension 32, with the Split-VAE model partitioning this into two separate
	latent spaces of dimension 16 each. The rectified linear unit (ReLU) was used as
	the activation function throughout. Layer normalization was applied to the
	Split-VAE architecture. A learning rate of 0.001 was used for both models, and
	all hyperparameters were fixed (after tuning) across the three OTDs considered.
	The layer dimensions for both VAE architectures are provided in
	\tbl{tbl-vae-min-threat-dimensions} in the Appendix for OTD $\dataset_3$. The
	layer dimensions for $\dataset_1$ and $\dataset_2$ are the same as $\dataset_3$
	except that the input size changes to the corresponding feature size
	$\nDimData$. To train the Split-VAE, we generated \emph{model} (noiseless) 
	trajectories by solving \eqnnt{lti-dyn-model} from the initial states defined
	by the first states of each of the trajectories in the OTD.}

\revnew{To assess the similarity of the generated dataset $\datasetGen =
	\{\datum_{i}^{g}\}_{i=1}^{\nGen}$ to $\dataset,$ we considered: 1) similarity of
	the first four statistical moments and 2) a Noise-aware Dynamics Residual Ratio
	(NDRR) defined as $\text{NDRR} = \frac{\|\dot{\linstate} - A
		\linstate\|^2}{\|\dot{\linstate}\|^2}.$ Lower values of this index indicates
	better conformance (smaller violation) of the system
	dynamics~\eqnnt{lti-dyn-model}. }

\revnew{\paragraph{Statistical Similarity:} Based on the statistical moments
	shown in \tbl{tbl-vae-lti-all},
	the Split-VAE exhibits superior performance in terms of variance similarity, 
	indicating a
	broader and more representative spread of generated features. In contrast, the
	S-VAE achieves a slightly better score in terms of the mean similarity, suggesting
	alignment with the average feature values of the reference distribution. The
	skewness and kurtosis indices are comparable across both models, implying
	similar symmetry and tail behavior in the feature distributions.}

\revnew{A qualitative analysis of the feature distributions projected
	along the top three principal components shown in \figf{fig-pca-vae-lti}
	 reveals further
	differences. The S-VAE-generated samples appear tightly clustered, indicating a
	lack of diversity in the latent space traversal. For OTD $\dataset_2$,
	this behavior becomes even more pronounced, where S-VAE outputs are
	predominantly confined to a narrow linear manifold, highlighting poor
	generalization and an inability to capture the full variability present in the
	training data. In contrast, Split-VAE samples are dispersed similar to the OTD.}

\begin{table}
	\centering
	\caption{\revnew{Statistical moments of VAE-generated datasets for the LTI system
			with $\nGen = 1000$.}}
	\label{tbl-vae-lti-all}
	\renewcommand{\arraystretch}{1.3}
	\resizebox{\textwidth}{!}{%
		\begin{tabular}{l ccc ccc ccc ccc}
			\toprule
			\textbf{} & \multicolumn{3}{c}{\textbf{Mean}} &
			\multicolumn{3}{c}{\textbf{Variance}} &
			\multicolumn{3}{c}{\textbf{Skewness}} &
			\multicolumn{3}{c}{\textbf{Kurtosis}} \\
			\cmidrule(lr){2-4} \cmidrule(lr){5-7} \cmidrule(lr){8-10} \cmidrule(lr){11-13}
			\multicolumn{13}{c}{For the system with OTD $\dataset_1$} \\ \cmidrule(lr){2-13} 
			$\dataset_1$ & -33.39 & -3.419 & -0.0748 & 284.2\engE{3} & 6544 & 553.3 & 0.0447 & 
			0.0209 & 0.0223 & 2.577 & 2.843 & 3.144 \\
			
			$\datasetGen_1$ (S-VAE) & \textbf{-75.19} & \textbf{-0.2468} & \textbf{0.0555} & 161.8 
			& 3.153 & \textbf{0.5151} & -0.5646 & 0.5180 & \textbf{0.4738} & 3.31 & 3.218 & 
			\textbf{3.458} \\
			
			$\datasetGen_1$ (Split-VAE) & 152.5 & 1.840 & -21.20 & \textbf{167.4\engE{3}} & 
			\textbf{6355} 
			& 2443 & \textbf{-0.1162} & \textbf{-0.0502} & -1.314 & \textbf{2.642} & 
			\textbf{2.686} & 6.456 \\ \cmidrule(lr){2-13}
			\multicolumn{13}{c}{For the system with OTD $\dataset_2$} \\ \cmidrule(lr){2-13} 
			$\dataset_2$ & -79.18 & -7.957 & 0.1403 & 403.9\engE{3} & 10.61\engE{3} & 7062 & 
			-0.0537 & 0.0106 & 	0.0792 & 2.661 & 2.671 & 2.886 \\
			
			$\datasetGen_2$ (S-VAE) & \textbf{-44.17} & \textbf{-26.88} & -3.360 & 
			\textbf{3.740\engE{6}} 
			& \textbf{187.4} & \textbf{48.40} & \textbf{-0.0325} & -0.8366 & 1.399 & 
			\textbf{2.770} & \textbf{3.601} & 4.957 \\
			
			$\datasetGen_2$ (Split-VAE) & 814.0 & 25.91 & \textbf{2.381} & 4.759\engE{6} & 
			32.83\engE{3} & 30.62\engE{3} 
			& -0.3731 & \textbf{0.2966} & \textbf{-0.3565} & 2.385 & 3.722 & \textbf{4.498} \\ 
			\cmidrule(lr){2-13}
			\multicolumn{13}{c}{For the system with OTD $\dataset_3$} \\ \cmidrule(lr){2-13} 
			$\dataset_3$ & 6.088 & -22.17 & -17.11 & 16.95\engE{6} & 579.3\engE{3} & 280.3\engE{3} 
			& 0.0210 &	-0.0659 & 0.0131 & 2.676 & 2.658 & 2.821 \\
			
			$\datasetGen_3$ (S-VAE) & \textbf{11.18} & \textbf{7.897} & \textbf{-0.0044} & 142.8 
			& 54.78 & 0.3941 & \textbf{0.2047} & -0.2188 & 0.4109 & 3.766 & 3.743 & 3.119 \\
			
			$\datasetGen_3$ (Split-VAE) & -987.9 & 71.55 & -67.71 & \textbf{6.851\engE{6}} & 
			\textbf{440.2\engE{3}} & \textbf{422.7\engE{3}} & 0.2203 & \textbf{0.0077} & 
			\textbf{-0.0258} & \textbf{2.689} & \textbf{2.611} & \textbf{2.548} \\
			\bottomrule
		\end{tabular}
	}
\end{table}

\revnew{\paragraph{Noise-aware Dynamics Residual Ratio (NDRR):} As shown in
	\tbl{tbl-lti-NDRR}, the Split-VAE consistently outperforms the S-VAE across all
	three (OTDs). Specifically, the additive noise present in the samples generated
	by the Split-VAE is closer to the total noise levels observed in the
	corresponding OTDs, indicating effective modeling of the inherent stochasticity
	in the system. 
	This suggests that the Split-VAE replicate the noise characteristics of the OTD,
	contributing to more realistic and diverse sample generation.}

\revnew{In summary, for this problem of dynamical systems high-dimensional state spaces,
	and small volume of training data ($\nData = 500$), the 
	purely data-driven S-VAE fails to generate datasets with statistical similarity
	to the OTD, whereas the proposed Split-VAE succeeds.}

\begin{table}[htbp]
	\centering
	\caption{NDRR with $\nData = 500$ and $\nGen = 1000$ for the LTI system.}
	\label{tbl-lti-NDRR}
	\renewcommand{\arraystretch}{1.3}
	\begin{tabular}{l r r r}
		\toprule 
		& \multicolumn{1}{c}{$\dataset_1$} 
		& \multicolumn{1}{c}{$\dataset_2$} 
		& \multicolumn{1}{c}{$\dataset_3$} \\
		\midrule
		OTD & 88.00 & 88.22 & 87.33 \\
		Split-VAE & \textbf{92.04} & \textbf{138.3} & \textbf{146.3} \\
		S-VAE & 110.8 & 148.8 & 257.6 \\
		\bottomrule
	\end{tabular}
\end{table}

\def\thisfigwidth{0.8\columnwidth}
\def\thisfigspace{0.0\columnwidth}

\begin{figure}
	\centering
	\subfigure[For the system with OTD $\dataset_1.$]{
		\includegraphics[width=\thisfigwidth]{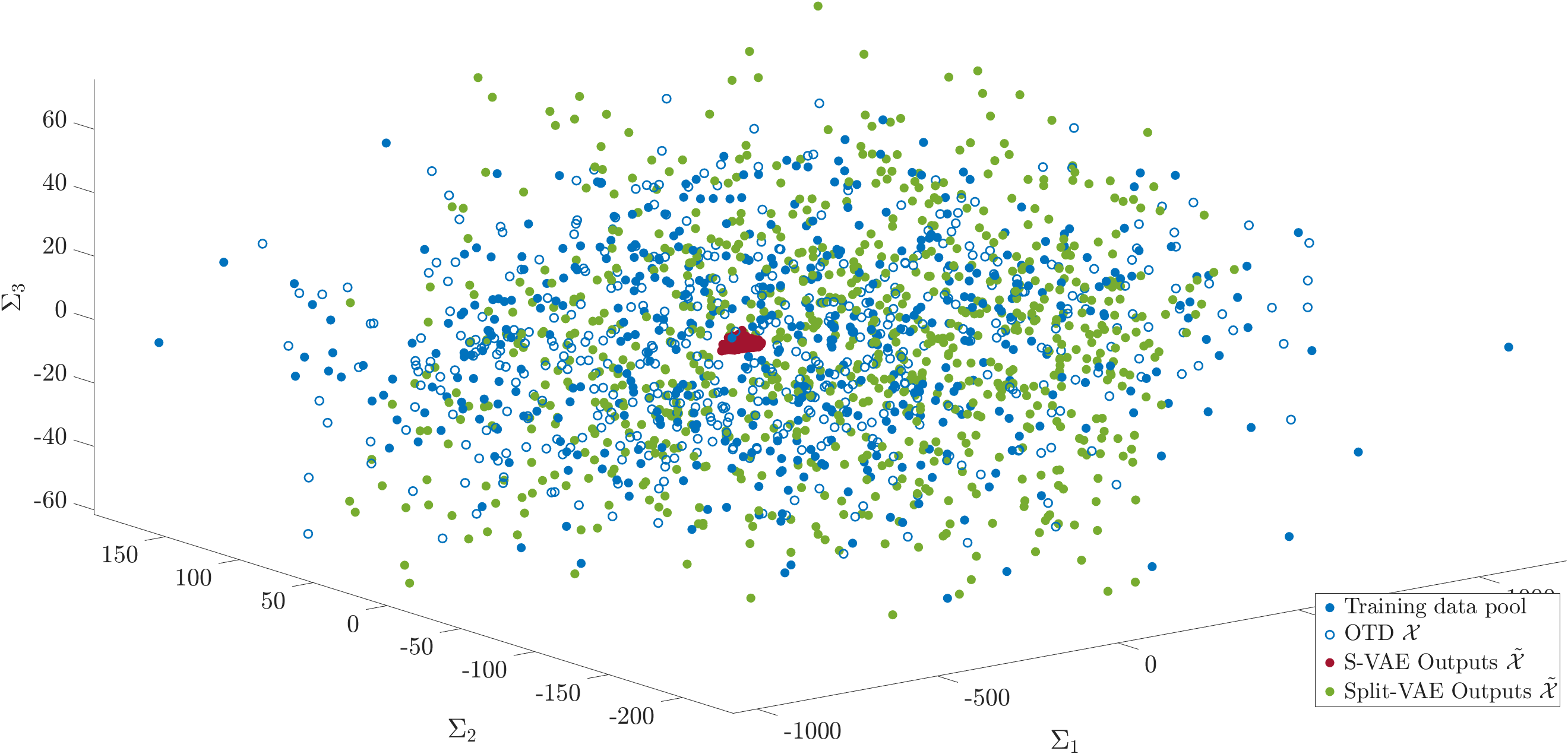}
	}
	\subfigure[For the system with OTD $\dataset_2.$]{
		\includegraphics[width=\thisfigwidth]{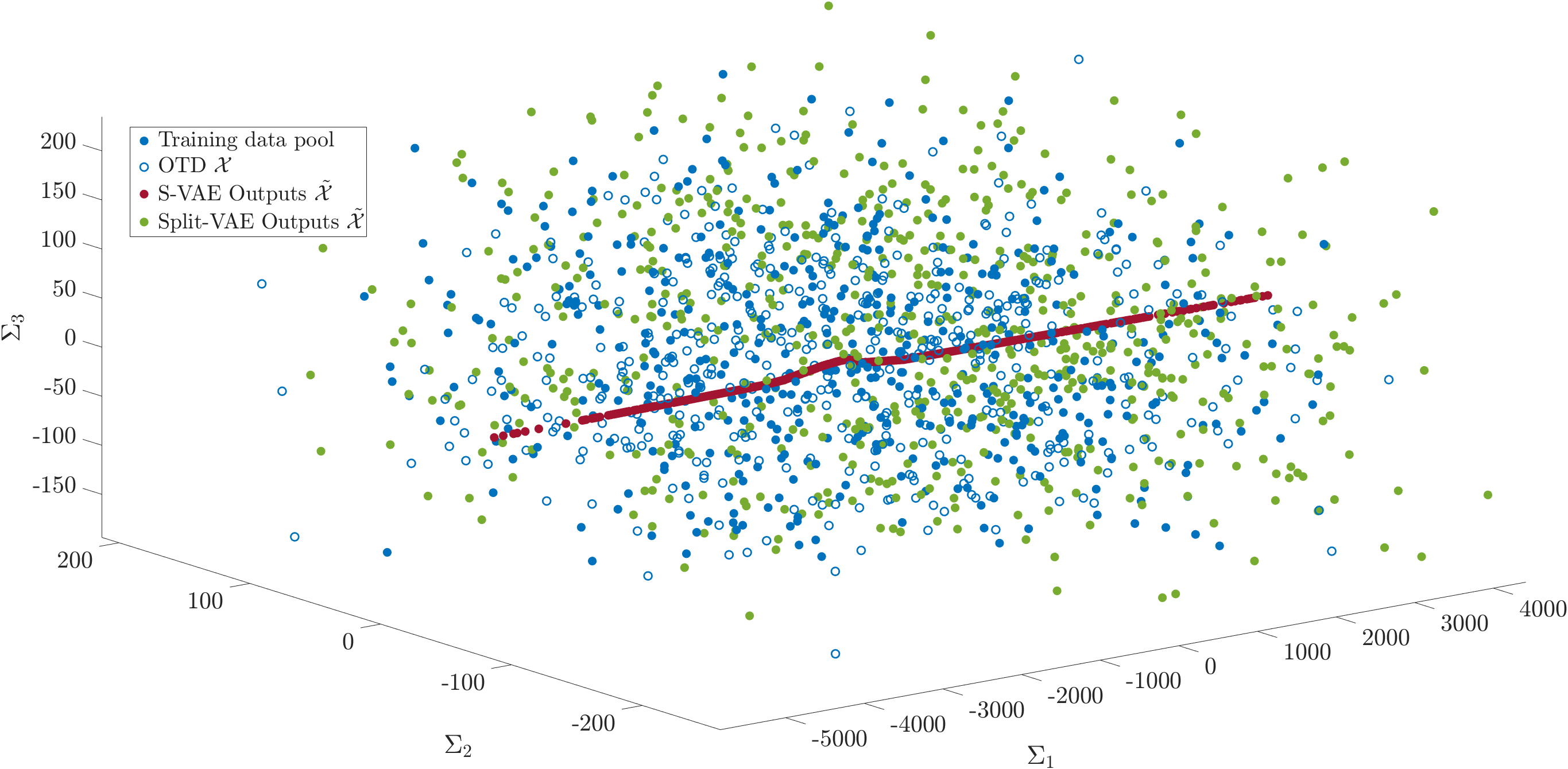}
	}
	\subfigure[For the system with OTD $\dataset_3.$]{
		\includegraphics[width=\thisfigwidth]{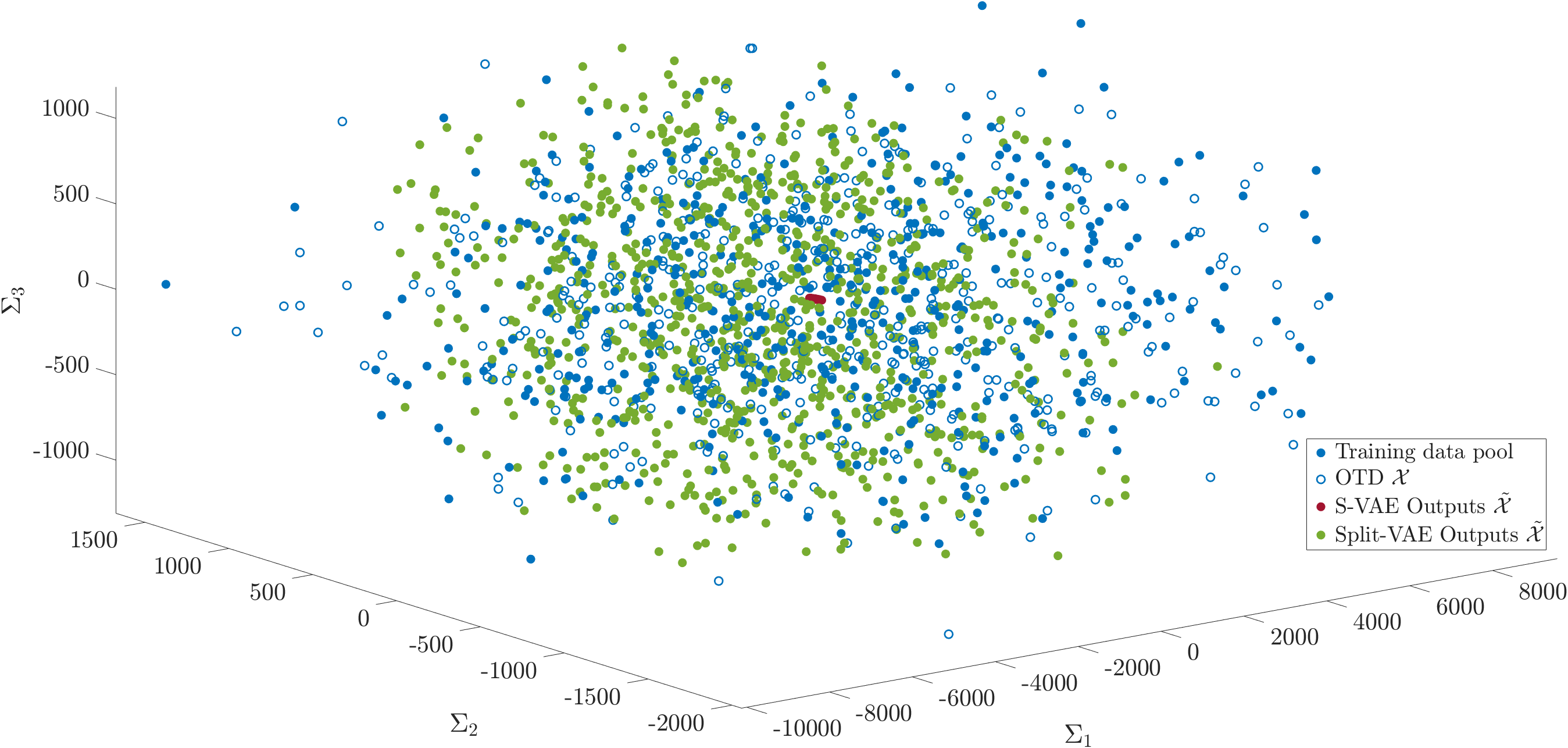}
	}
	\caption{Scatter plots of first three principal components of data points in the OTD and 
		generated datasets for S-VAE and Split-VAE with $\nData = 500$ and $\nGen = 1000$ for the 
		LTI dynamical systems.}
	\label{fig-pca-vae-lti}
\end{figure}

\subsection{\revnew{Summary of Findings}}

\revnew{We developed and studied the following generative models: GAN, Z-GAN1, Z-GAN2,
S-VAE, Z-VAE, and Split-VAE. We evaluated these models across three problems:
the Zermelo minimum-time navigation problem, the minimum threat exposure
problem, and a high-dimensional linear time-invariant (LTI) system. For the
Zermelo navigation problem, we employed the GAN, Z-GAN1, Z-GAN2, S-VAE, and
Z-VAE models. For the minimum threat exposure problem, we used S-VAE and
Split-VAE, while the high-dimensional LTI problem was analyzed using S-VAE and
Split-VAE as well. In all cases, the newly proposed models showed improved
performance over their baselines (S-GAN and S-VAE).}

\revnew{The Z-GAN1 and Z-GAN2 models introduced the use of physical constraints during
training. However, due to training instabilities, their performance was not
satisfactory across all evaluation metrics. More precisely, we observed mode
collapse during the training of these GAN models. This phenomenon is a
well-known challenge associated with the instability of GAN
training~\cite{ahmad2024understanding}. A critical factor is the need to
maintain a balance between the learning dynamics of the generator and the
discriminator. When the discriminator becomes too dominant—typically by learning
faster than the generator—the generator tends to produce a limited set of
outputs that can satisfactorily fool the discriminator, rather than capturing
the diversity of the underlying data distribution. This behavior is particularly
common in low-data regimes~\cite{karras2020training}. To address this, we
transitioned to a VAE-based architecture~\cite{park2020began}, leading to the
development of Z-VAE, which can be viewed as the VAE analog of the Z-GANs. On
the Zermelo problem, both VAE-based models outperformed the GAN-based models by
not succumbing to mode collapse, with Z-VAE surpassing S-VAE by successfully
integrating physics-based constraints. This demonstrated that incorporating
problem-specific physical knowledge can significantly improve learning outcomes.}

\revnew{For the minimum threat exposure problem, we assumed the true system dynamics
were partially unknown due to unknown parameters in the \emph{objective
function}. In this problem. the Split-VAE architecture led to better
generalization and improved performance over S-VAE. It is important to note that
both the model and observed data were optimal for their respective cost
structures, and the mismatch was treated as unknown or unmodeled dynamics.}

\revnew{A similar analysis was performed on a dataset derived from a high-dimensional
LTI system perturbed by additive noise, which was interpreted as representing
unmodeled dynamics. In this case, too, the Split-VAE consistently outperformed
S-VAE, indicating that separating clean and noisy components in the latent space
improves robustness to such perturbations.}

\revnew{
In summary, our results suggest that when governing equations or physical
constraints are known, incorporating them into training can enhance performance.
When full knowledge of the dynamics is unavailable, approximate models can still
be effectively leveraged. However, simply adding a governing equation residual term
the training loss functions may not suffice. An architectural change,
such as the proposed Split-VAE architecture, is needed to improve
performance on noisy or real-world data. These findings support the broader
conclusion that utilizing either physical constraints or approximate models,
even if imperfect, can guide learning and improve robustness in data-scarce or
noise-dominated settings.}

%

\section{Conclusion}\label{sec-con}

We studied generative artificial neural network models for two
optimally controlled systems, namely, minimum-time and minimum-threat
navigation. For these systems, we developed new GAN and VAE
architectures that incorporated the governing equations --
specifically, necessary conditions derived from variational optimal
control theory -- into their training. In the GAN architecture, these
equations were incorporated as an additional discriminator. In the
VAE architecture, these equations were used to produce ideal
trajectories mapped to one subspace of the latent vector space. We
compared our models against standard, i.e., purely data-driven,
variants of these architectures. We were unable to resolve mode
collapse issues with the GAN models, and neither our proposed GAN
model nor the standard variant provided satisfactory generative
performance. However, our proposed VAE models significantly
outperformed the standard VAE models for both systems. Specifically,
we found that, for a fixed large volume of training data, our proposed
VAE models always outperformed the standard VAE models in terms of
statistical similarity and satisfaction of the governing equations,
both. Furthermore, for small volumes of training data, our proposed
models provided satisfactory generative performance, whereas the
standard VAE models were unable to do so.

\section*{Funding Sources}

This research was sponsored by the DEVCOM Analysis Center and was accomplished under 
Cooperative Agreement Number W911NF-22-2-0001. The views and conclusions contained 
in this document are those of the authors and should not be interpreted as representing
the official policies, either expressed or implied, of the Army Research Office 
or the U.S. Government. The U.S. Government is authorized to reproduce and distribute
reprints for Government purposes notwithstanding any copyright notation herein.

\appendix
\section*{Appendix}

\begin{table}[h]
	\caption{Layer dimensions for GAN models.}
	\label{tbl-gan-dimensions}	
	\centering
	\begin{tabular}{r r r r r r r r r r r}
		\toprule
		& Input & H1 & H2 & H3 & H4 & H5 & H6 & H7 & H8 & Output \\
		\midrule
		$\gen$ & 20 & 64 & 100 & 225 & 324 & 400 & 441 & 625 & 900 & 175 \\
		$\dsc$ & 50 & 900 & 625 & 441 & 400 & 324 & 225 & 100 & 25 & 1 \\ 
		\bottomrule
	\end{tabular}
\end{table}

\begin{table}[h]
	\centering
	\caption{Layer dimensions for the S-VAE and Z-VAE models.}
	\label{tbl-layer-dimensions}
	\begin{tabular}{r r r r r r r r r r r}
		\toprule[1pt]
		& Input & H1 & H2 & H3 & H4 & H5 &H6 & Output \\
		\midrule[1pt]
		S-VAE $\enc$ & 400 & 324 & 225 & 196 & 125 & 100 & 81 & 32  \\
		$\dec$& 32 & 81 & 100 & 125 & 196 & 225 & 324 & 400 \\ 
		\midrule[1pt]
		Z-VAE $\enc$ & 400 & 225 & 196 & 125 & 100 & 81 & & 32  \\
		$\dec$ & 32 & 81 & 100 & 125 & 196 & 225 & & 400  \\ 
		\bottomrule[1pt]
	\end{tabular}
\end{table}

\begin{table}[h]
	\centering
	\caption{Layer dimensions for the S-VAE and Split-VAE models for the minimum threat 
	problem.}
	\label{tbl-vae-min-threat-dimensions}
	\begin{tabular}{r r r r r r r r}
		\toprule[1pt]
		& Input & H1 & H2 & H3 & H4 & H5 & Output \\
		\midrule[1pt]
		S-VAE $\enc$ & 2057 & 225 & 196 & 125 & 100 & 81 & 64  \\
		$\dec$ & 64 & 81 & 100 & 125 & 196 & 225 & 2057 \\ 
		\midrule[1pt]
		Split-VAE $\enc$ & 2057 & 625 & 400 & 225 & & &20,20   \\
		$\dec$ & 20,20 & 225 & 400 & 625 & & &2057 \\ 
		\bottomrule[1pt]
	\end{tabular}
\end{table}

\def\thisfigwidth{0.23\columnwidth}
\def\thisfigspace{0.01\columnwidth}
\begin{figure}
	\centering
	\subfigure{\includegraphics[width=\thisfigwidth]{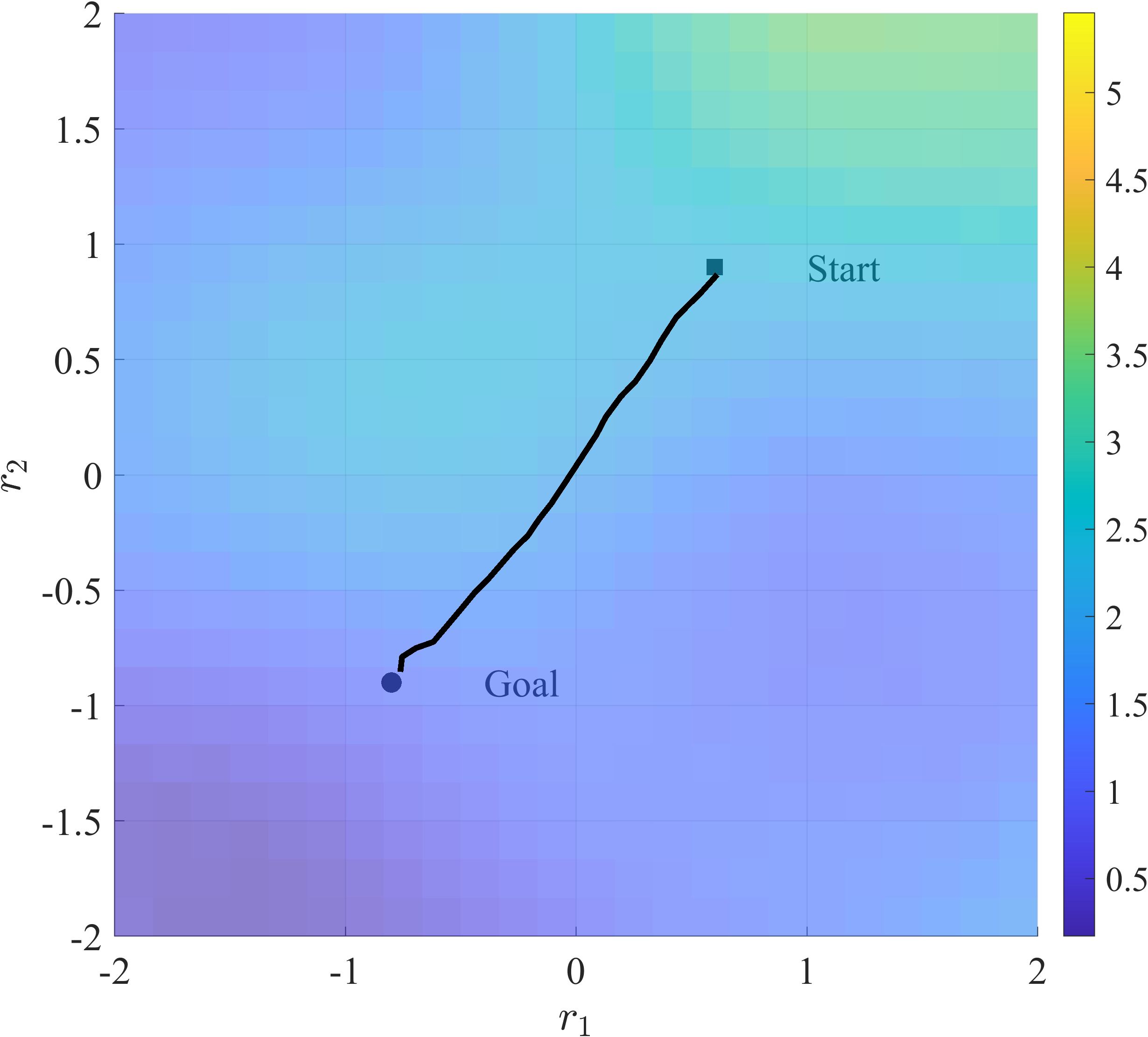}}
	\hspace{\thisfigspace}
	\subfigure{\includegraphics[width=\thisfigwidth]{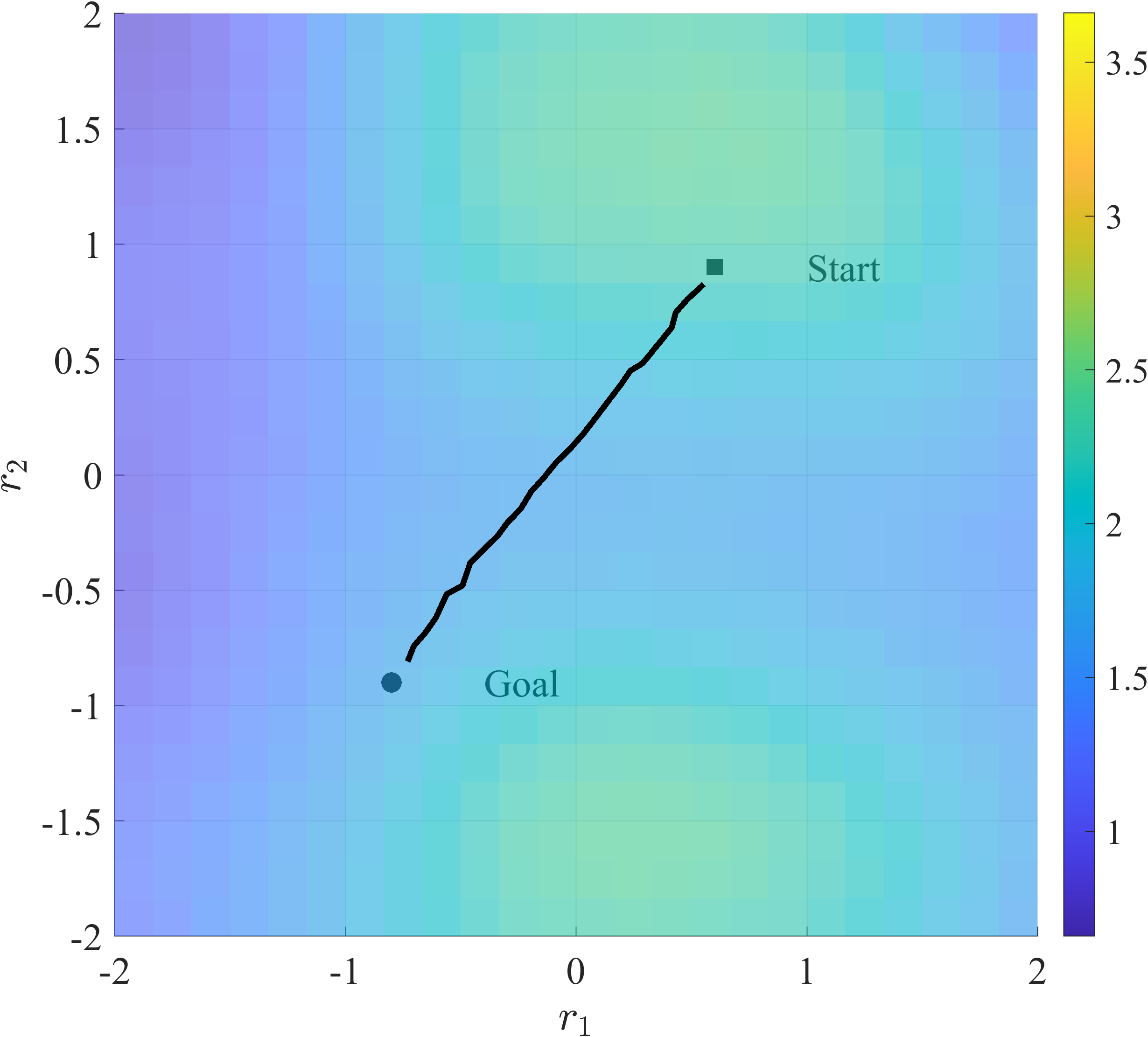}}
	\hspace{\thisfigspace}
	\subfigure{\includegraphics[width=\thisfigwidth]{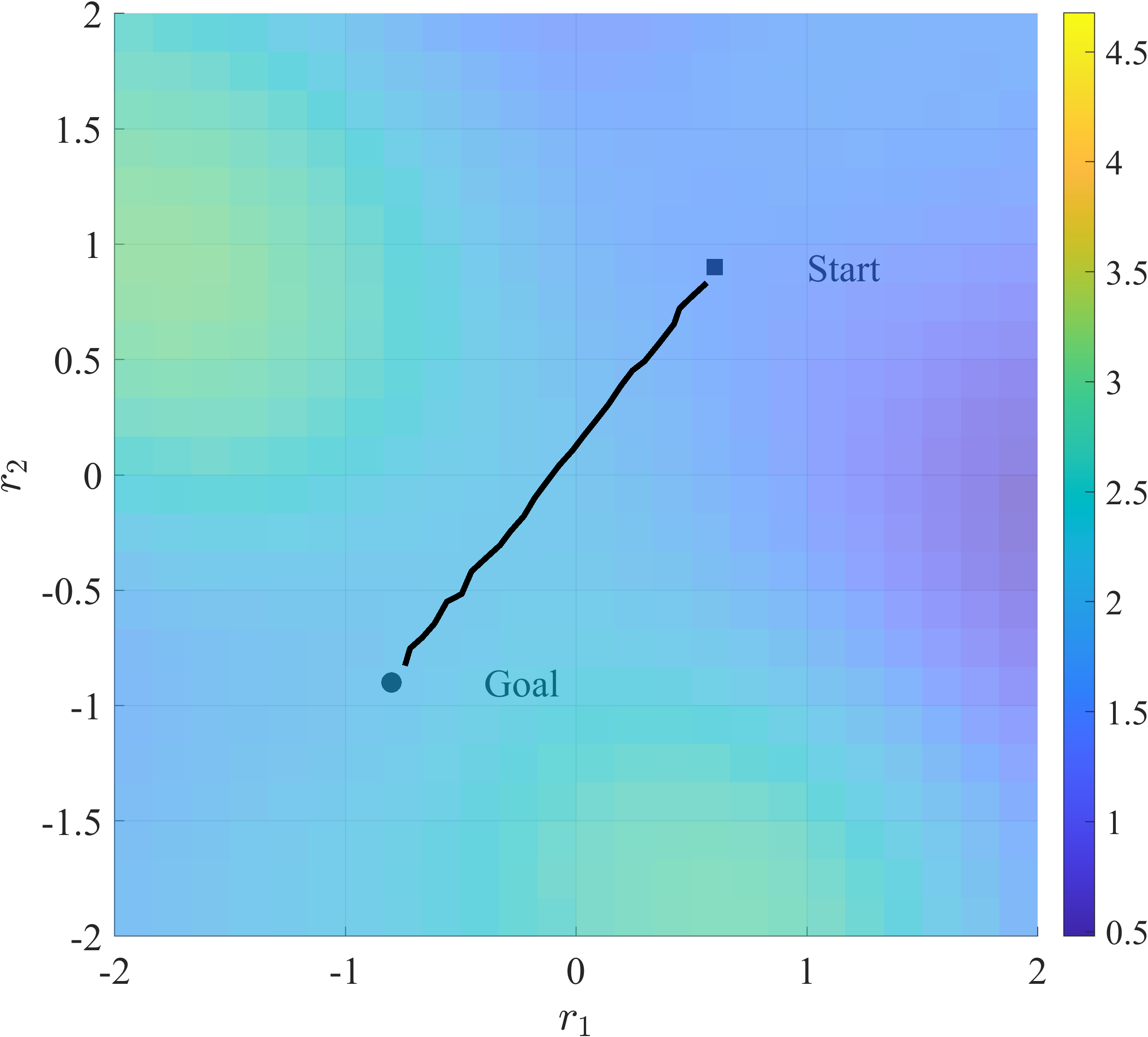}}
	\hspace{\thisfigspace}
	\subfigure{\includegraphics[width=\thisfigwidth]{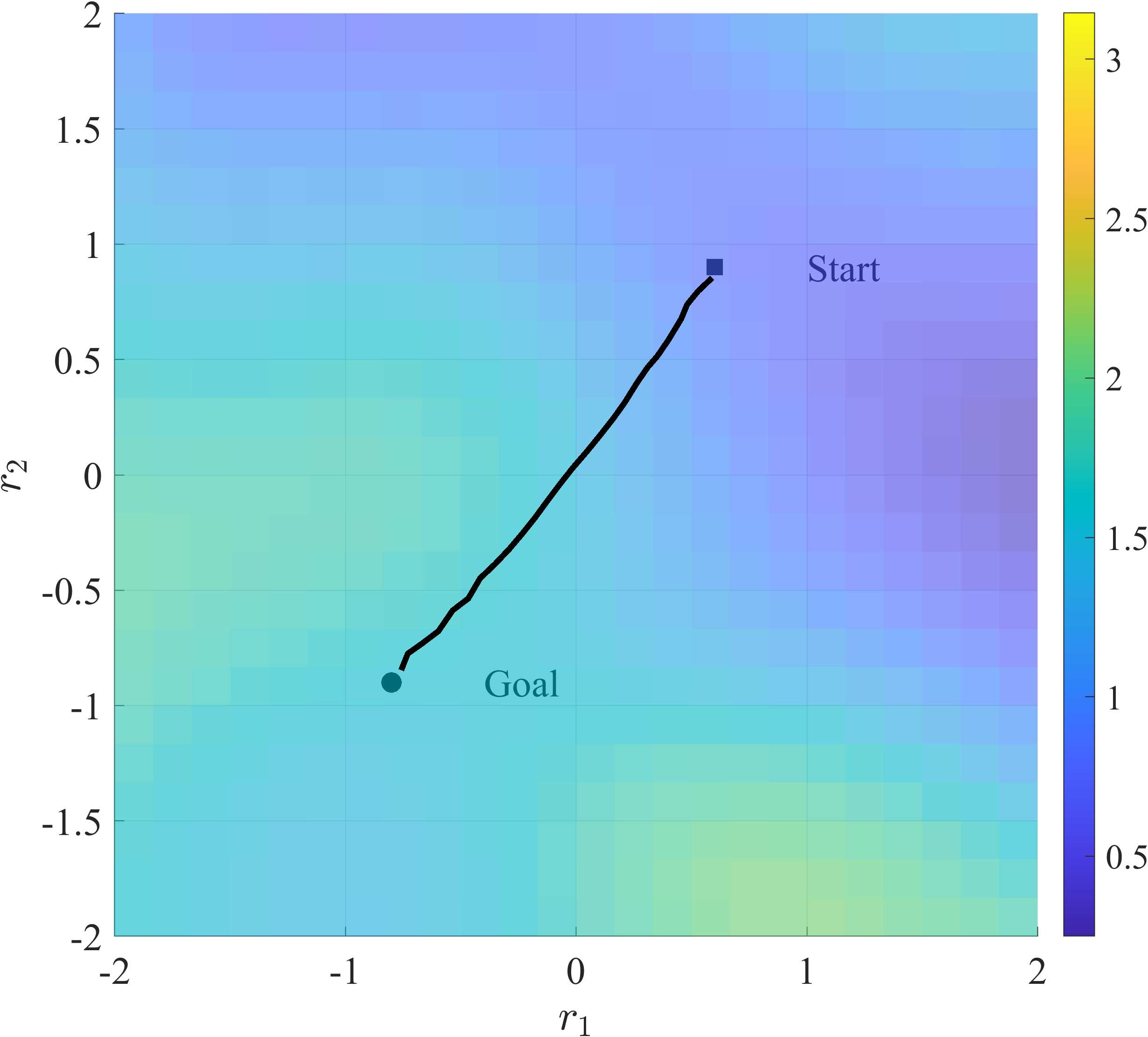}}
	\caption{Sample outputs of the Split-VAE for 200 training samples for $\lambda =5$.}
	\label{fig-splitvae-200_5}
\end{figure}


\def\thisfigwidth{0.23\columnwidth}
\def\thisfigspace{0.01\columnwidth}
\begin{figure}
	\centering
	\subfigure{\includegraphics[width=\thisfigwidth]{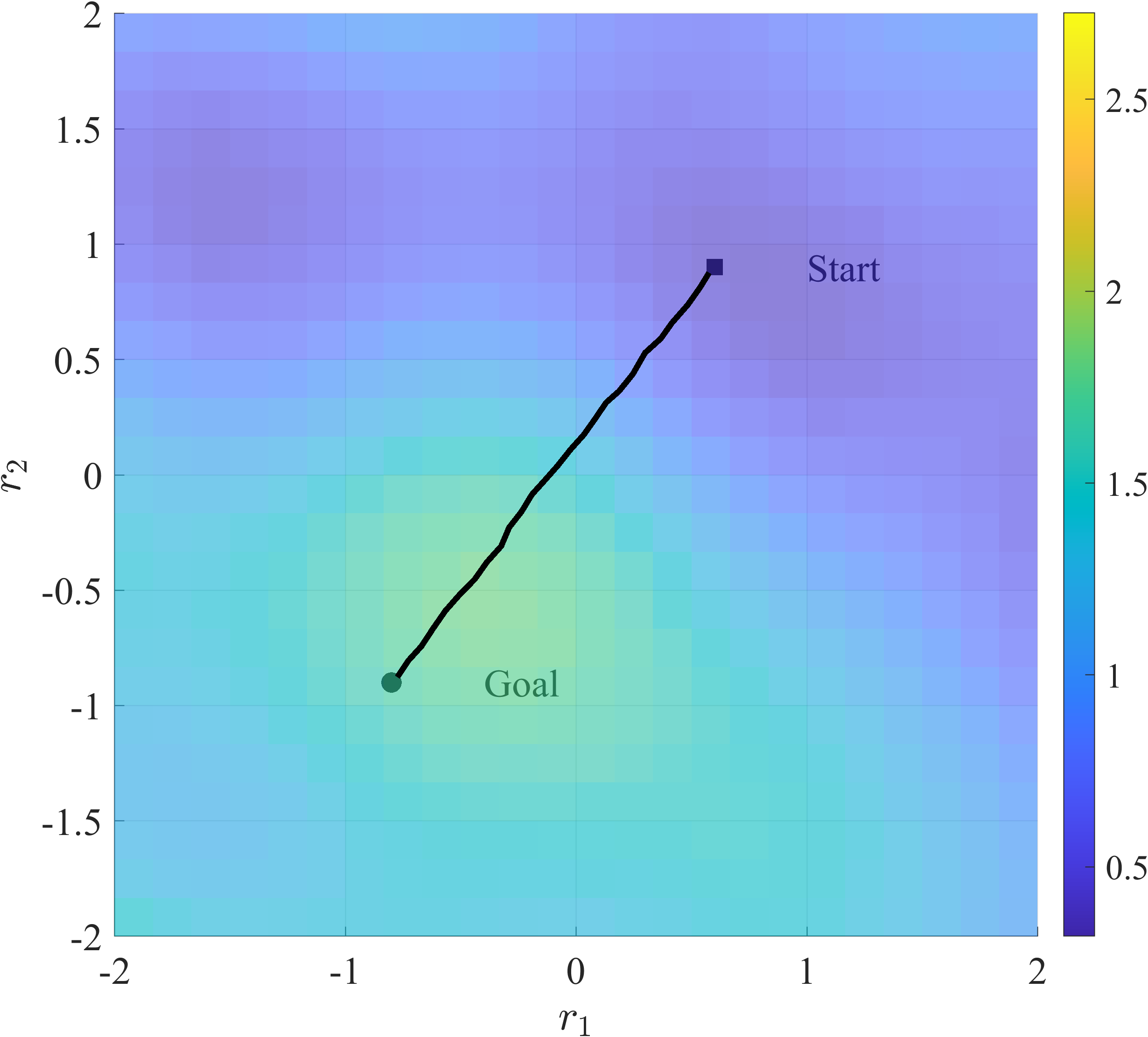}}
	\hspace{\thisfigspace}
	\subfigure{\includegraphics[width=\thisfigwidth]{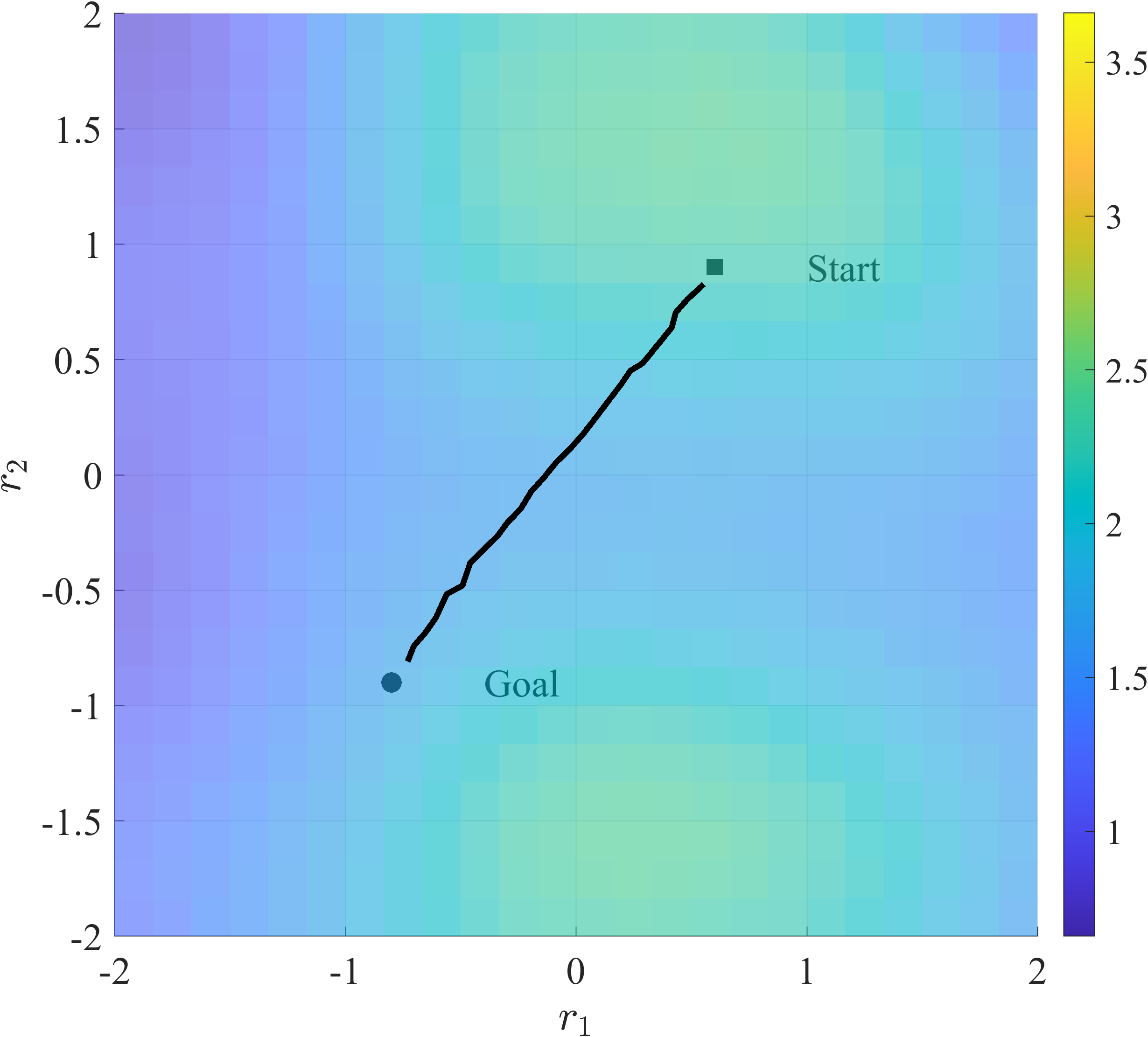}}
	\hspace{\thisfigspace}
	\subfigure{\includegraphics[width=\thisfigwidth]{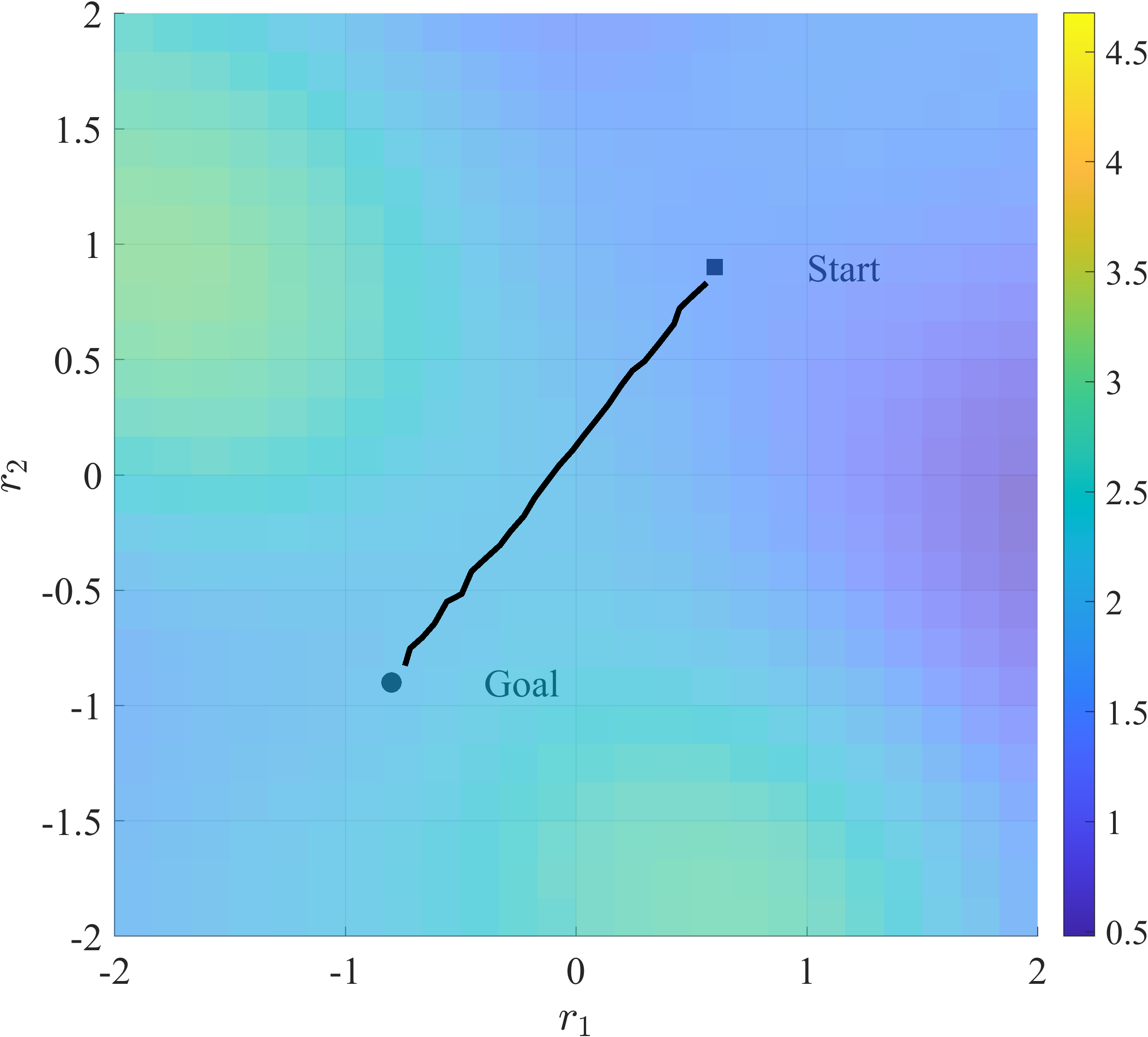}}
	\hspace{\thisfigspace}
	\subfigure{\includegraphics[width=\thisfigwidth]{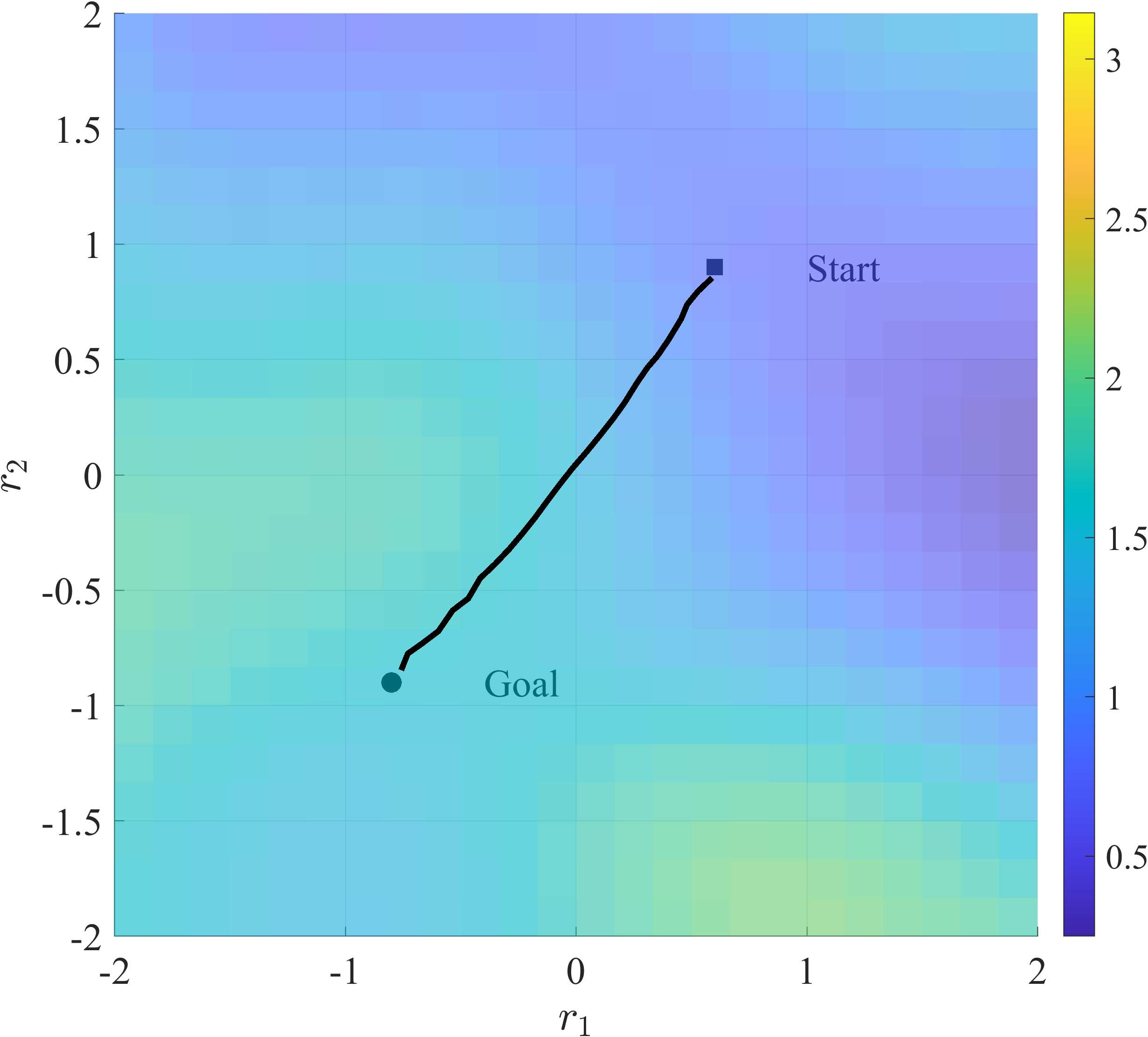}}
	\caption{Sample outputs of the Split-VAE for 200 training samples for $\lambda =10$.}
	\label{fig-splitvae-200_10}
\end{figure}



\bibliographystyle{ksfh_nat}
\bibliography{References}

\end{document}